\documentclass[traditabstract,longauth]{aa}  

\usepackage{graphicx}
\usepackage{txfonts}
\usepackage{multirow}
\usepackage{hyperref}
\hypersetup{colorlinks,citecolor=blue}
\usepackage{amstext}

\begin{document}

   \title{The miniJPAS survey quasar selection III: Classification with artificial neural networks and hybridisation}
   \titlerunning{Classification with artificial neural networks and hybridisation in the AEGIS field}

    \author{G. Mart\'inez-Solaeche \inst{\ref{1}}
    \and Carolina Queiroz \inst{\ref{2},\ref{3}} \and
    R. M. Gonz\'alez Delgado  \inst{\ref{1}} \and
    Nat\'alia V. N. Rodrigues \inst{\ref{3}} \and
    \and R. Garc\'ia-Benito \inst{\ref{1}} \and
    Ignasi P\'erez-R\`afols \inst{\ref{4},\ref{5}} \and
    L. Raul Abramo \inst{\ref{3}} \and 
    Luis D\'iaz-Garc\'ia \inst{\ref{1}} \and
    Matthew M. Pieri \inst{\ref{6}} \and
    Jon\'as Chaves-Montero \inst{\ref{7}} \and
    A. Hern\'an-Caballero \inst{\ref{8}} \and
    J. E. Rodr\'iguez-Mart\'in \inst{\ref{1}} \and
    Silvia Bonoli \inst{\ref{7},\ref{10}} \and
    Sean S. Morrison \inst{\ref{6},\ref{10}} \and
    Isabel M\'arquez \inst{\ref{1}} \and
    J. M.~V\'ilchez  \inst{\ref{1}} \and
    C.~L\'opez-Sanjuan  \inst{\ref{9}} \and
    A.~J.~Cenarro  \inst{\ref{9}} \and
    R.~A.~Dupke  \inst{\ref{12},\ref{13},\ref{14}} \and
    A.~Mar\'in-Franch \inst{\ref{9}} \and
    J.~Varela \inst{\ref{9}} \and
    H.~V\'azquez~Rami\'o \inst{\ref{9}} \and
    D.~Crist\'obal-Hornillos \inst{\ref{8}} \and
    M.~Moles \inst{\ref{8},\ref{1}} \and
    J.~Alcaniz \inst{\ref{12}} \and
    N.~Benitez \inst{\ref{1}} \and
    J.A. Fernández-Ontiveros \inst{\ref{8}} \and
    A.~Ederoclite \inst{\ref{8}} \and
    V.~Marra \inst{\ref{15},\ref{16}} \and
    C.~Mendes~de~Oliveira \inst{\ref{17}} \and
    K.~Taylor \inst{\ref{18}}}

    \institute{Instituto de Astrof\'isica de Andaluci\'a (CSIC), PO Box 3004, 18080 Granada, Spain (\email{gimarso@iaa.es}) \label{1} \and 
    Departamento de Astronomia, Instituto de Física, Universidade Federal do Rio Grande do Sul (UFRGS), Av. Bento Gonçalves, 9500, Porto Alegre, RS, Brazil \label{2} \and
    Departamento de Física Matemática, Instituto de Física, Universidade de São Paulo, Rua do Matão, 1371, CEP 05508-090, São Paulo, Brazil \label{3} \and
    Sorbonne Université, Université Paris Diderot, CNRS/IN2P3, Laboratoire de Physique Nucléaire et de Hautes Energies, LPNHE, 4 Place Jussieu, F-75252 Paris, France \label{4} \and
    Institut de Física d’Altes Energies (IFAE), The Barcelona Institute of Science and Technology, 08193 Bellaterra (Barcelona), Spain \label{5} \and
    Aix Marseille Univ, CNRS, CNES, LAM, Marseille, France \label{6} \and
    Donostia International Physics Center, Paseo Manuel de  Lardizabal 4, E-20018 \label{7} \and
    Centro de Estudios de F\'isica del Cosmos de Arag\'on (CEFCA), Plaza San Juan, 1, 44001 Teruel, Spain \label{8} \and
    Centro de Estudios de F\'isica del Cosmos de Arag\'on (CEFCA), Unidad Asociada al CSIC, Plaza San Juan, 1, E-44001 Teruel, Spain \label{9} \and
    Donostia-San Sebastian, Spain
    Ikerbasque, Basque Foundation for Science, E-48013 Bilbao, Spain \label{10} \and
    Department of Astronomy, University of Illinois at Urbana-Champaign, Urbana, IL 61801, USA \label{11} \and
    Observat\'orio Nacional, Rua General Jos\'e Cristino, 77, São Crist\'ovão, 20921-400, Rio de Janeiro, Brazi  \label{12} \and
    Department of Astronomy, University of Michigan, 311 West Hall, 1085 South University Ave., Ann Arbor, USA \label{13} \and
    Department of Physics and Astronomy, University of Alabama, Box 870324, Tuscaloosa, AL, USA \label{14} \and
    INAF, Osservatorio Astronomico di Trieste, via Tiepolo 11, 34131 Trieste, Italy \label{15} \and
    IFPU, Institute for Fundamental Physics of the Universe, via Beirut 2, 34151, Trieste, Italy \label{16} \and
    Universidade de SÃ£o Paulo, Instituto de Astronomia, Geof\'isica e Ci\^encias Atmosf\'ericas, R. do Matão 1226, 05508-090, São Paulo,Brazil \label{17} \and 
    Instruments4, 4121 Pembury Place, La Cañada-Flintridge, Ca 91011, USA \label{18}}
    
   \date{}

 
  \abstract{This paper is part of large effort within the J-PAS collaboration that aims to classify point-like sources in miniJPAS, which were observed in 60 optical bands over $\sim$ 1 deg$^2$ in the AEGIS field. We developed two algorithms based on artificial neural networks (ANN) to classify objects into four categories: stars, galaxies, quasars at low redshift ($z < 2.1)$, and quasars at high redshift ($z \geq 2.1$). As inputs, we used miniJPAS fluxes for one of the classifiers (ANN$_1$) and colours for the other (ANN$_2$). The ANNs were trained and tested using mock data in the first place. We studied the effect of augmenting the training set by creating hybrid objects, which combines fluxes from stars, galaxies, and quasars. Nevertheless, the augmentation processing did not improve the score of the ANN. We also evaluated the performance of the classifiers in a small subset of the SDSS DR12Q superset observed by miniJPAS. In the mock test set, the f1-score for quasars at high redshift with the ANN$_1$ (ANN$_2$) are $0.99$ ($0.99$), $0.93$ ($0.92$), and $0.63$ ($0.57$) for $17 < r \leq 20$, $20 < r \leq 22.5$, and $22.5 < r \leq 23.6$, respectively, where $r$ is the J-PAS rSDSS band. In the case of low-redshift quasars, galaxies, and stars, we reached  $0.97$ ($0.97$), $0.82$ ($0.79$), and $0.61$ ($0.58$);  $0.94$ ($0.94$), $0.90$ ($0.89$), and $0.81$ ($0.80$); and $1.0$ ($1.0$), $0.96$ ($0.94$), and $0.70$ ($0.52$) in the same r bins. In the SDSS DR12Q superset miniJPAS sample, the weighted f1-score reaches 0.87 (0.88) for objects that are mostly within $20 < r \leq 22.5$. We find that the most common confusion occurs between quasars at low redshift and galaxies in mocks and miniJPAS data. We discuss the origin of this confusion, and we show examples in which these objects present features that are shared by both classes. Finally, we estimate the number of point-like sources that are quasars, galaxies, and stars in miniJPAS.}


   \maketitle
%

\section{Introduction}
The new era in modern astronomy is closely connected to the era of big data. Astronomical observations  produce increasingly larger amounts of data. The new generation of surveys, such as the Dark Energy Spectroscopic Instrument \citep[DESI;][]{2013arXiv1308.0847L}, the Large Synoptic Survey Telescope \citep[LSST;][]{2019ApJ...873..111I} or the Square Kilometer Array \citep[SKA;][]{2009IEEEP..97.1482D}, will observe of the order of millions or even billions of objects. In particular, the Javalambre Physics of the Accelerating Universe Astrophysical Survey \citep[J-PAS;][]{2014arXiv1403.5237B} will observe thousands of deg$^2$ in the northen sky in the upcoming years  in 54 narrow-band filters, detecting more than 40 million objects.  Consequently, it is necessary to automatise all the tasks as much as possible so as to process the astronomical information faster and more efficiently. The identification and classification of astronomical objects is certainly the first step prior to any further scientific analysis. 
\par Traditionally, photometric surveys identified galaxies and stars based on their morphological structure and colour properties \citep[see e.g.][]{2010MNRAS.404...86B,2011MNRAS.412.2286H,2012ApJ...746..128S,2019A&A...622A.177L}. Typically, galaxies are extended objects, while point-like sources are mainly either stars or quasi-stellar objects (QSOs). Nevertheless, the lack of spatial resolution for the most distant and faint galaxies causes them look very similar to point-like sources. Furthermore, the colour space in multi-band photometric surveys becomes increasingly complex as the number of filters increases, which requires more sophisticated algorithms to fully exploit all the information encoded in the surveys. \par In past years, machine-learning (ML) algorithms have been used in many applications within the astronomical field, from the estimation of photometric redshifts \citep{2017MNRAS.465.1959C,2019A&A...621A..26P,2021arXiv211112118R}, the identification of low-metallicity stars \citep{2019A&A...622A.182W}, the determination of the star formation rate \citep{2019MNRAS.486.1377D,2019A&A...622A.137B}, the classification of morphological types in galaxies \citep{2018MNRAS.476.3661D}, and the identification of causality in galaxy evolution \citep{2022arXiv220107814B} to the measurement of the equivalent widths of emission lines in photometric data \citep{MS21}. The problem of source identification in photometric surveys has also been addressed by ML either to distinguish between point-like and extended sources \citep{2011AJ....141..189V,2015MNRAS.453..507K,2017MNRAS.464.4463K,2019MNRAS.490.3952B,2021A&A...645A..87B} or even between galaxies, stars, and QSOs \citep{2016A&A...596A..39K,2019AJ....157....9B,2020A&A...633A.154L,XIAOQING2021303,2021MNRAS.508.2039H}.

\par The goal of the present paper is to classify the objects detected with the miniJPAS survey \citep{2021A&A...653A..31B} into stars, galaxies, QSOs at high redshift ($z \geq 2.1$), and QSOs at low redshift ($z \leq 2.1$) by using artificial neural networks (ANN). The threshold at $z = 2.1$ corresponds to the limit at which QSOs show the Lyman-$\alpha$ emission line within the J spectra. The miniJPAS survey is part of the J-PAS project\footnote{http://www.j-pas.org/}, which detected more than $60\,000$ objects within the  All-wavelength Extended Groth Strip International survey \citep[AEGIS;][]{2007ApJ...660L...1D} using 56 narrow-band J-PAS filters ($ \sim145 $ \AA $ $) and the four $ugri$ broad-band filters. The separation of 100 \AA $ $ among filters makes the J-PAS filter system equivalent to obtaining a low-resolution spectrum with $ R \sim 60 $ (J spectrum hereafter). These unique characteristics enable observation and analysis of galaxies and QSOs in continuous redshift ranges, $0 \lesssim z \lesssim 1$ and $0.5 \lesssim z \lesssim 4$, respectively \citep{2021A&A...653A..31B}. Different studies have proved the capability of J-PAS to address several topics within the astrophysical field, for instance, the evolution of the stellar population properties of galaxies up to $z \sim 1$ \citep{2021arXiv210213121G}, the properties of the nebular emission lines of galaxies down to $z \leq 0.35$ \citep{2022A&A...661A..99M}, the measurement of black hole virial masses for the QSO population \citep{2021arXiv211101180C}, or the study of galaxy properties within galaxy clusters \citep{2022A&A...666A.160R} and groups \citep{2022arXiv220705770G}. Unfortunately, the data available for spectroscopically confirmed sources in the miniJPAS area are not sufficient to train and test ML algorithms for our purpose. Therefore, we employed mock data developed by \cite{10.1093/mnras/stac2962}, and used the sources identified spectroscopically within the AEGIS field by the Sloan Digital Sky Survey \citep[SDSS;][]{2000AJ....120.1579Y} in the DR12Q superset catalogue \citep{2017A&A...597A..79P} as truth table.
\par Modern deep ANN generally perform better than traditional methods. They remain poorly calibrated most of the time, however \citep{2014arXiv1412.6572G,2017arXiv170604599G}. The probabilities associated with the predicted label classes may be overly confident because they do not correspond to true likelihoods. Consequently, objects are classified as part of one class or another with high probability regardless of the prediction accuracy. Realistic probability distributions are particularly important for spectroscopy follow-up programs that typically prioritise the observation of high-probability objects of some particular class. Furthermore, some objects are indeed dual in nature. Although we consider as QSOs only galaxies with an extremely luminous active galactic nucleus (AGN), there are objects in which a significant fraction of the detected light comes from the stars in the host galaxy. In this scenario, ML algorithms should ideally provide a high probability in both classes. 
\par In a recent paper, \cite{2017arXiv171009412Z} proposed an original idea called \texttt{mix up} to enlarge the dataset and improve the generalisation of the trained model, thus increasing the robustness to adversarial examples. Latter on, \cite{2019arXiv190511001T} showed that \texttt{mix up} or hybridisation, as we prefer to call it, also improves the calibration and predictive uncertainty of deep neural networks. In this work, we enlarge our training set by mixing features from stars, galaxies, and QSOs, and we study the effect of hybridisation on the performance and calibration of the models. The ML classifiers used in this work will be combined with other ML algorithms. In \cite{10.1093/mnras/stac2836}, we trained convolutional neural networks (CNNs) by proposing different approaches to incorporate photometry uncertainties as inputs in the training phase. In this work, we focus our attention on the galaxy-QSO degeneracy and the ability of ANN to estimate realistic probability density distributions (PDF). In \cite{10.1093/mnras/stac2836}, we study other relevant aspects, such as the stellar types that are more frequently confused with the QSO population, the J-PAS feature importance, or the stability of the CNNs predictions with respect to minor changes in the training set. In P\'erez-R\`afols et al. (in prep. a), we additioanlly adapt the code SQUEZE \citep{2020MNRAS.496.4931P} to work with J-PAS data. This code is based on optical emission line identification to separate between QSOs and non-QSOs and also estimates the redshift of QSOs. Ultimately, all these codes will be merged in a combined algorithm (P\'erez-R\`afols et al. in prep. b) so as to classify the miniJPAS sources more efficiently and to provide a high-redshift QSO target list for a spectroscopic follow-up with the WEAVE multi-object spectrograph survey \citep{2014SPIE.9147E..0LD,2022arXiv221203981J}, 
which is planning to carry out a Lyman-$\alpha$ forest and metal line absorption survey \citep{2016sf2a.conf..259P}.  
\par This paper is organised as follows. In Sect. \ref{sec:miniJPAS_mocks} we present miniJPAS data, and we briefly summarise the processes employed in the construction of the mock catalogue. In Sect. \ref{sec:classifiers} we describe the main characteristics of the classifiers in detail and relate how data augmentation was employed through hybridisation. We indicate the performance metrics used for testing purposes in the paper in Sect. \ref{sec:per_metric}, and we show the main results obtained in the paper in Sect. \ref{sec:results} and Sect. \ref{sec:minijpas_catalogue}. Finally, we summarise and conclude in Sect. \ref{sec:conclusions}. Throughout the paper, all magnitudes are presented in the AB system \citep{1983ApJ...266..713O}.

\section{MiniJPAS survey and mocks}\label{sec:miniJPAS_mocks}
The miniJPAS survey includes data from four pointings scanning $\sim~1$~deg$^2$ along the AEGIS field. The photometric system includes 56 bands, namely 54 narrow-band filters in the optical range plus two medium bands, one in the near-UV (uJAVA band), and the other in the NIR (J1007 band). With a separation of $\sim 100 \ {\AA}$,  each narrow-band filter has a full width at half maximum (FWHM) of $ \sim 145 $ \AA, whereas the FWHM of the uJAVA band is 495 \AA ,$ $ and J1007 is a high-pass filter. Additionally, four broad bands $u$,$g$,$r$, and $i$ were used to complement the observations. These were carried out with the $2.55$ m telescope at the Observatorio Astrofísico de Javalambre, a facility developed and operated by CEFCA, in Teruel (Spain). The data were acquired using the pathfinder instrument, a single CCD direct imager ($9.2k \times 9.2k$, $10\mu m$ pixel) located at the centre of the $2.55$ m telescope FoV with a pixel scale of 0.23 arcsec pix$^{-1}$, vignetted on its periphery, providing an effective FoV of 0.27 deg$^2$. The $r$ band was chosen as the detection band, and the reference image is in the dual-mode catalogue. This image was used to define the position and sizes of the apertures from which the rest of the photometry was extracted. The miniJPAS survey is $99\%$ complete up to $r \le 23.6$ mag for point-like sources, and it detected more than  $60\,000$ objects\footnote{http://archive.cefca.es/catalogues/minijpas- pdr201912} \citep{2021A&A...653A..31B}. In this paper, we only analyse objects with \texttt{FLAGS$=0$} and \texttt{MASK\_FLAGS$=0$} (46441 in total), that is, they are free from detection issues such as contamination from bright stars and light reflections in the telescope or in its optical elements. We refer to \cite{2021A&A...653A..31B} for details on the flagging scheme. Removal of flagged sources from the catalogue decreased our sample size, but did not introduce any bias because the fraction of sources that are flagged is independent of their magnitude \citep{2021A&A...654A.101H}.
\par The algorithms presented in this work were trained and tested on mock data \citep{10.1093/mnras/stac2962}. The J spectra of galaxies, stars, and QSOs were simulated by convolving SDSS spectra included in the SDSS DR12Q superset catalogue \citep{2017A&A...597A..79P} with the transmission profiles of J-PAS photometric system (synthetic fluxes). The SDSS DR12Q superset contains all objects targeted as QSOs from the final data release of the Baryon Oscillation Spectroscopic Survey \citep[BOSS;][]{2013AJ....145...10D}. Therefore, it also contains galaxies and stars whose broad-band colours are compatible with those from QSOs. Since the SDSS sample is complete only up to $r$ $\sim$ 20.5, fainter objects were generated by adding random fluctuations (noise) to the synthetic fluxes. The mock catalogue includes several noise models that mimic the observed signal-to-noise-ratio (S/N) in miniJPAS for the \texttt{APER\_3} magnitudes. The \texttt{APER\_3} magnitude contains the light integrated within an aperture of three arcsec. Thus, in order to account for the missing light emitted by each source, we implemented an aperture correction that was based on the \texttt{APER\_6} magnitude, as detailed in \cite{10.1093/mnras/stac2962}. The main results presented in this paper are not affected by the noise model choice. We used model 1, which assumes that the noise distribution of miniJPAS point-sources in each filter is well described by a single Gaussian distribution \citep{10.1093/mnras/stac2962}. We considered SDSS spectra to be noise free. Although the S/N is lower at short and long wavelengths, the error in the mocks is dominated by miniJPAS observations, which reach fainter objects. As we did not simulate miniJPAS images, the photometry in the mocks is not affected by any image-processing effect. This choice makes our mocks different from observations to a certain extent. For example, miniJPAS observations were not carried out with 56 filters simultaneously. Instead, the observations were taken in trays of 14 CCDs, thus different observational conditions were presented in the SED of individual objects. Taking these effects into consideration would have made the mock-building process very complex and tedious. Additionally, there is no guarantee that these modifications would have improved the performance of the algorithms. In the future, as soon as J-PAS begins to observe the sky, we will have J-PAS data for objects that were previously observed by other spectroscopic surveys such as WEAVE. We will then be in a position to re-train our models with observed data and reduce the existing gap between simulated and real observations
\par  P\'erez-R\`afols et al. (in prep. b) will compare the performance of each algorithm in the mock test sample with different noise models, and we will classify sources in miniJPAS for each one of them. Galaxies follow the magnitude-redshift distribution of SDSS and DEEP3 \citep{2011ApJS..193...14C,2012MNRAS.419.3018C} found in miniJPAS. QSOs follow the luminosity function of \cite{2016A&A...587A..41P}, and stars are distributed according to the Besançon model of stellar population synthesis of the Galaxy \citep{2003A&A...409..523R} and the SDSS miniJPAS spectroscopic sample. 
\par The number of stars, galaxies, and QSOs in training set are balanced to prevent biases in the classifiers towards over-represented classes. The same applies for the test set and the validation set, which were used to fine-tune the hyper-parameter of the classifiers. Additionally, another test sample, the 1-deg$^2$ test set, includes the expected number of point-like sources within the miniJPAS area. This sample was generated to provide a more direct comparison with miniJPAS observations. Finally, we used a sub-sample of miniJPAS observations as a true table. This sample is called the SDSS test sample and includes only the objects observed by the SDSS DR12Q superset. Unfortunately, the SDSS test sample does not contain a sufficient number of objects, and it is biased toward bright objects. Therefore, it can only give us a small impression of the actual performance of the algorithms in real data. In Table \ref{Tab:mocks} we summarise the number of objects contained in all samples. Further details of how these synthetic data were created can be found in \cite{10.1093/mnras/stac2962}.

\begin{table}

\caption{\small{Number of objects in each data set}}
\begin{tabular}{lllll}
\hline
\\
\vspace{0.2cm}
\textbf{Sample} & \textbf{Galaxies} &  \textbf{Stars} &  \textbf{QSOs} & \textbf{All} \\ \hline \\
\textbf{Training}  & $10^5$ & $10^5$ &  $10^5$  & $3 \times 10^5$ \vspace{0.1cm} \\ 
\textbf{Validation}  & $10^4$ & $10^4$ &  $10^4$& $3 \times 10^4$ \vspace{0.1cm} \\ 
  \textbf{Test}  & $10^4$ & $10^4$ &  $10^4$ & $3 \times 10^4$ \vspace{0.1cm} \\ 
  \textbf{1-deg$^2$}  & $6410$ & $2190$ &  $510$ & $9110$ \vspace{0.1cm} \\ 
\textbf{SDSS test}  & $40$ & $117$ &  $115$ & $272$ \\ 

 \hline \vspace{0.1cm}
\end{tabular}

\label{Tab:mocks}
\end{table}

\section{Star, galaxy, and QSO classifier}\label{sec:classifiers}
In this section, we describe in detail how the ML classifiers were developed. Although our main focus is on ANN, we also developed an RF classifier in order to compare the performance of the two algorithms. They were designed to distinguish between four different classes: stars, galaxies, QSOs at high redshift ($z \geq 2.1$), and QSOs at low redshift ($z < 2.1$), referred to as QSO-h and QSO-l, respectively. 
\subsection{Artificial neural networks}\label{subsec:ANN}
The ANN were coded with the {\tt Tensorflow} \citep{tensorflow2015-whitepaper} and {\tt Keras} libraries \citep{chollet2015keras} in {\tt Python}. The ANN has eight hidden layers with 200 neurons each. As a regularisation technique, we used weight constraints and imposed a maximum value of two in each neuron (kernel constraint). We also removed $15 \%$ of the neurons in each layer. We used the rectified linear unit (ReLU) as our activation function \citep{Relubibcode}. Weights were initialised with the He initialisation strategy \citep{geron2019hands}. The loss function we employed is the cross entropy. 
\par We trained two models that use two different sets of inputs, called ANN$_1$ and ANN$_2$. The inputs of ANN$_1$ (59 in total) were relative fluxes, that is, the flux in each filter was divided by the flux in the $r$ band. Since the miniJPAS dual-mode catalogue used the $r$-band for detection, this normalisation is well defined for all the objects. The inputs of ANN$_2$ were the colours measured with respect to the $m_{AB}$ in the $r$ band, plus the normalised magnitude in this band (60 inputs in total),
\begin{equation}
\begin{split}
    \text{ANN}_1^j  = \hspace{0.1cm} & \frac{f_{\lambda}^j}{f_{\lambda}^{\text{r}}} \\ 
    \text{ANN}_2^{j} = \hspace{0.1cm} & m_{AB}^j - m_{AB}^r ,\hspace{0.5cm} \text{ANN}_2^{r} =  \frac{m_{AB}^{r} - \text{max}(m_{AB}^r)}{\text{min}(m_{AB}^r) - \text{max}(m_{AB}^r)}
\end{split}
\label{eq:mABnorm}
,\end{equation}
where $m_{AB}$ and  $f_{\lambda}$ stand for the magnitude and the flux in the $j$th filter, respectively, and  $\text{min}(m_{AB}^r)$ and $\text{max}(m_{AB}^r)$ are the minimum and maximum values of the magnitude in the r band within our training set. Both sets of inputs capture the shape of the spectrum, but the ANN$_2$ also includes information about the observed luminosity of each source, which anchors the SED to a particular magnitude.
\par Objects in the dual-mode catalogue might be undetected in some bands (non-detection). This happens when the S/N values for these bands are very low and the measured fluxes are null or negative after the sky background subtraction. In the mock catalogues, a non-detection (ND) follows the pattern observed in miniJPAS. For specific details of how ND are modelled, we refer to \cite{10.1093/mnras/stac2962}. We set the inputs of the ANN$_2$ to zero when fluxes are negative because the colours are otherwise undefined. This approach in sub-optimal in the sense that colours might be zero if fluxes are identical, thus it is difficult for the ANN to identify NDs. Another option is to set NDs to an arbitrary number. Unfortunately, no improvement has been observed with this approach. In the case of ANN$_1$, we allowed the inputs to be below zero because ANN$^j_1$ values are positive by definition. In principle, the ANN$_1$ should be able to model the sky background  better because it can identify the regime of low emission in which fluxes are close to zero or even negative.
\subsection{Random forest}\label{subsec:RF}
Random forest (RF) is an ensemble machine-learning algorithm that uses multiple decision trees to predict an outcome. Multiple trees (a forest) are built using a random subset of features and observations from the training data (bootstrapped sample). For each bootstrapped sample, a decision tree is grown to the maximum depth or until a stopping criterion is reached. The effectiveness of each split in a decision tree can be quantified by the information gain, which is a measure of the expected reduction in impurity as defined by either \textit{entropy} or the \textit{gini} impurity function. The intrinsic randomness of the RF helps reduce the risk of overfitting and creates a more robust model than single decision trees. After building several trees, the algorithm takes the average or majority vote of the predictions from each tree to produce a final prediction for each data point.
\par We built an RF classifier that uses the same features as ANN$_1$ and ANN$_2$ in order to compare them. The algorithm was implemented with the \texttt{scikit-learn} python package with the default setting\footnote{\url{https://scikit-learn.org/stable/modules/generated/sklearn.ensemble.RandomForestClassifier.html}}.

\subsection{Data augmentation via hybridisation}\label{subsec:hyb}
Data augmentation has been proven to be an excellent tool to increase the size of the training sample, and consequently, the performance of ML algorithms when only limited training samples are available \citep{DAG2019}. Rotation, translation, or scaling are among the most popular techniques for image classification \citep{2022arXiv220408610Y}. In the case of non-image features such as the J spectra, the most common manner to perform data augmentation is via Gaussian noise. However, the benefit of this technique in our training sample would be limited because it was already used to generate objects at different magnitudes bins in the construction of the mock catalogue. Thus, we adapted the \texttt{mix up} technique proposed in \cite{2017arXiv171009412Z} to our classifiers, which aim to distinguish between four classes. This technique allowed us to enlarge the training set by mixing features from different classes, generating a new training set composed only of hybrid objects. The new set of hybrid objects ($y^H_{i}$) and their respective fluxes were generated as a linear 
combination of individual objects in the original training set,
\begin{align}
& \mathbf{y^H_i}  =  \alpha_i \mathbf{y_i} + \sum_{j=1}^{4}c_{ij}(1-\alpha_j)(1-\delta_{ij}) \mathbf{y_j} \\
& \mathbf{f^H_i(\lambda)}  = \alpha_i \mathbf{f_i (\lambda)} + \sum_{j=1}^{4} c_{ij}(1-\alpha_j)(1-\delta_{ij}) \mathbf{f_j (\lambda),} 
\label{eq:hyb}
\end{align}
where $\mathbf{y_1}$ ($\mathbf{f_1(\lambda)}$), $\mathbf{y_2}$ ($\mathbf{f_2(\lambda)}$), $\mathbf{y_3}$ ($\mathbf{f_3(\lambda)}$), and, $\mathbf{y_4}$ ($\mathbf{f_4(\lambda)}$) are the vectors (fluxes) of each one of the classes (stars, galaxies, QSO-l, and QSO-h, respectively), and $\alpha_j$ is the mixing coefficient, which varies between zero and one according to an exponential distribution function that depends on $\beta$, the scale parameters that control the level of mixing,
\begin{equation}
\alpha_j(x;\beta)~=~1~-~(1/\beta) \exp(-x/\beta)
,\end{equation}
where $x$ is a positive random variable between 0 and 1. As $\beta$ increases, $\alpha_j$ is more likely to become smaller, thus more mixing (1 - $\alpha_j$) is performed between classes. Finally, in Eq. \ref{eq:hyb}, $\delta_{ij}$ is the Kronecker delta, and $c_{ij}$ are the luminosity coefficients, given by
\begin{equation}
c_{ij} = \frac{N_j}{N_{\text{tot}} - N_{i}}
,\end{equation}
where $N_i$ is the number of objects belonging to class $i$ within each magnitude bin in the original training set, and $N_{\text{tot}} = N_{1} + N_{2} + N_{3} + N_{4}$. The luminosity coefficients modulate how much of the mixing (1 - $\alpha_j$) comes from each class as a function of their relative number in different magnitude bins. For instance, if we generate an hybrid galaxy ($\mathbf{y^H_2}$) at magnitude $m_{AB}^{\text{r}} = 22.5$, Eq. \ref{eq:hyb} becomes
\begin{equation}
\mathbf{f^H_2(\lambda)} = \alpha_2 \mathbf{f_2(\lambda)} + (1-\alpha_2)(c_{21} \mathbf{f_1(\lambda)} + c_{23} \mathbf{f_3(\lambda)} + c_{24} \mathbf{f_4(\lambda)}).
\end{equation}
For low values of $\beta$, $\alpha_2$ is near one with a high probability. Therefore, the new hybrid galaxy is still a galaxy, but it has some of level of contamination from the other classes. The probability of not being a galaxy ($1-\alpha_2$) is distributed among the other classes by taking their relative amounts at $m_{AB}^{\text{r}} = 22.5$ into account. Since stars are less frequent at this magnitude, $c_{21}$ is close to zero, and the new hybrid galaxy is mixed mainly with QSO-l and QSO-h. In order to compute the $c_{ij}$ coefficients, we split the training set into rSDSS magnitude bins that contained roughly $20\,000$ objects. In this way, only objects with a similar brightness were mixed together. After hybridisation, the proportion of each class in the training set remained the same. That is to say, even though features of different objects were mixed together, each object conserves most of the original features, and hence they still have a high probability to belong to the original class. We enlarged the training set five times with $\beta = 0.1 $ (we discuss this choice in Sec. \ref{subsec:test_mock}). We caution that hybridisation does not mix objects following a physical recipe, but is rather a mathematical transformation of the data. Since hybridisation was implemented at the level of fluxes, Eq. \ref{eq:mABnorm} still needs to be applied to the resulting hybrid fluxes in order to standardised the input data for ANN$_1$ and RF$_1$, and RF$_2$ and ANN$_2$.

\subsection{Training strategy}\label{subsect:tra-strag}
The intrinsic randomness of the training procedure usually leads to solutions that are not optimal. Weights and biases are drawn from a distribution function that generates the initial state. Therefore, each time that the training is performed, the algorithm converges to a different local minimum of the loss function. Furthermore, the training set itself is augmented in a random manner via hybridisation. The hybrid space is filled in a slightly different way in each realisation. In the limit case where the hybrid set is much larger than the original, this effect would be negligible. However, a huge training set is less practical to handle and more difficult to train than a smaller set. For all these reasons, we followed the committee approach \citep{10.5555/525960}, that is, we trained several ANNs and computed the median to provide a final classification. Then, we re-normalised the output probabilities to ensure that the sum was one. In order to determine the optimal number of ANN or committee members needed, we started with two, and we added an additional member at each step until the results did not improve for the validation sample in terms of the $f_1$ score (see Sect. \ref{sec:per_metric} for a definition of this metric). We determined that eight members are enough to reach convergence.   
\subsection{Performance metrics}\label{sec:per_metric}
In this section, we provide a detailed discussion of the metrics used to evaluate the performance and robustness of the classifiers. Each metric offers unique insights into the ability of the ML algorithms to classify sources. By examining these metrics, we can better understand the strengths and limitations of the classifiers.
\subsubsection{Confusion matrix}
The confusion matrix is especially useful in the context of a multi-label classification problem. The actual classification of each object is shown in the columns of the matrix, while the predicted objects lie in the rows. Therefore, in the best (ideal) case scenario, the matrix would be purely diagonal, and every prediction would coincide with the actual classification. Non-diagonal terms indicate which classes are confused and provide valuable information for improving the training set and fine-tuning the hyperparameters of the model. 
\subsubsection{f1-score}
Unlike the confusion matrix, the f1-score yields one single scalar for each one of the classes. It finds a compromise between purity (precision) and completeness (recall),
\begin{align}
\text{Purity} = & \frac{TP}{TP+FP}, \hspace{0.5cm} \text{Completeness} = \frac{TP}{TP+FN} \\
f_1 \hspace{0.1cm} {\text{score}} = & \frac{2 \cdot\text{Purity} \cdot \text{Completeness} }{\text{Purity}+\text{Completeness,}}
\end{align}
where $TP$, $FP,$ and, $FN$ are the true-positive rate, the false-positive rate, and the false-negative rate, respectively. FP appears as non-diagonal terms in the columns of the confusion matrix, while TN lies on the rows. In the case of an unbalanced test set in which one or more classes are underrepresented, the average performance of the model can be estimated using the weighted $f_1$ score,
\begin{equation}
f^{W}_1 \hspace{0.1cm} {\text{score}}  = \sum_{i=0}^{N_{\text{class}}}\frac{n_i f_1^i \hspace{0.1cm} \text{score}}{n_{\text{obj}}}
,\end{equation}
where $n_i$ is the number of objects belonging to the $i$th class in the test set, and $n_{\text{obj}}$ is the total number of objects. 
\subsubsection{Expected calibration error}
A well-calibrated probabilistic classifier is able to predict probabilities that coincide on average with the fraction of objects that truly belong to a certain class (accuracy). We assume that we took 100 objects with a probability of $10\%$ of being a star. When the classifier is well calibrated, about 10 of them should be stars. If there are more, our classifier is under-confident. If there are less, then the classifier is over-confident. Probability calibration curves are normally employed to display this relation, where we bin the probability estimates and plot the accuracy versus the mean probability in each bin. Let $B_{mj}$ be the set of objects whose predicted probabilities of being class $j$ fall into bin $m$. The accuracy and confidence of $B_{mj}$ are defined as 
\begin{equation}
acc(B_{mj}) = \frac{N_{mj}}{|B_{mj}|}
\end{equation}

\begin{equation}
conf(B_{mj}) = \frac{1}{|B_{mj}|} \sum_{i \in B_{mj}} P_{ij}
,\end{equation}
where $P_{ij}$ is the probability of being class $j$ for the $i$th object, and $N_{mj}$ is the number of objects of class $j$ within bin $m$. The expected calibration error (ECE) is then defined as
\begin{equation}
\text{ECE} = \frac{1}{N_c} \sum_{j=0}^{N_c} \sum_{m=0}^{N_b} \frac{|B_{mj}|}{N_b}|acc(B_{mj}) - conf(B_{mj})| 
,\end{equation}
where $N_c$ is the number of classes, and $N_b$ is the number of bins. The lower the value of the ECE, the better the calibration of the model. However,
the output of the ANN only represents true probabilities under the assumption that our mock sample and the miniJPAS observations do not differ essentially. In order to have a better estimate of the ECE of the model, we would need to compute this metric in a sufficiently large true table, which is not yet possible. In the future, when more data will be gathered, we will be able to employ this metric on observations and evaluate the ECE of the ANN properly. In the remainder of the paper, we refer to the outputs of the ANN as probabilities, recalling that they are simply a proxy of the true probabilities. 
\par
\subsubsection{Entropy}
The entropy is a measurement of disorder. In the context of ML, the entropy of a classifier prediction can tell us how uncertain the classifier is. The entropy of the $i$th object can be written as follows:
\begin{equation}
S_i = -\sum_{j=0}^{N_c} P_{ij} \log_{2} (P_{ij} + \epsilon) 
,\end{equation}
where $P_{ij}$ is the probability that the $i$th object is class $j$, $N_c$ is the number of classes, and $\epsilon$ is an arbitrarily small number ($10^{-14}$) to avoid the divergence of the logarithm when the probability for a given class is zero. The entropy is maximum ($\log_{2} N_c$) when each class has a probability of $1/N_{c}$ and zero when the probability of belonging to a particular class is one.

\section{Results}\label{sec:results}
In this section, we test the performance of the algorithms on the mock test sample (Sect. \ref{subsec:test_mock}). We discuss in detail the effect of augmenting the data through hybridisation and we compare the differences between classifiers. Finally, we evaluate the classification obtained with SDSS objects observed by miniJPAS in the AEGIS field.
\subsection{Test sets}\label{subsec:test_mock}
\begin{figure}
    \centering
        \includegraphics[width=\hsize]{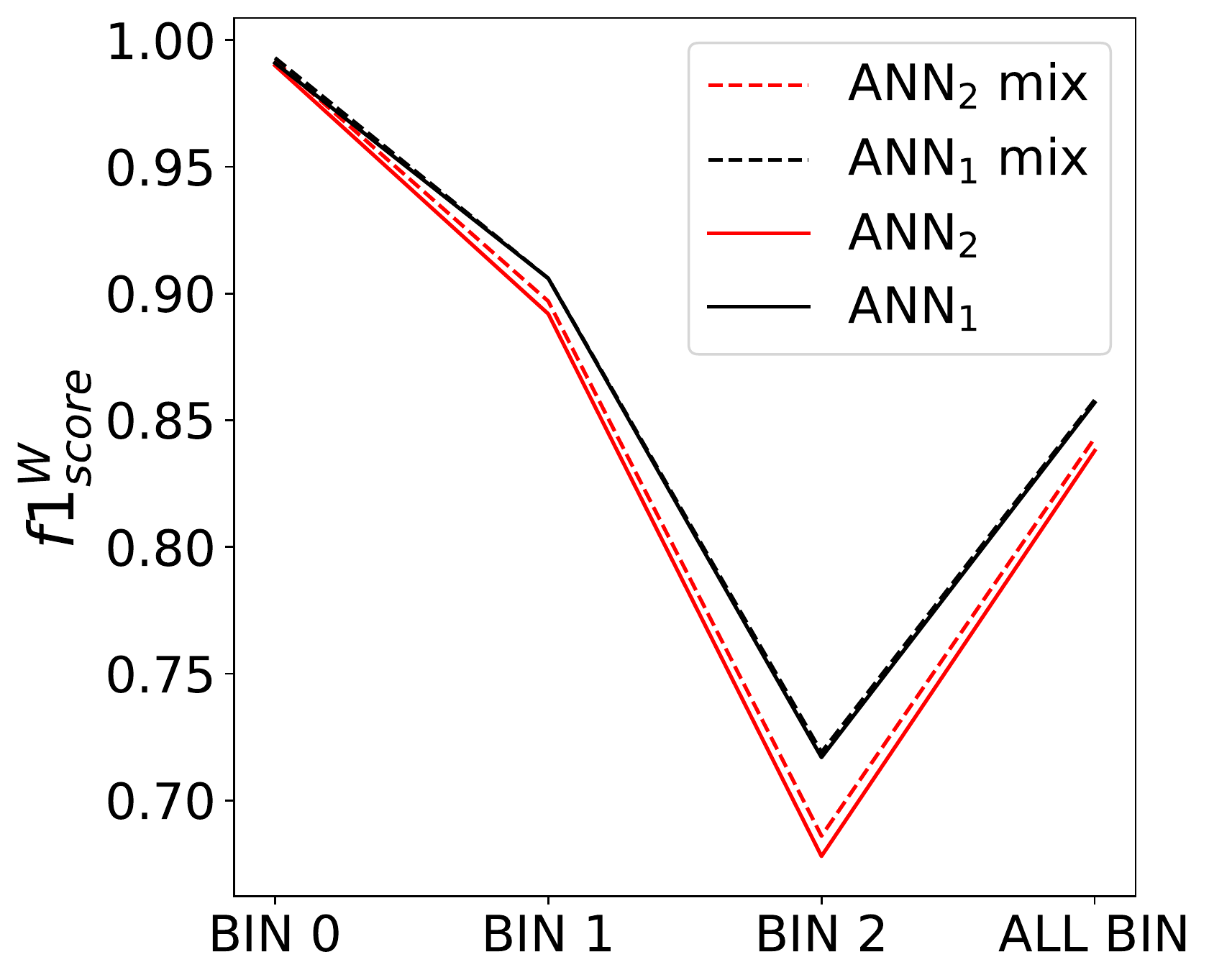}
        \caption{{$f^{W}_1$ score for different magnitude bins as defined in Eq. \ref{eq:BIN}, and the score for the full sample (ALL BIN). Dashed (solid) lines represent the models trained with the hybrid (original) training set. ANN$_2$ and ANN$_2$ mix are trained with colours, while ANN$_1$ and ANN$_1$ mix are trained with fluxes (see Sect. \ref{subsec:ANN}).}}
\label{fig:f1_score_mock}
\end{figure}

\begin{figure}
    \centering
        \includegraphics[width=\hsize]{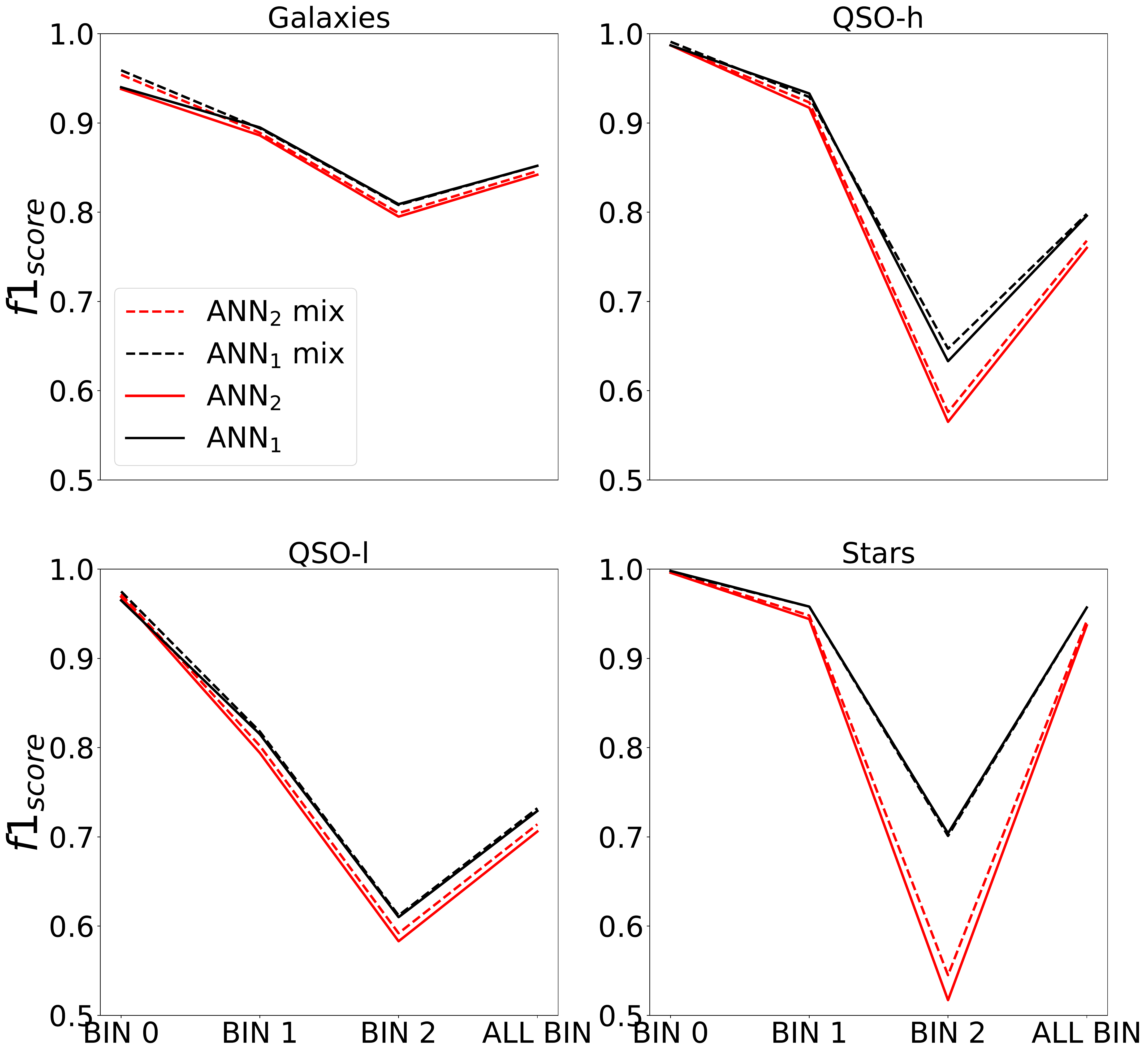}
        \caption{{$f_1$ score for each of the classes galaxies, QSO-h, QSO-l, and stars as a function of the magnitude bins defined in Eq. \ref{eq:BIN}, and score for the full sample (ALL BIN). Dashed (solid) lines represent the models trained with the hybrid (original) training set. ANN$_2$ and ANN$_2$ mix are trained with colours, while ANN$_1$ and ANN$_1$ mix do with fluxes (see Sect. \ref{subsec:ANN})}}
\label{fig:f1_score_mock_classes}
\end{figure}

The performance of a classifier changes as a function of the magnitude of the objects. Fainter objects are more difficult to classify because their S/N is lower. In order to quantify the potential bias of the classifiers at different magnitudes, we therefore split the validation and the test samples into three different bins according to the $r$-band magnitude,
\begin{align}
\text{BIN 0} \hspace{0.05cm} : & \hspace{0.2cm} 17 < r \leq 20 \nonumber \\ 
\text{BIN 1} \hspace{0.05cm} : & \hspace{0.2cm} 20 < r \leq  22.5   \label{eq:BIN} \\ 
\text{BIN 2} \hspace{0.05cm} : & \hspace{0.2cm} 22.5 < r \leq  23.6. \nonumber  
\end{align}
The number of objects in the test sample for BIN 0, 1, and 2 are 4987, 13357, and 9488, respectively. In Fig. \ref{fig:f1_score_mock} we show the $f^{W}_1$ score for each of the magnitude bins defined above, including the average performance for the full sample (ALL BIN). We compare the score of the ANN trained with the hybrid set (ANN$_1$ mix and ANN$_2$ mix) and with the original training set (ANN$_1$ and ANN$_2$). In Fig. \ref{fig:f1_score_mock_classes} the $f_1$ score is also shown for each of the classes. Overall, both classifiers (ANN$_{1}$ and ANN$_{2}$) are very similar, with small differences in each magnitude bin for each class. The ANN$_{1}$ classifier is slightly better probably not only because of the representation of the data (relative fluxes), but also because it captures the sky background better. \par As expected, the accuracy of the classifiers decreases for fainter objects. The performance obtained with the hybrid set is very similar to the original training set, suggesting that the latter already contains all the variance needed, and more examples do not necessarily imply a better performance (but see the next section for a more detailed discussion).


\par The confusion matrices as a function of the magnitude bin for the ANN$_{1}$ model are shown in Fig. \ref{fig:CM_ANN1_mock}. The matrices from the remaining models are provided in Appendix \ref{app:CM}. QSO-l and galaxies are more difficult to distinguish, especially at the faint end, where the data are noisier. These objects do not belong to independent classes. Sometimes, the host galaxy of an AGN might contribute strongly to the SED. In Seyfert galaxies, the observed spectrum is usually a combination of the light coming from the AGN and the stellar populations within the galaxy. Therefore, we expect confusions between QSO-l and galaxies more often than between any of the other classes. Finally, in the faintest magnitude bin, $31.4 \%$ of QSO-h are classified as QSO-l, and $18.9 \%$ of the stars are confused with QSO-l. 

\begin{figure*}
    \centering
        \includegraphics[width=6cm,height=6.3cm]{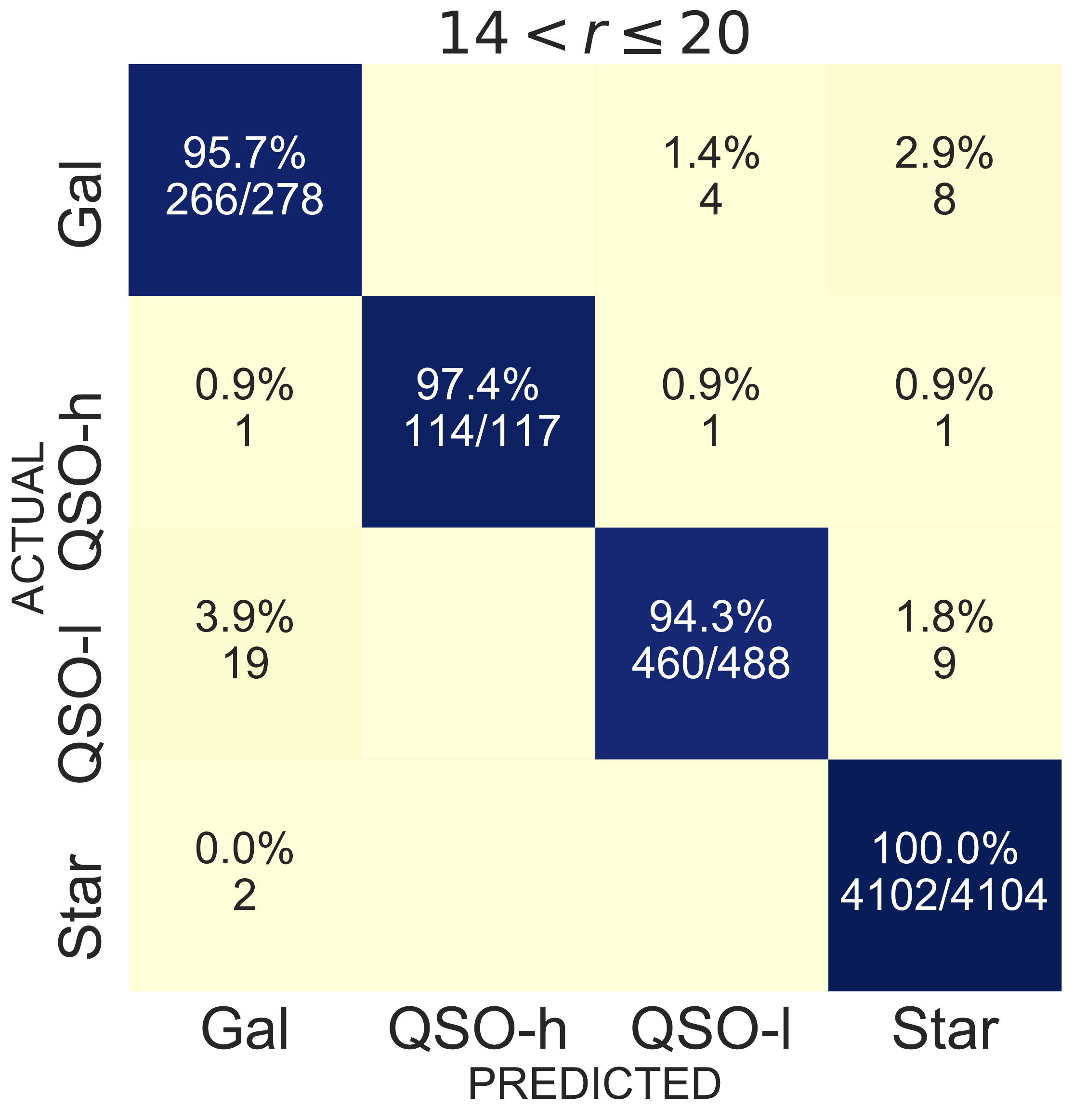}
        \includegraphics[width=5.6cm,height=6.3cm]{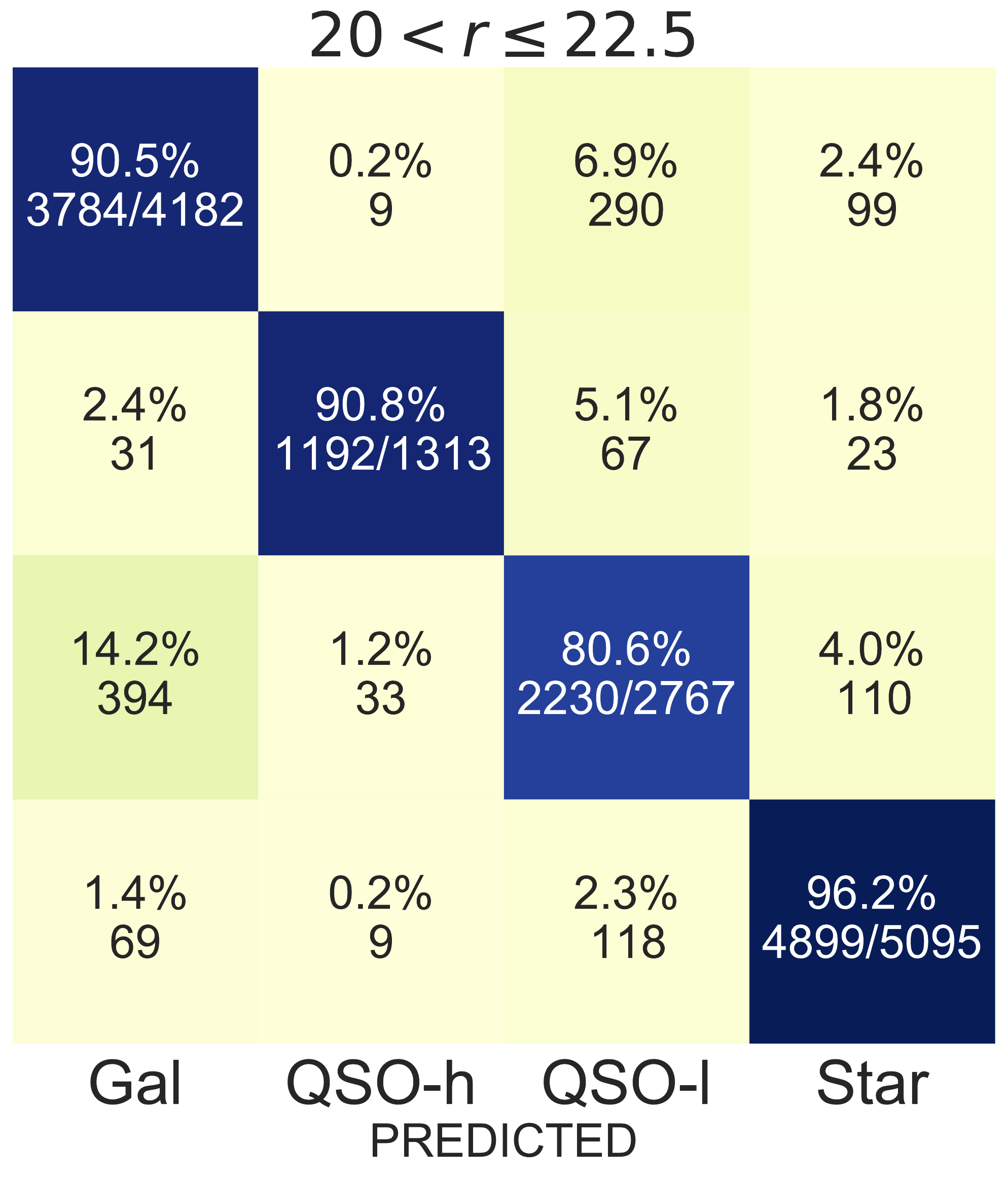}
        \includegraphics[width=5.6cm,height=6.3cm]{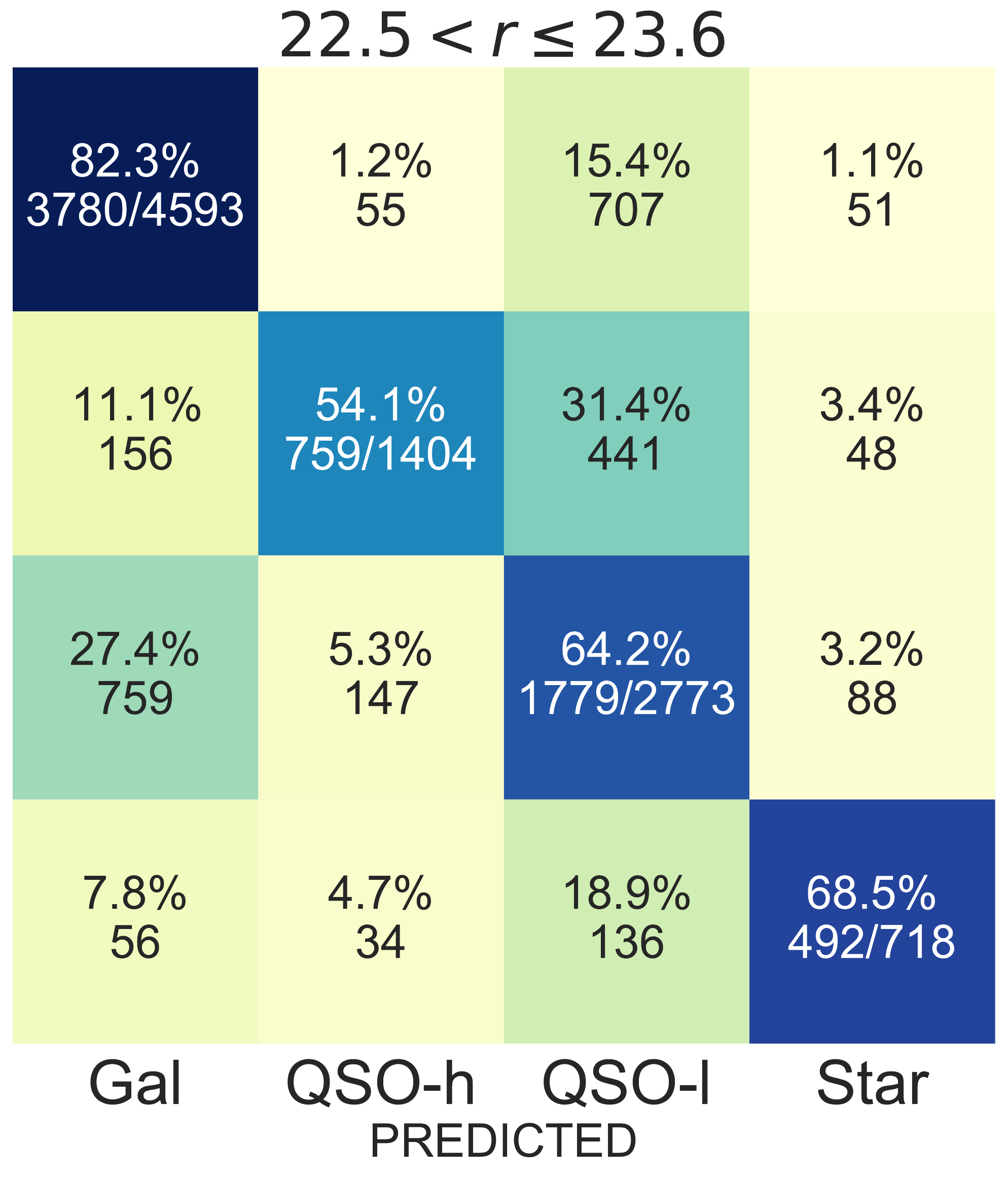}
        \caption{{Confusion matrices obtained with the ANN$_{1}$ in the test sample}.}
\label{fig:CM_ANN1_mock}
\end{figure*}
\par
\begin{figure*}
\includegraphics[width=5.9cm,height=5.6cm]{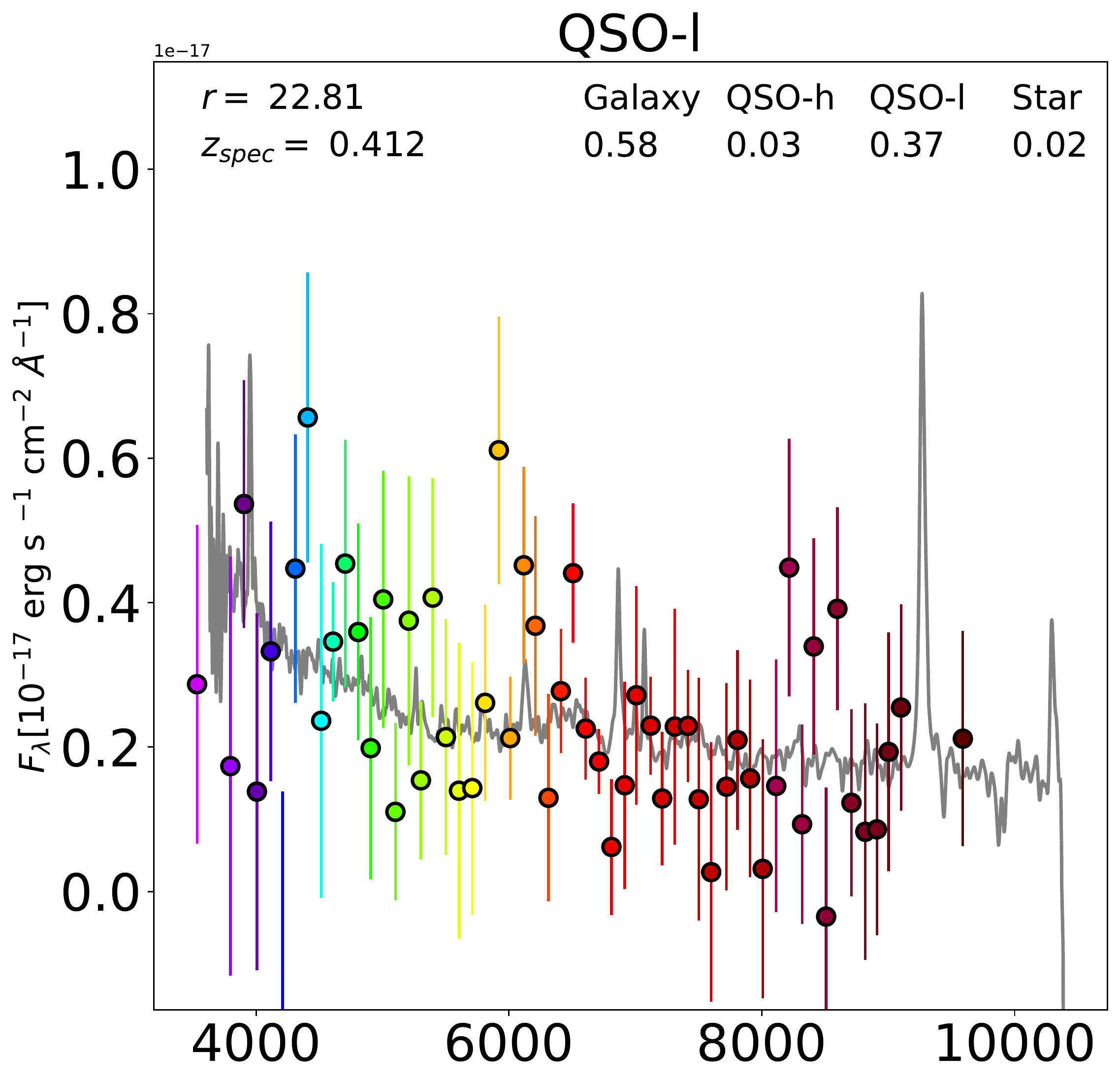}
\includegraphics[width=5.9cm,height=5.6cm]{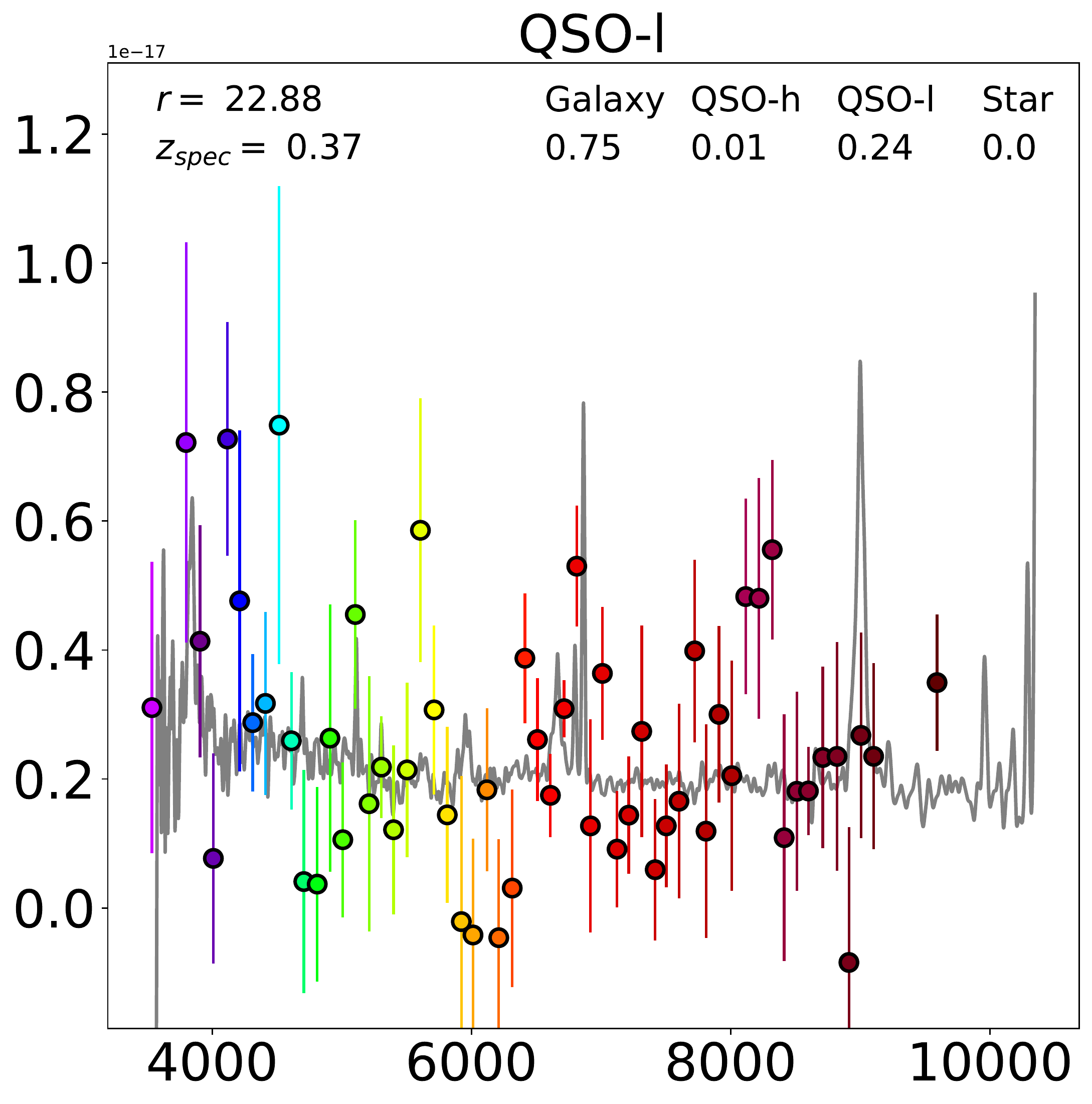}
\includegraphics[width=5.9cm,height=5.6cm]{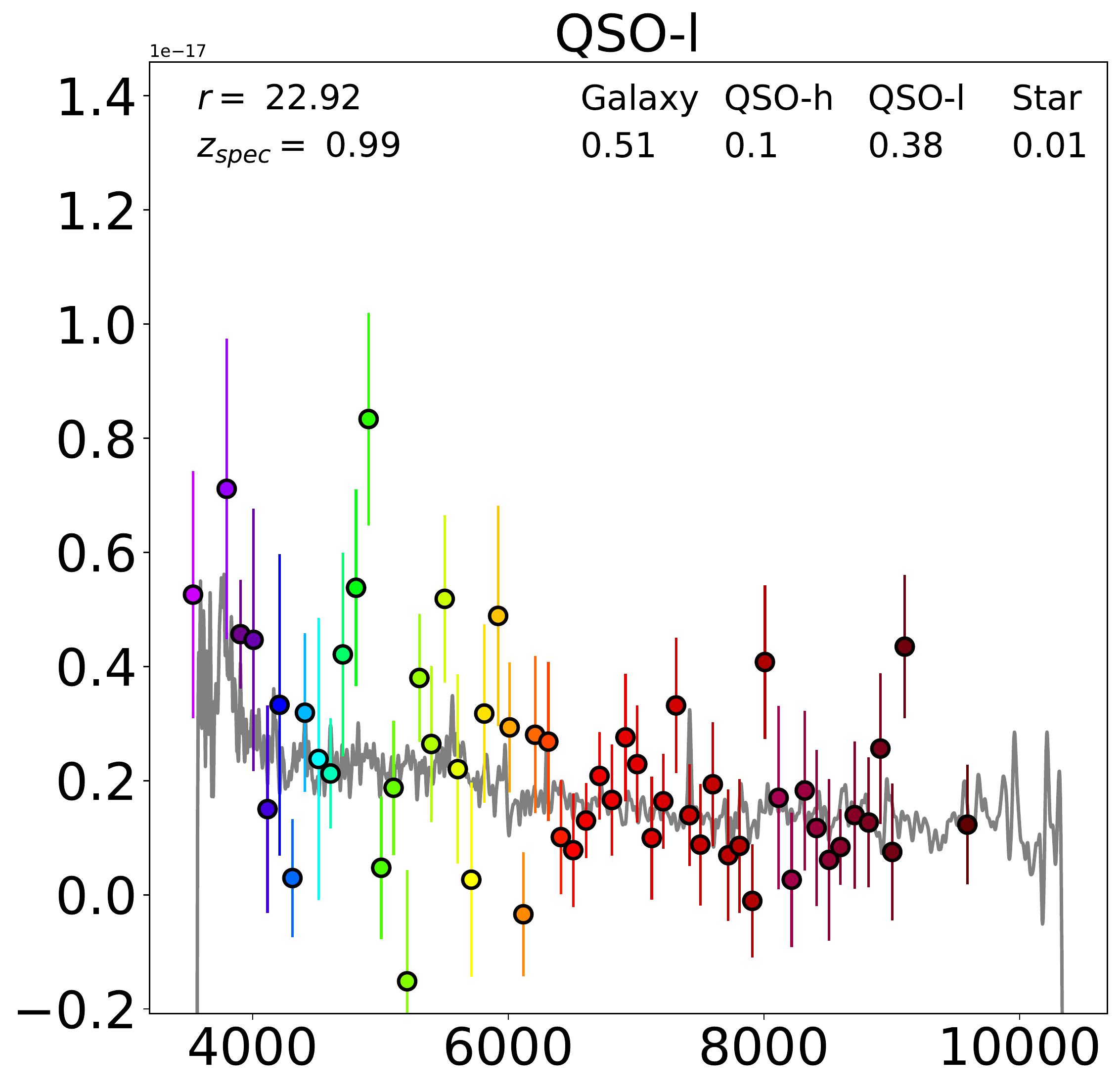}
\includegraphics[width=5.9cm,height=5.6cm]{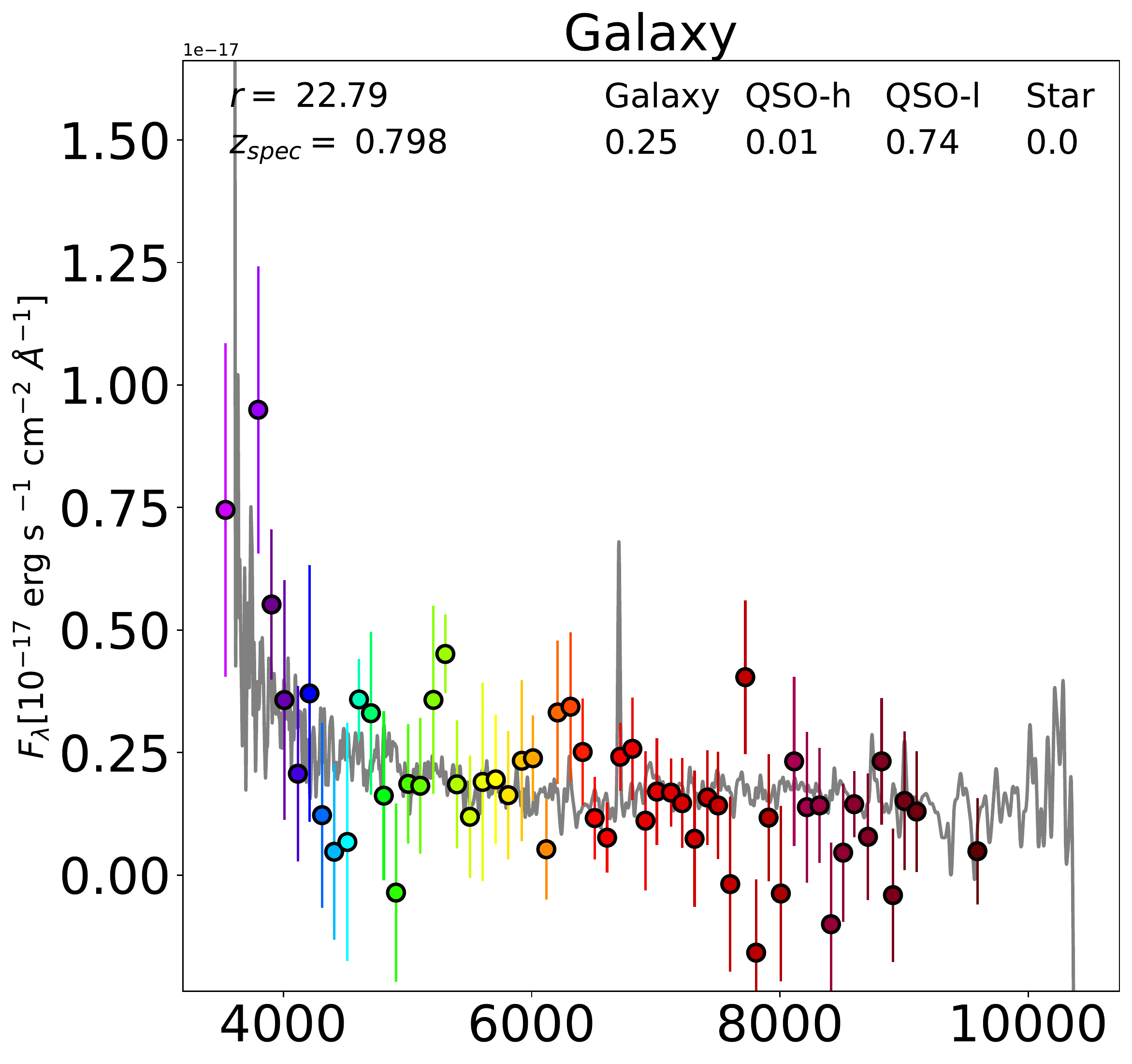}
\includegraphics[width=5.9cm,height=5.6cm]{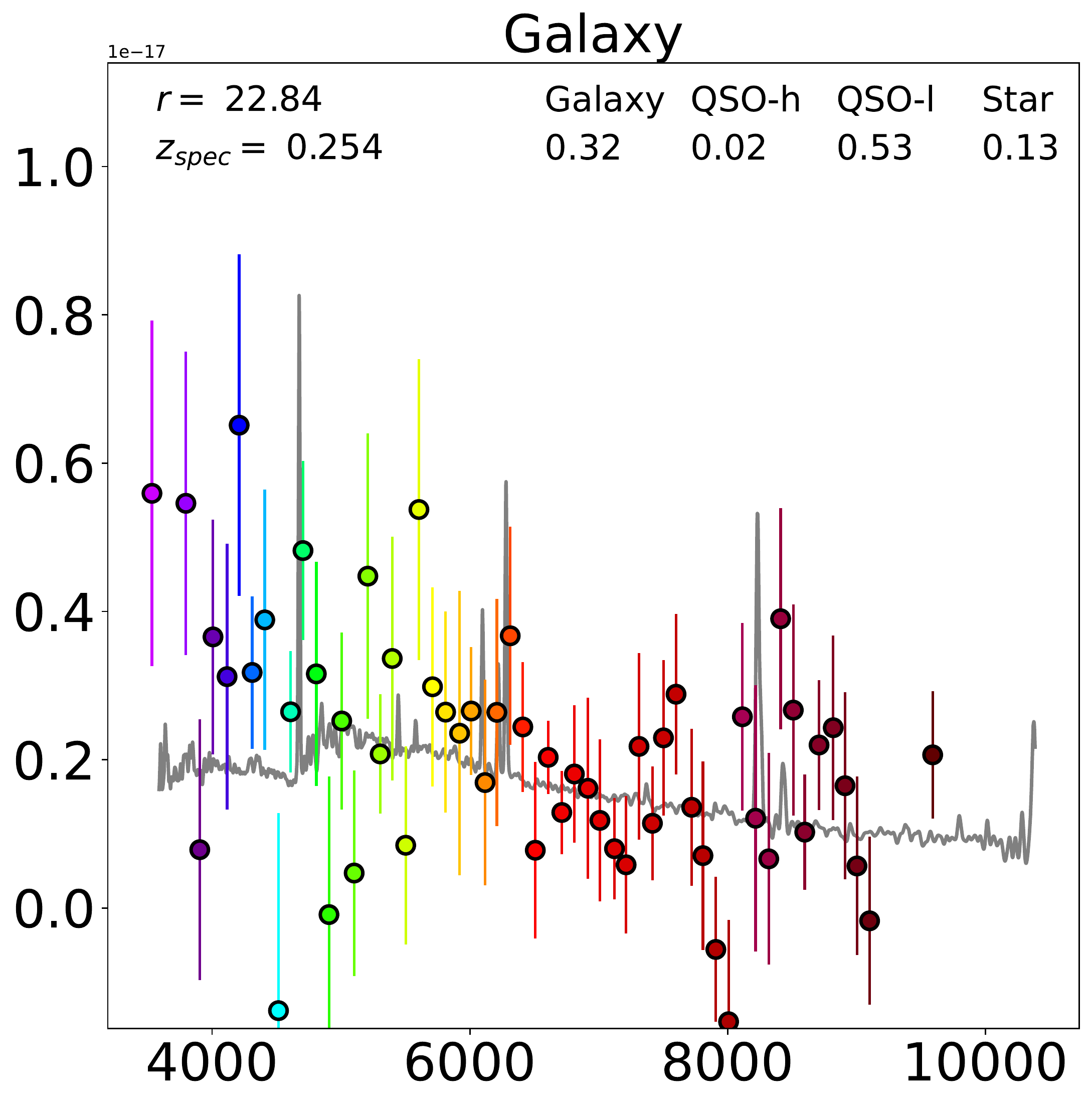}
\includegraphics[width=5.9cm,height=5.6cm]{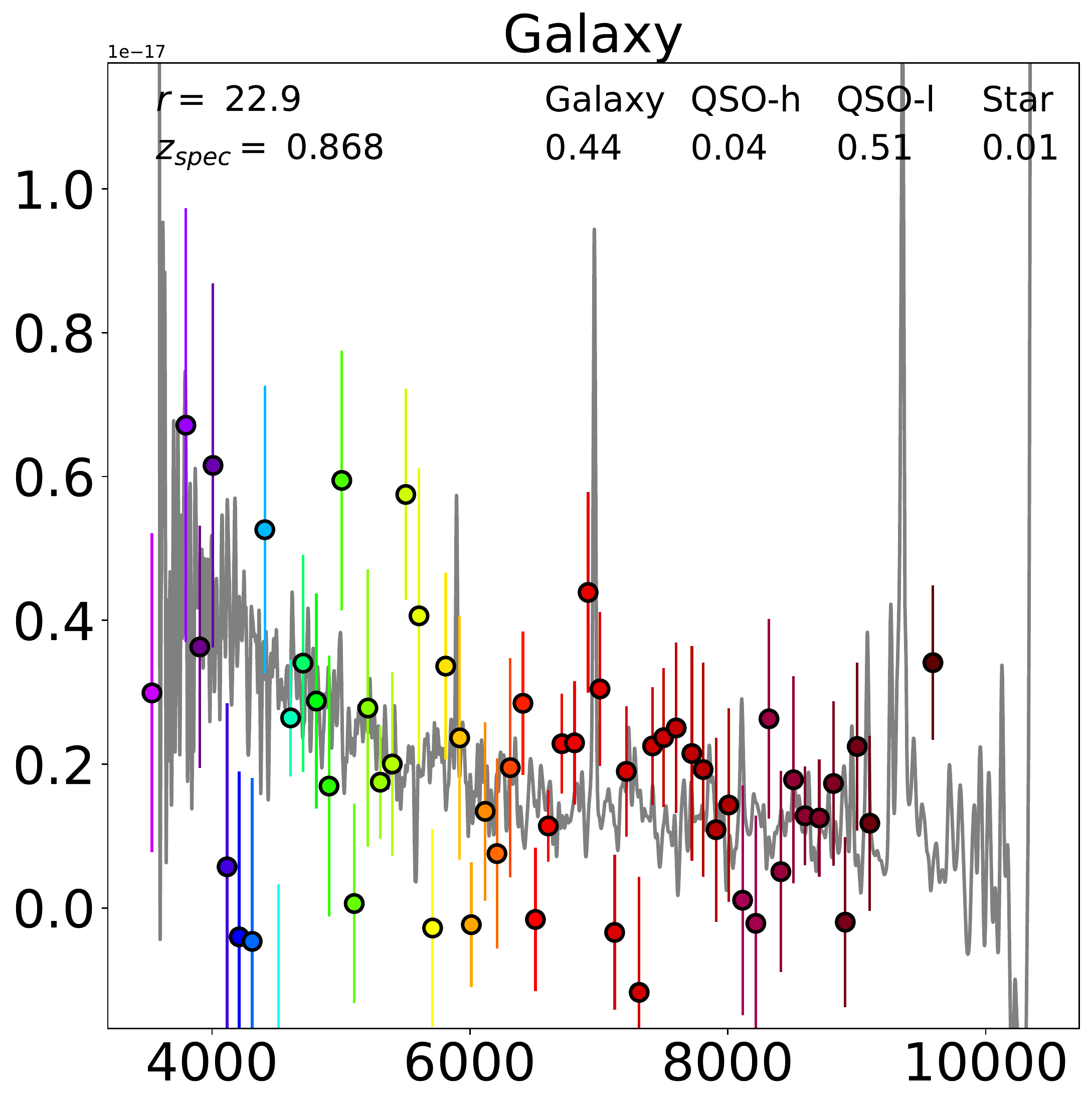}
\includegraphics[width=5.9cm,height=5.6cm]{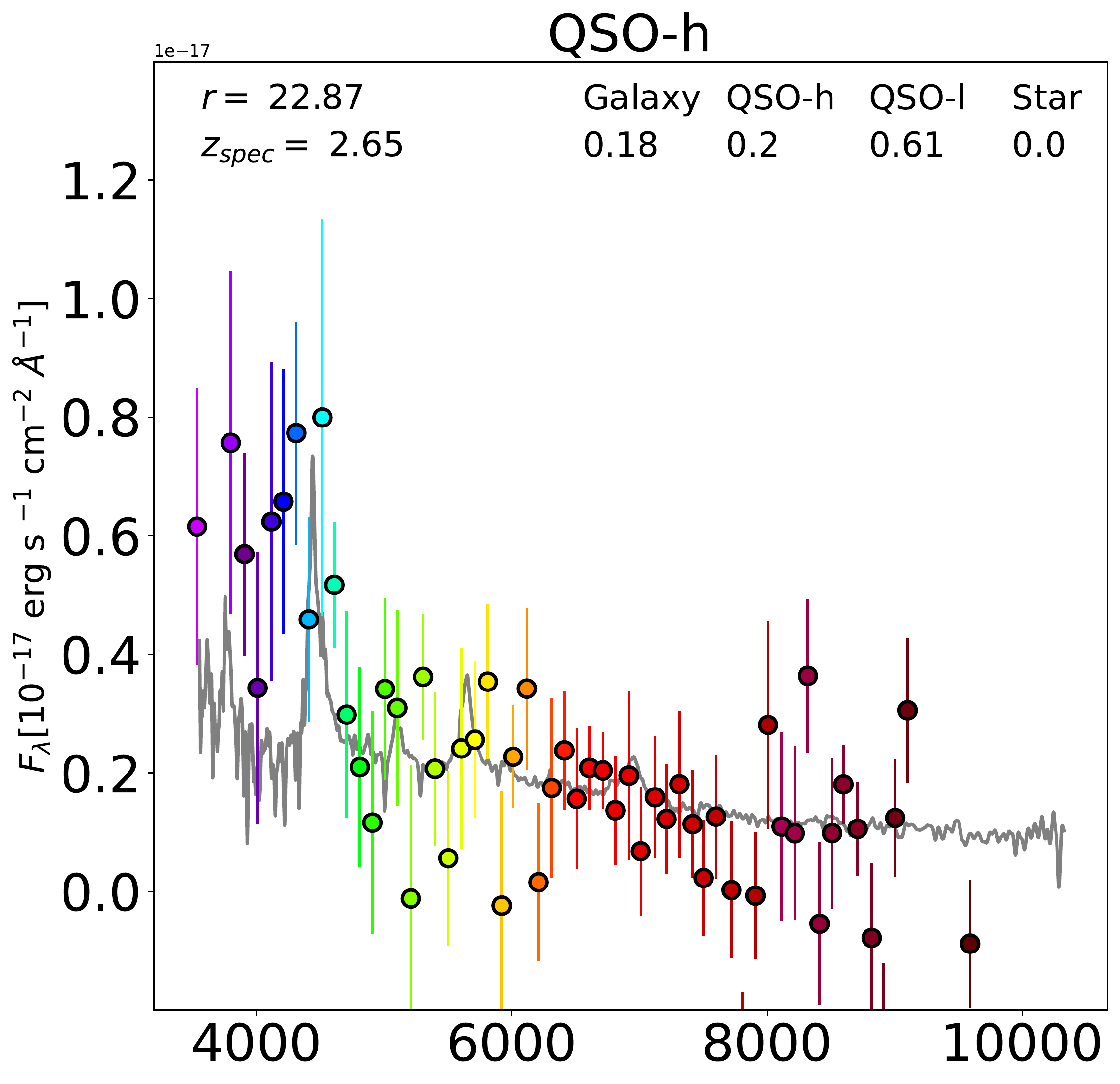}
\includegraphics[width=5.9cm,height=5.6cm]{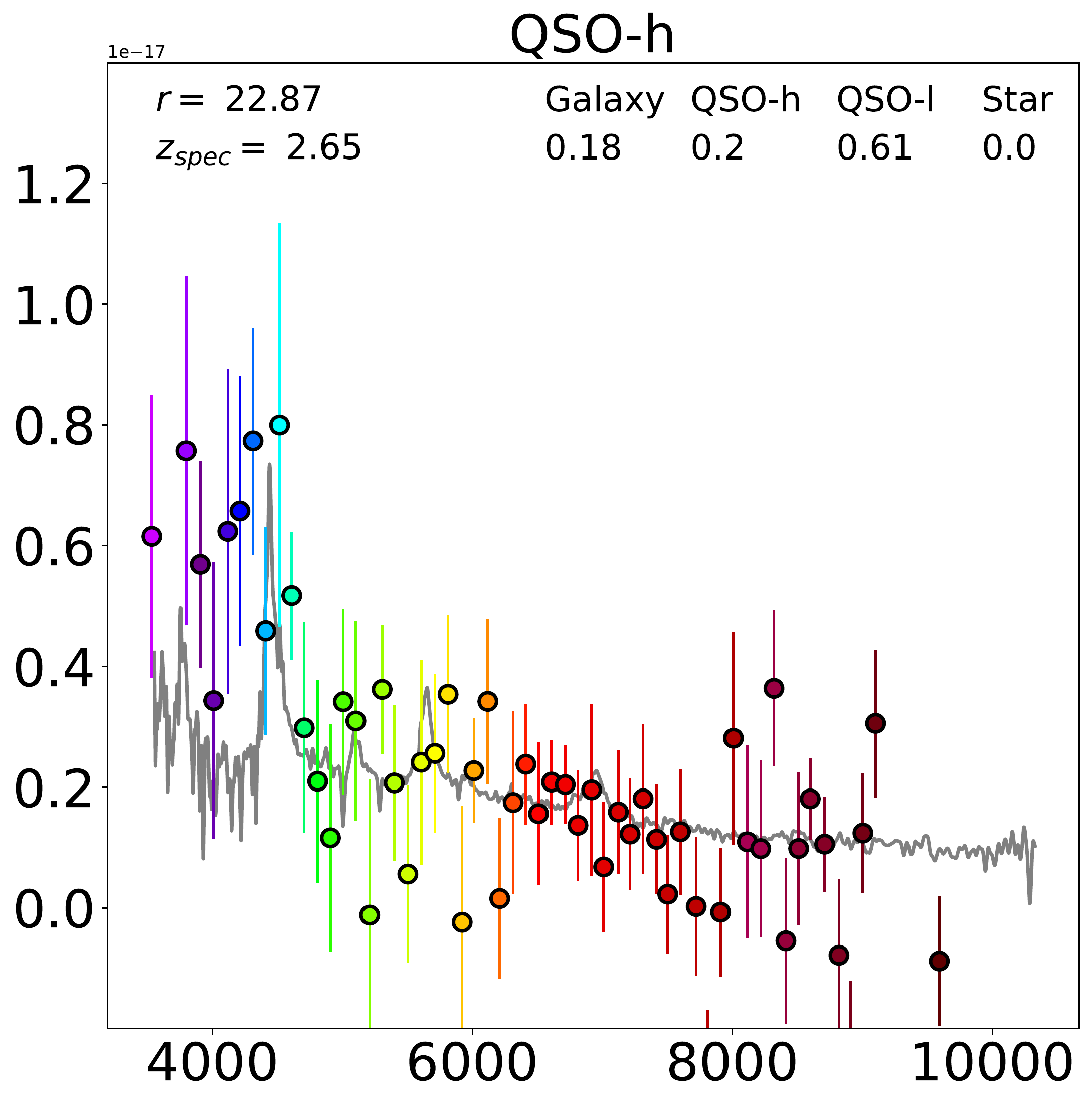}
\includegraphics[width=5.9cm,height=5.6cm]{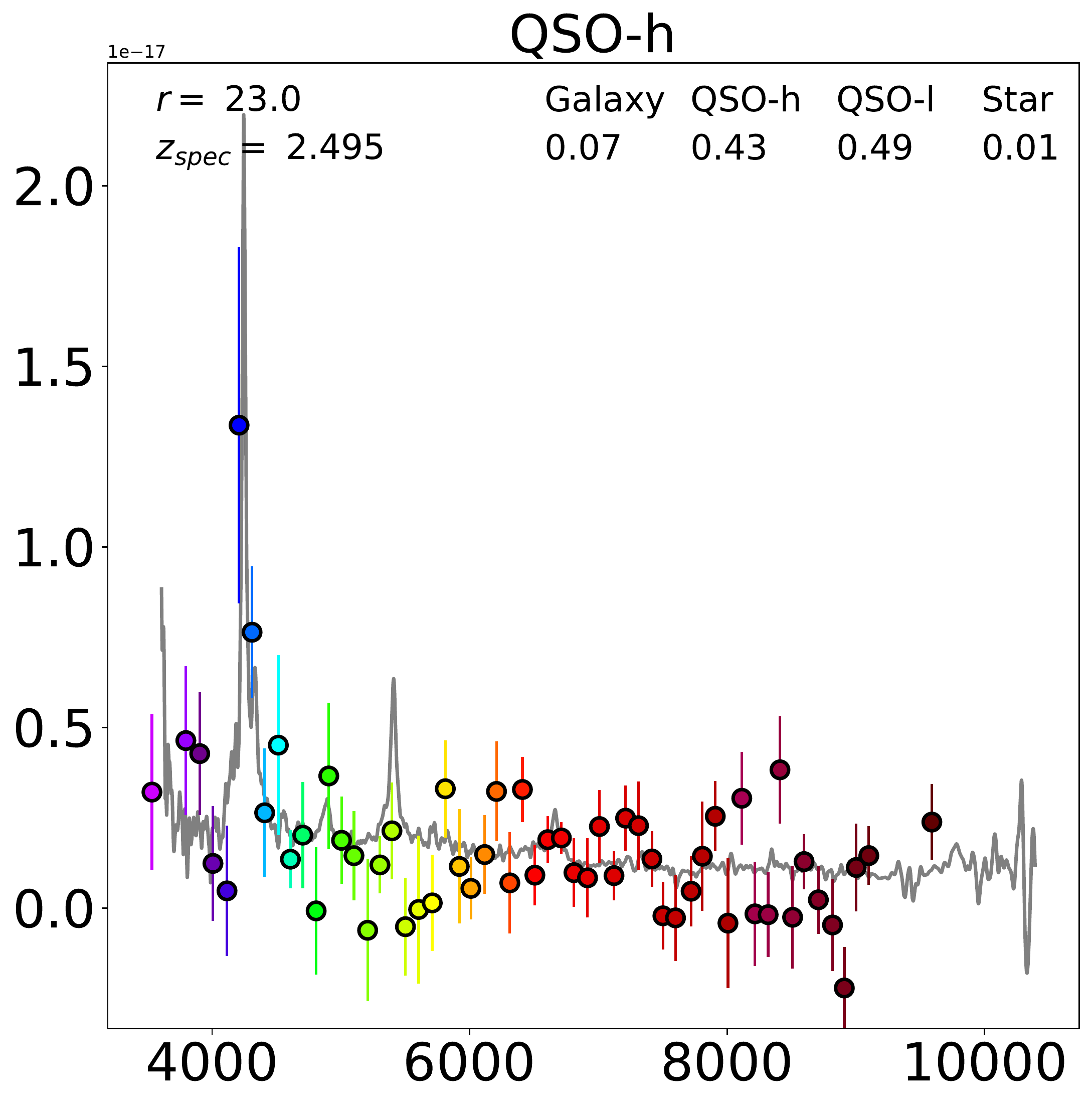}
\includegraphics[width=5.9cm,height=5.6cm]{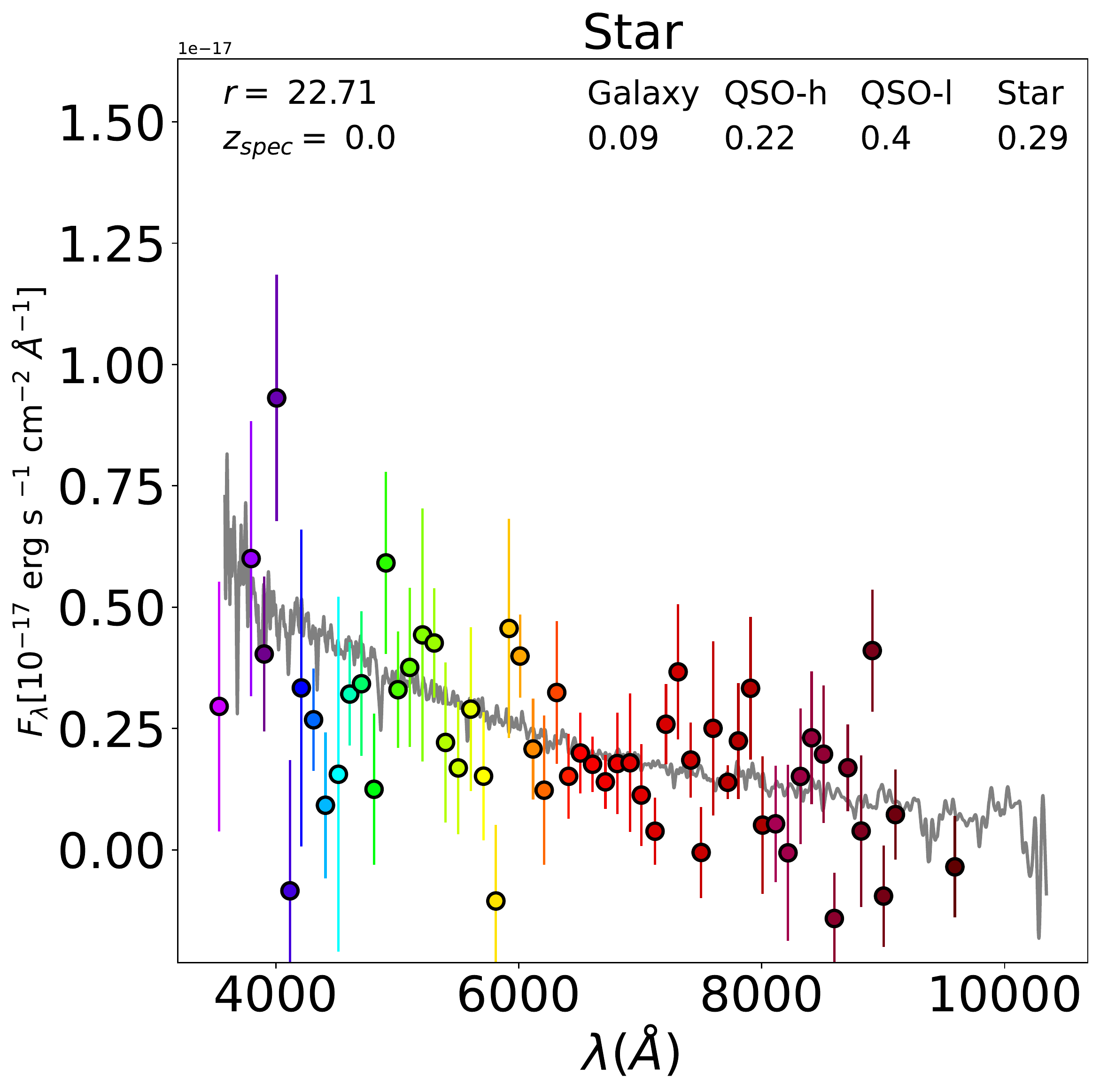}
\includegraphics[width=5.9cm,height=5.6cm]{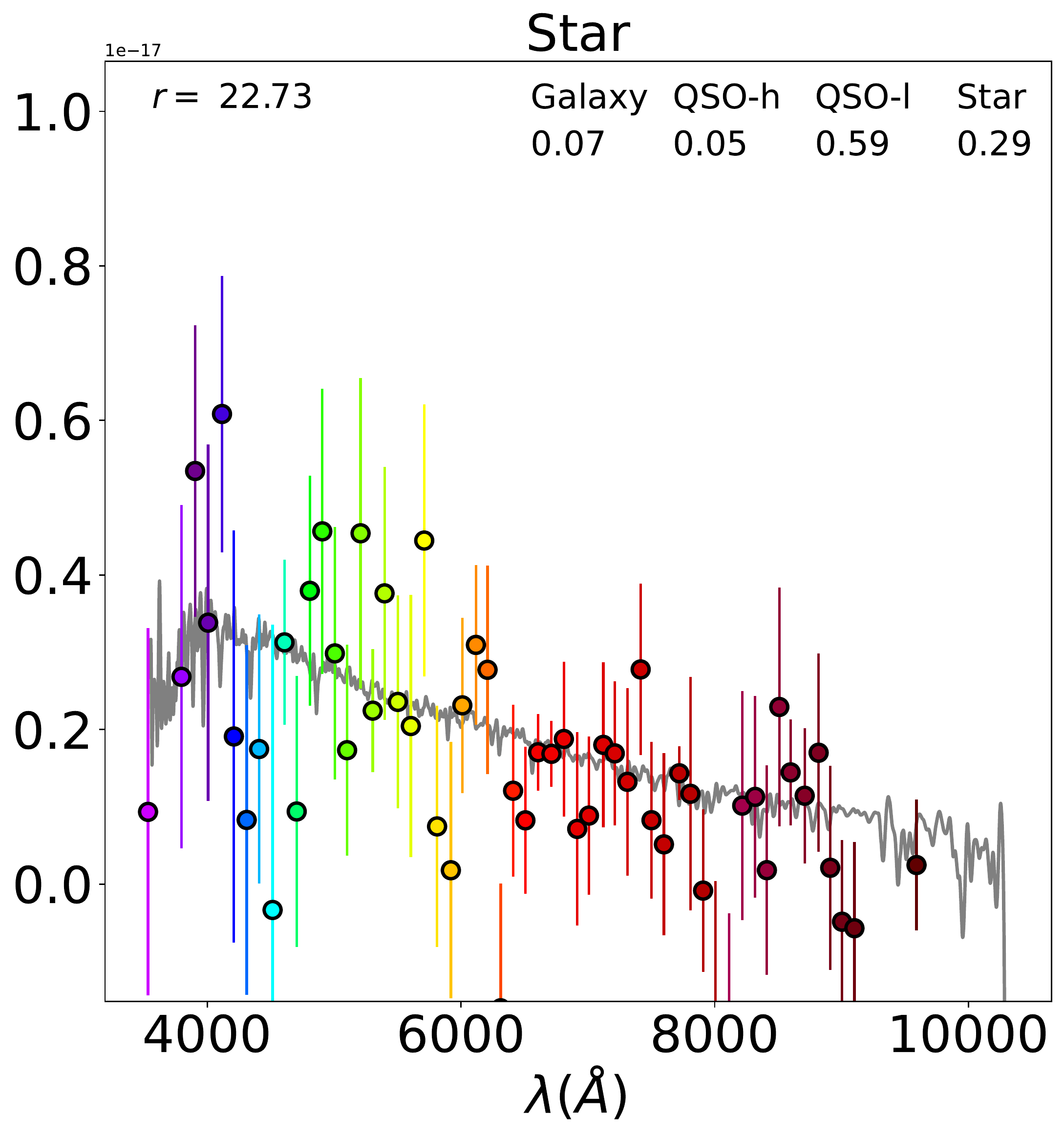}
\includegraphics[width=5.9cm,height=5.6cm]{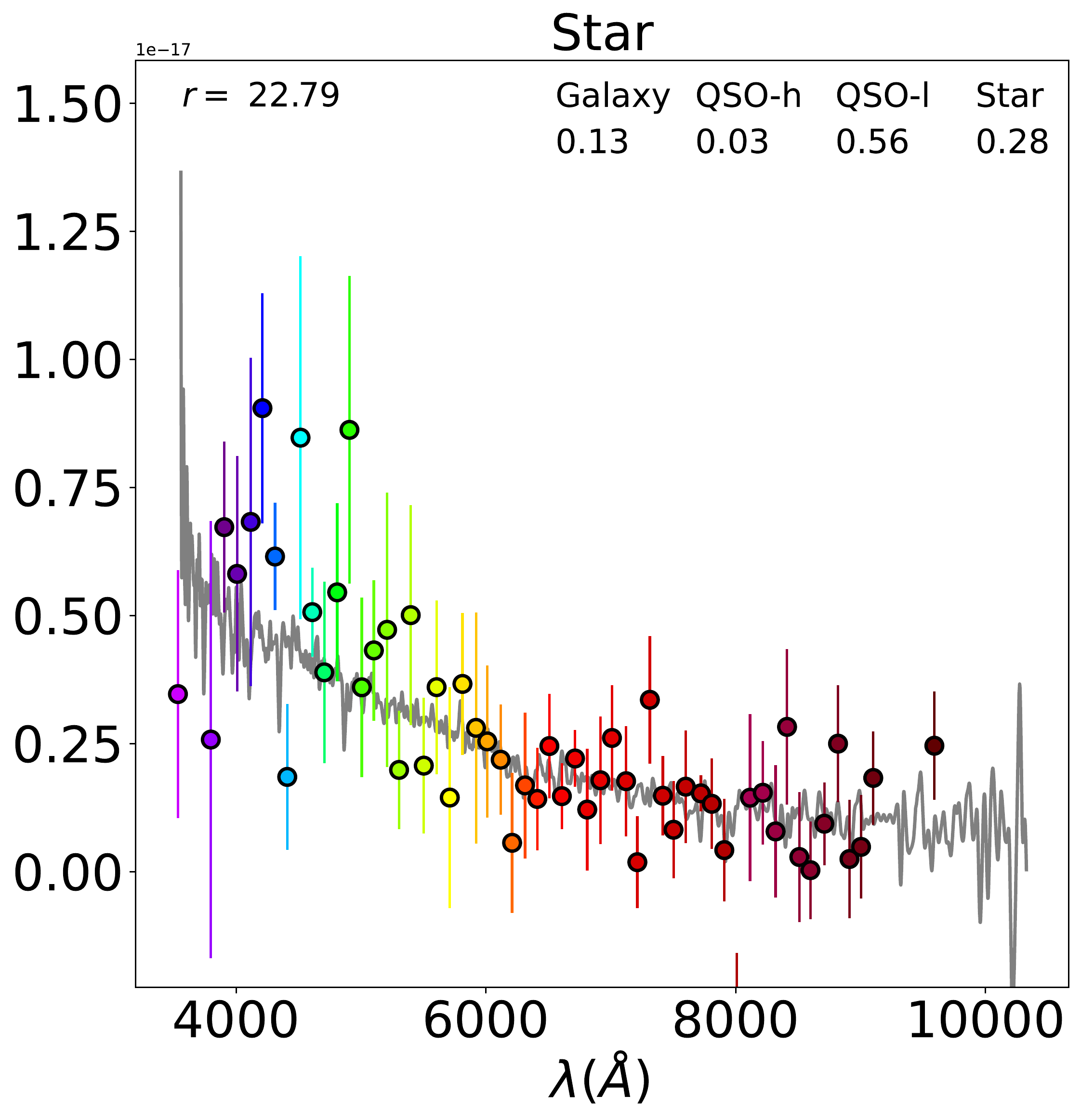}
\caption{{Examples of the most typical misclassification of objects in the mock test sample. The first row shows QSO-l classified as galaxies, the second row shows galaxies classified as QSO-l, the third row shows QSO-h classified as QSO-l, and the fourth row shows stars classified as QSO-l. From left to right, the objects are fainter. In each panel, we indicate the AB magnitude in the $r$ band, the redshift (top left), and the probabilities yielded by the ANN$_1$ classifier for each one of the classes (top right). The SDSS spectra from which the test sample is generated are shown in the background in grey.}}
\label{fig:mixobj_test}
\end{figure*}

 In Fig. \ref{fig:mixobj_test} we show examples of the most common missclassifications. The first row is composed of QSO-l that were classified as galaxies. In the second row, we show galaxies that were identified as QSO-l, the third row corresponds to QSO-h confused with QSO-l, and the last row shows stars classified as QSO-l. Even though we were unable to correctly predict the class of these objects, the second most likely class usually coincides with the actual class. Furthermore, it is important to emphasise that the objects shown in Fig. \ref{fig:mixobj_test} would be very difficult to  identify via visual inspection even for a human expert without seeing the spectrum. ML algorithms are indeed pushing the limits beyond the human capability. 

\par
\begin{figure}
    \centering
        \includegraphics[width=\hsize]{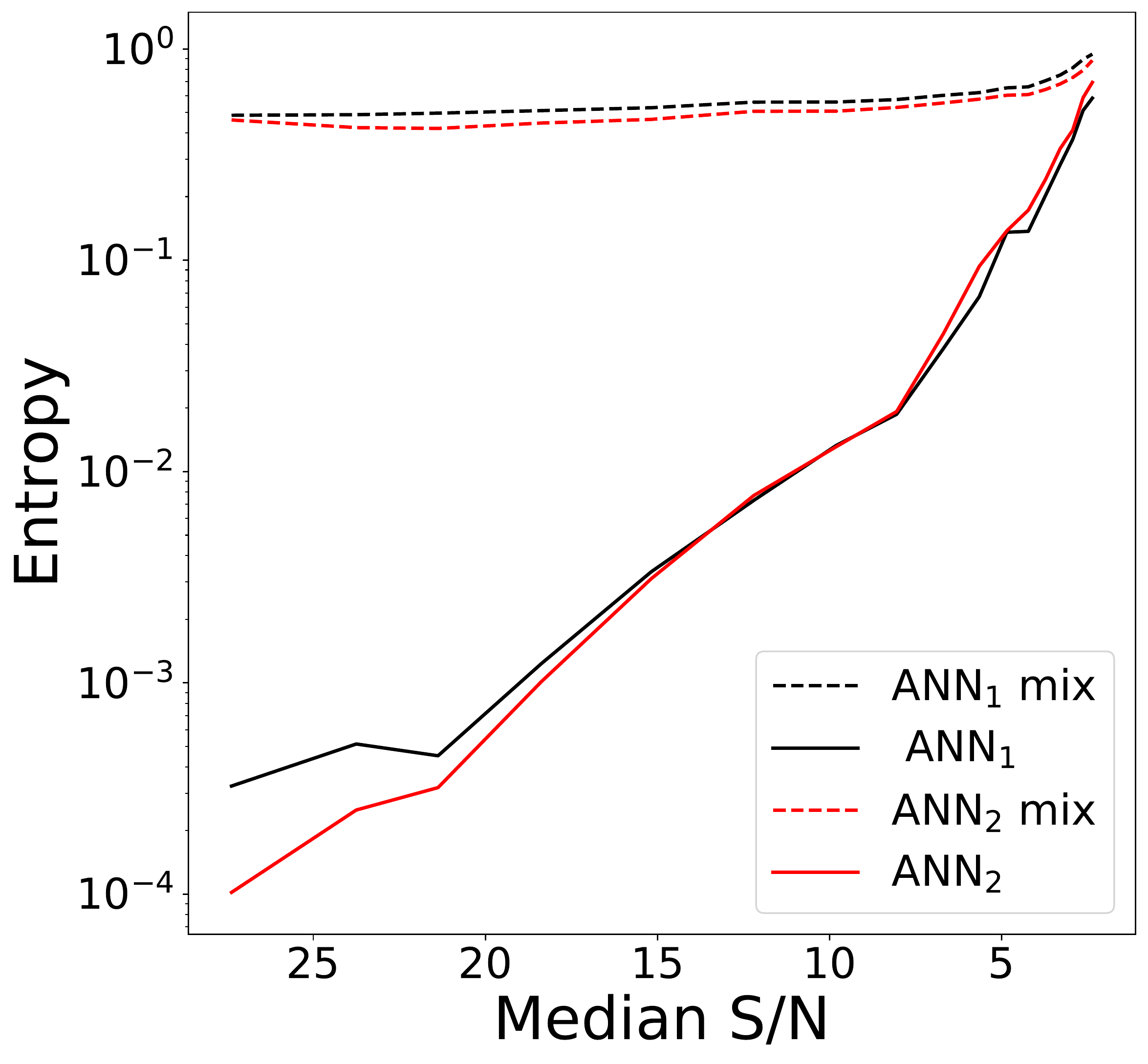}
        \caption{{Median entropy as a function of the median S/N in bins of $1000$ objects}.}
\label{fig:entropy}
\end{figure}

 It it expected that the ANN predictions for low S/N objects are more uncertain than those with high S/N. In Fig. \ref{fig:entropy} we show the median entropy as a function of the median S/N in bins of $1000$ objects. While the predictions are highly certain in the high S/N limit with the ANN$_1$ and ANN$_2$ classifiers, the entropy obtained by the ANNs trained in the hybrid sets remains almost constant from an S/N of 25 to 10 with a value of $\sim 0.4,$ and then it increases slightly. The lowest entropy obtained with the ANN$_1$ mix and ANN$_2$ mix classifiers is governed by the mixing coefficient ($\beta$) used to generate the hybrid set in Eq. \ref{eq:hyb}, and it coincides with the median entropy of the hybrid classes in the training set. In Sect. \ref{subsec:hyb} we chose $\beta$ to be $0.1$. This choice was arbitrary, but it does not have a major impact on the performance of the algorithms. Mainly the entropy (and probabilities) of the classifier solution is affected. On the one hand, as $\beta$ increases, the entropy of the hybrid training set increases up to a maximum value at which all objects reach equal probabilities. For these values, we expect the performance to be degraded because the class information completely vanished. On the other hand, for very low values of $\beta$, we recover the original training set, thus no hybridisation is performed in practice. In essence, no optimal value for $\beta$ exists that might improve the classifier performance and confidence simultaneously.
\par
\begin{figure*}
        \includegraphics[width=7.4cm,height=6.1cm]{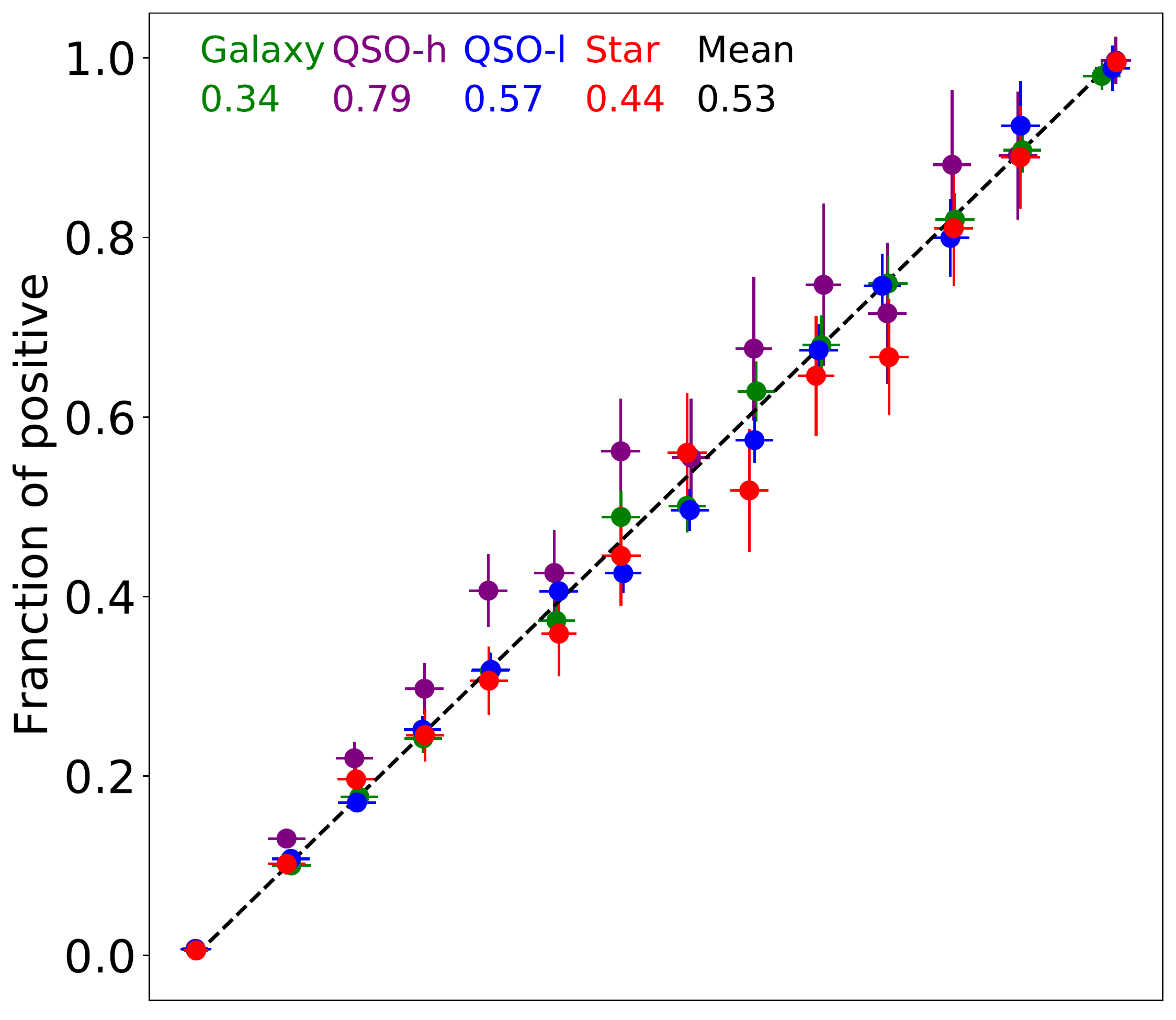}
        \includegraphics[width=0.08cm,height=6.1cm]{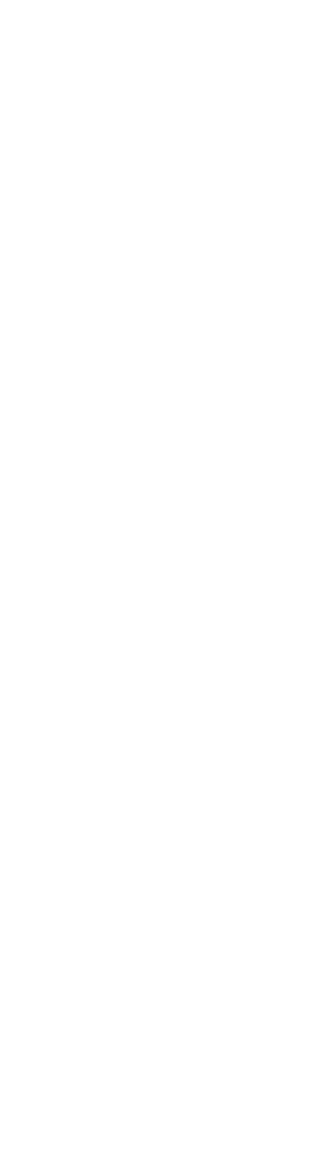}
        \includegraphics[width=6.5cm,height=6.1cm]{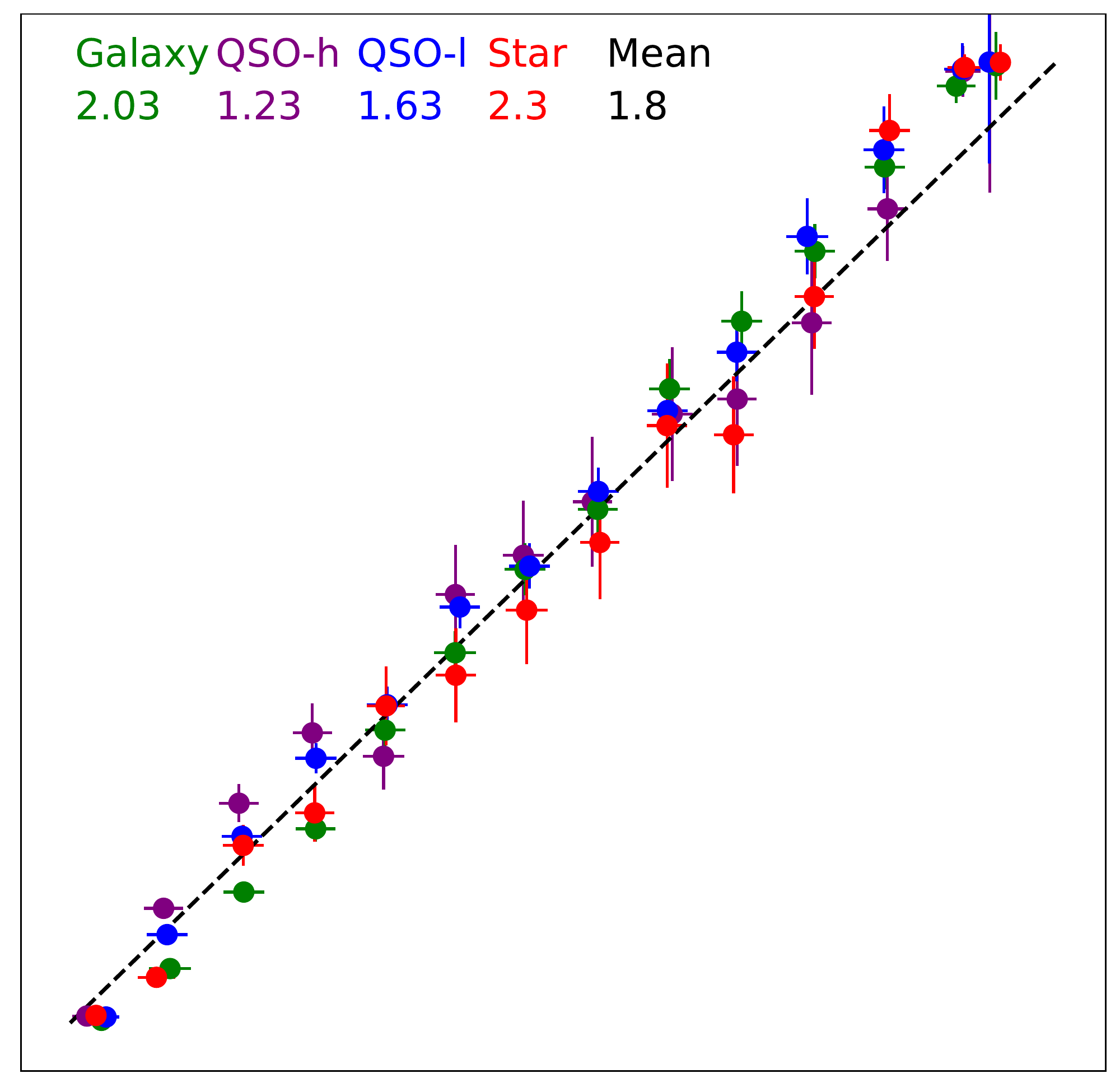}

        \includegraphics[width=7.4cm,height=6.5cm]{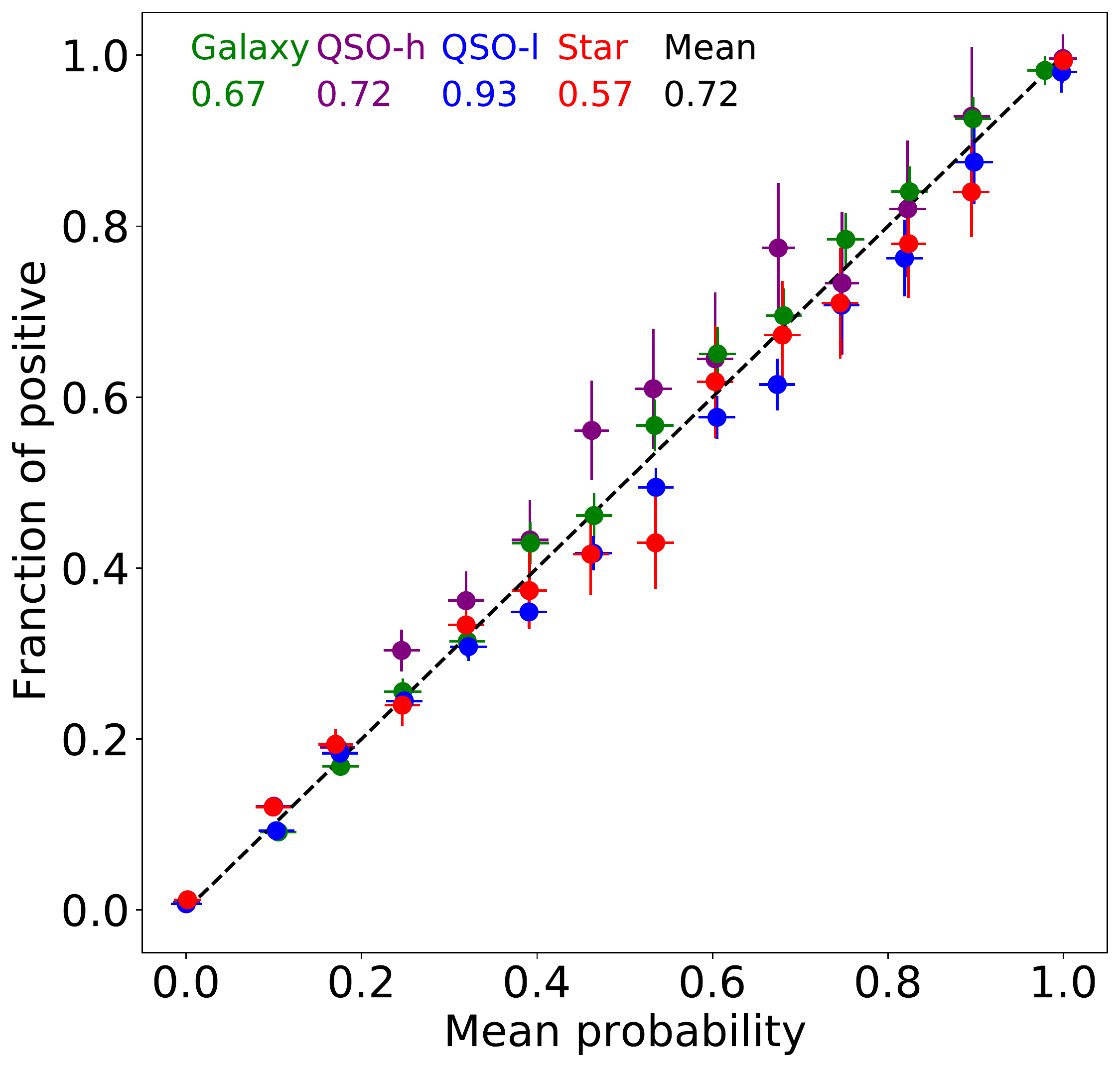}
        \includegraphics[width=0.07cm,height=6.5cm]{Images/barw.png}
        \includegraphics[width=6.5cm,height=6.5cm]{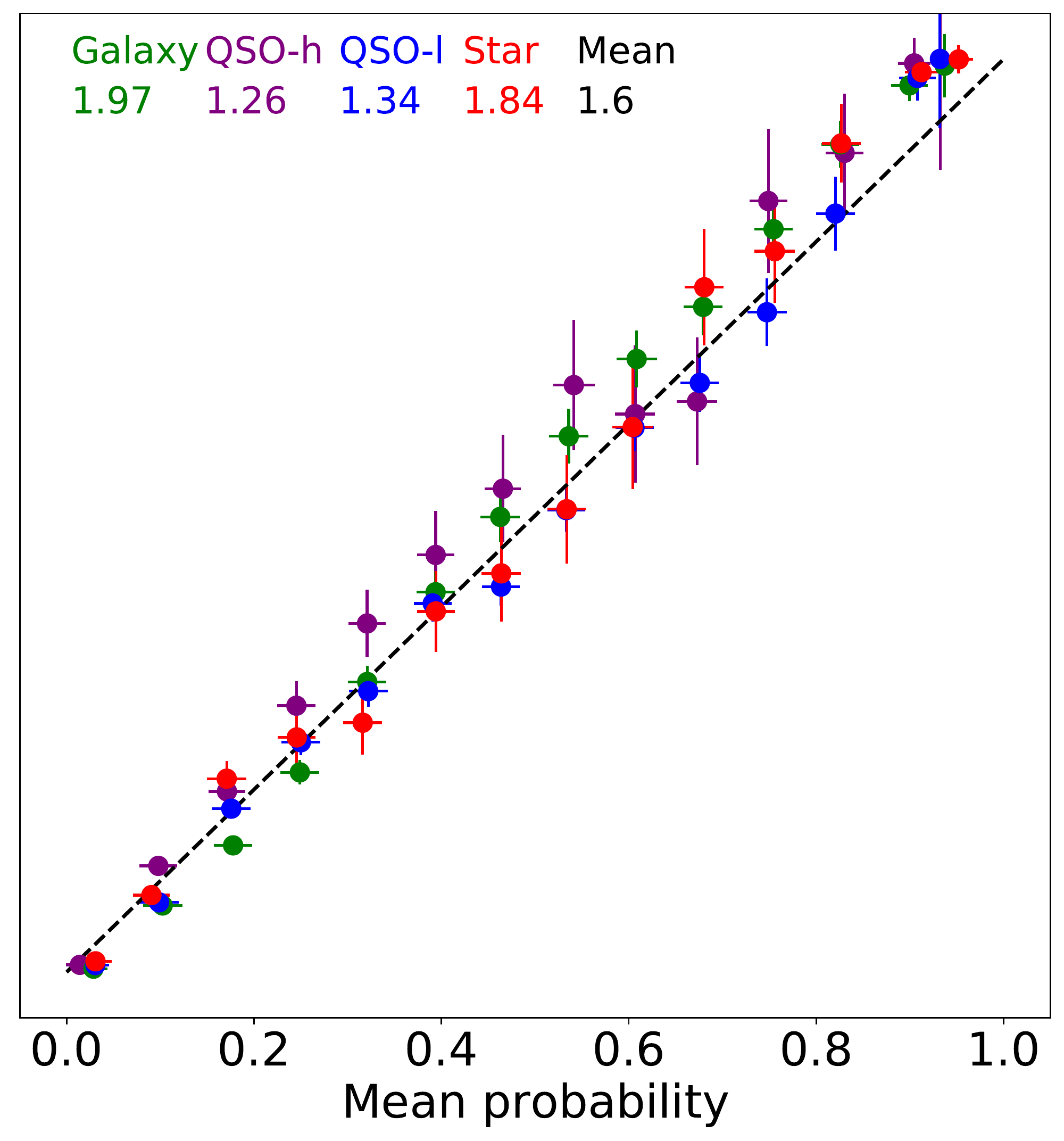}

        \caption{{Fraction of positives for each one of the classes as a function of the mean predicted probability. The ECE for galaxies, QSO-h, QSO-l, stars, and the mean ECE are shown at the top left side of each panel. The error bar represents the standard deviation in each one of the bins.  The left panels are trained with the original training set, and the right panel used the hybrid set. The top (bottom) panels show the results of ANN$_1$ (ANN$_2$).}}
\label{fig:ECE_image}
\end{figure*}

 In Fig. \ref{fig:ECE_image} we show the fraction of positive detection for each one of the classes as a function of the mean probability obtained with the ANNs in the mock test sample (r $\leq$ 23.6). In the top (bottom) left panel, we show the results of the ANN$_1$ (ANN$_2$) predictions trained with the original training set, and in the right panel, we show the predictions obtained with the hybrid set. The ECEs for galaxies, QSO-h, QSO-l, and stars are shown in the top left panel. Training with hybrid classes has a negative impact on the calibration. Once again, the ECE is a function of the mixing coefficient: as $\alpha$ increases, the ECE increases. Overall, the ANN$_1$ is slightly better calibrated than the ANN$_2$, but ANN$_2$ mix is better than ANN$_1$ mix. 
\par
\begin{figure}
    \centering
        \includegraphics[width=\hsize]{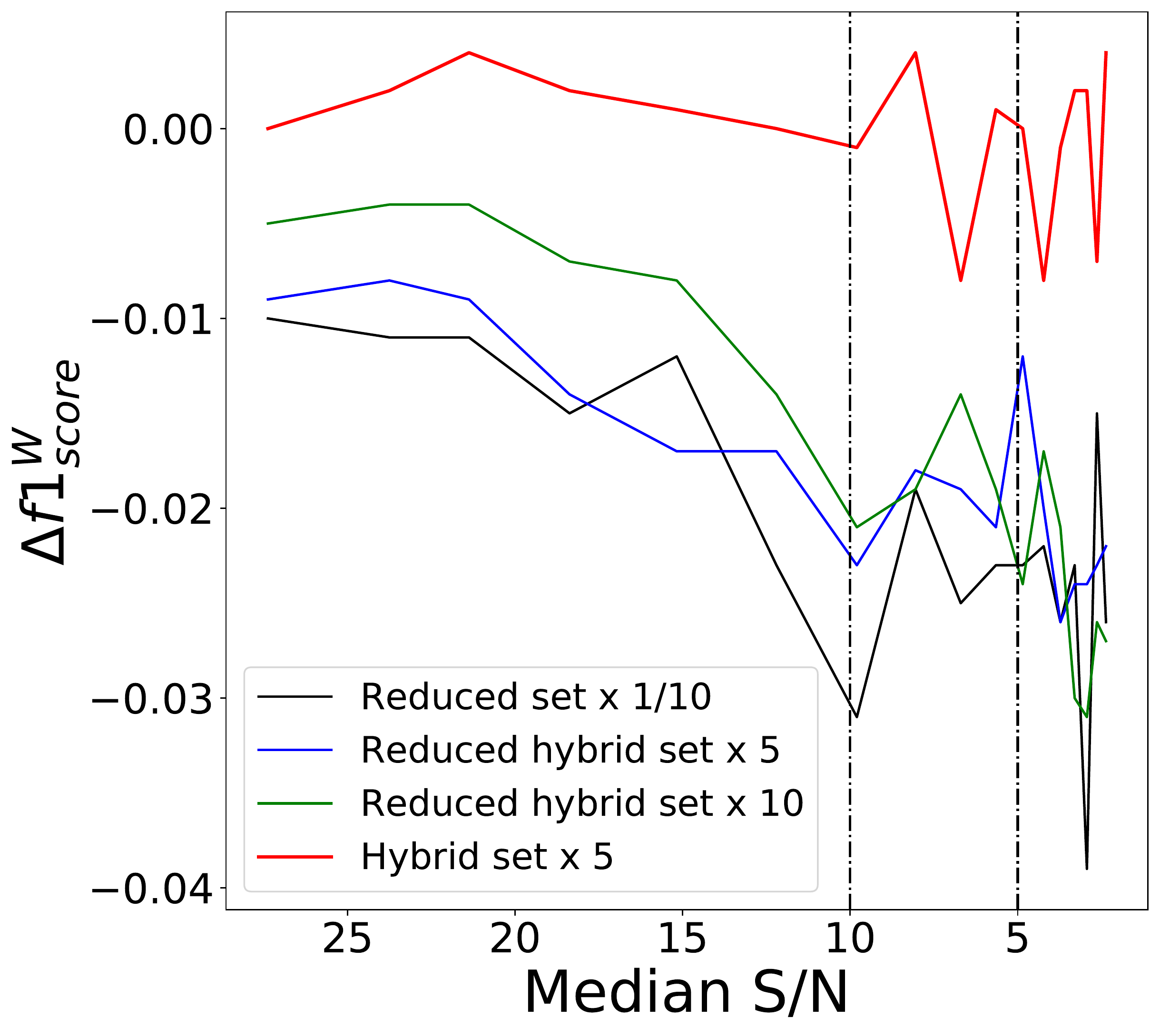} 
        \caption{Difference between the f$^W_1$ score obtained with the ANN$_1$ trained in the original training set and the reduced set (ten times smaller), the hybrid set (five time larger), the reduced hybrid set x 5 (five times larger than the reduced set), and the reduced hybrid set x 10 (ten times larger than the reduced set) as a function of the median S/N. Dashed vertical lines indicate an S/N of 10 and 5.}
\label{fig:f1_score_SN_reduced}
\end{figure}

 It is worth considering whether hybridisation improves the performance of the ANNs when the training set is smaller. The original training set in the mock catalogue is composed of $300\, 000$ objects, and the hybrid set is five times larger. We assumed a training set that was ten times smaller than the original set (reduced set). After applying hybridisation, we generated two new training sets that were five and ten times larger than the reduced set, respectively, known as the reduced hybrid set x5 and the reduced hybrid set x 10. We then compared the performance of ANN$_{1}$ in the mock test sample. In Fig. \ref{fig:f1_score_SN_reduced} we show the difference between the f$^W_1$ score in each one of the mentioned training sets and the f$^W_1$ score obtained in the original training set as a function of the median S/N. We do not observe a significant improvement that might justify the use of hybridisation, at least in the form we implemented in Eq. \ref{eq:hyb} for this particular data set.
\par
\begin{figure}
    \centering
        \includegraphics[width=\hsize]{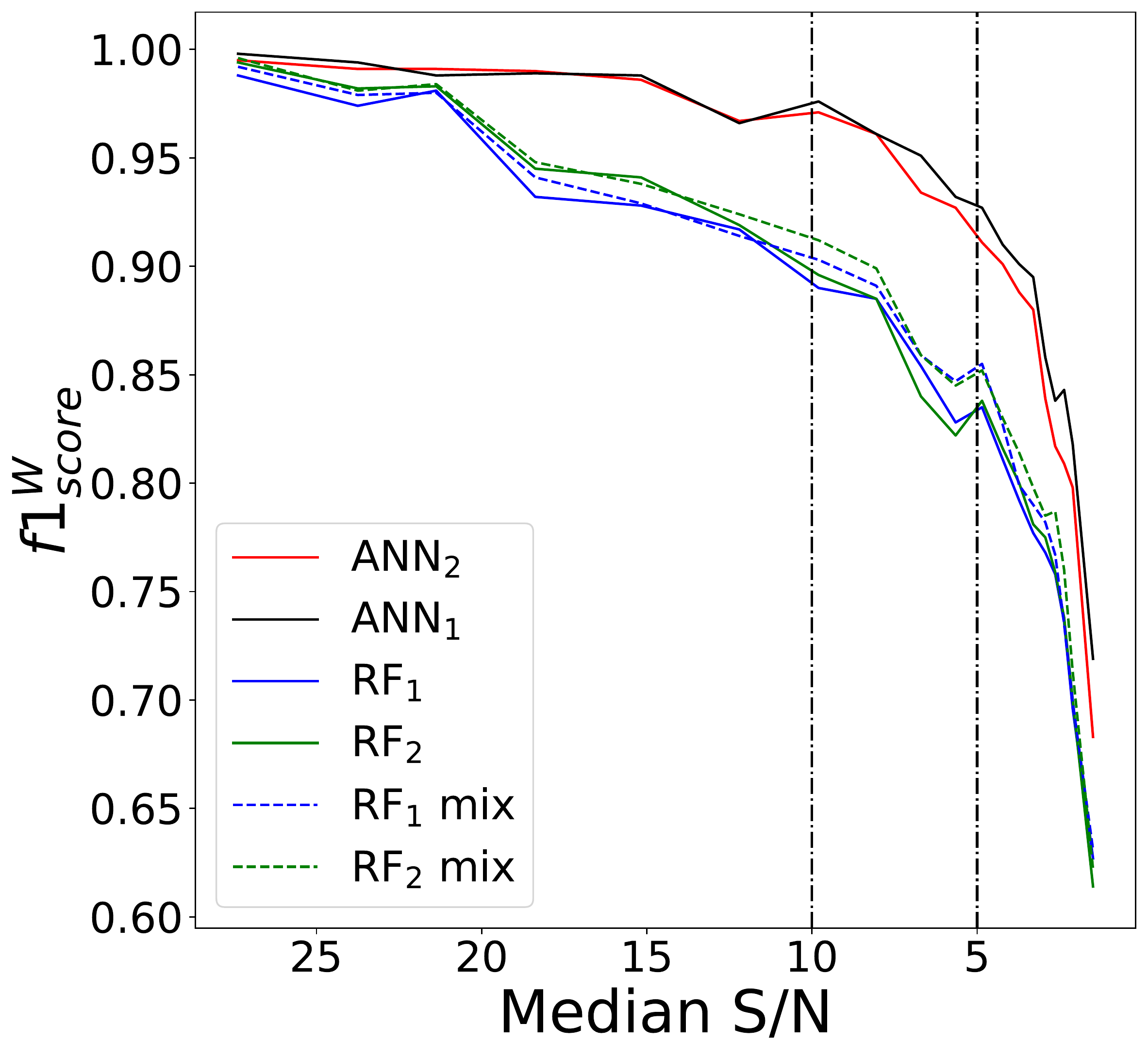} 
        \caption{f$^W_1$ score obtained with ANN$_{1}$ and ANN$_{2}$ as a function of the median S/N. Dashed vertical lines indicate an S/N of 10 and 5.}
\label{fig:f1_score_SN}
\end{figure}

 Finally, we compared the performance of ANNs with the results of RF classifier. In Fig. \ref{fig:f1_score_SN} we show the weighted f$^W_1$ score as a function of the median S/N in the observed filters for all classifiers. Each bin contains roughly 1000 objects. We do not show the results of ANN$_1$ mix and ANN$_2$ mix because they are very similar with respect to their non-hybrid counterparts. The ANNs perform better and score 10 $\%$ higher on average than the RF classifiers. It is remarkable that even with a median S/N of 5, the f$^W_1$ score reaches 0.9. We do not observe a significant difference between RF$_1$ and RF$_2$ or ANN$_1$ and ANN$_2$ , which suggests that both representation of the data encode essentially the same information. Hybridisation seems to improve the performance of the RF slightly. Nevertheless, the ECE increases from 3.89 (3.82) up to 4.23 (4.19) for RF$_1$ and RF$_2$, respectively. The ECE obtained with the original training set is much larger that what we found with ANN. Thus, the RF classifier tends to be under-confident in its predictions.

\subsection{SDSS versus miniJPAS}\label{subsec:miniJPAS-SDSS}
\begin{figure}
    \centering
        \includegraphics[width=\hsize]{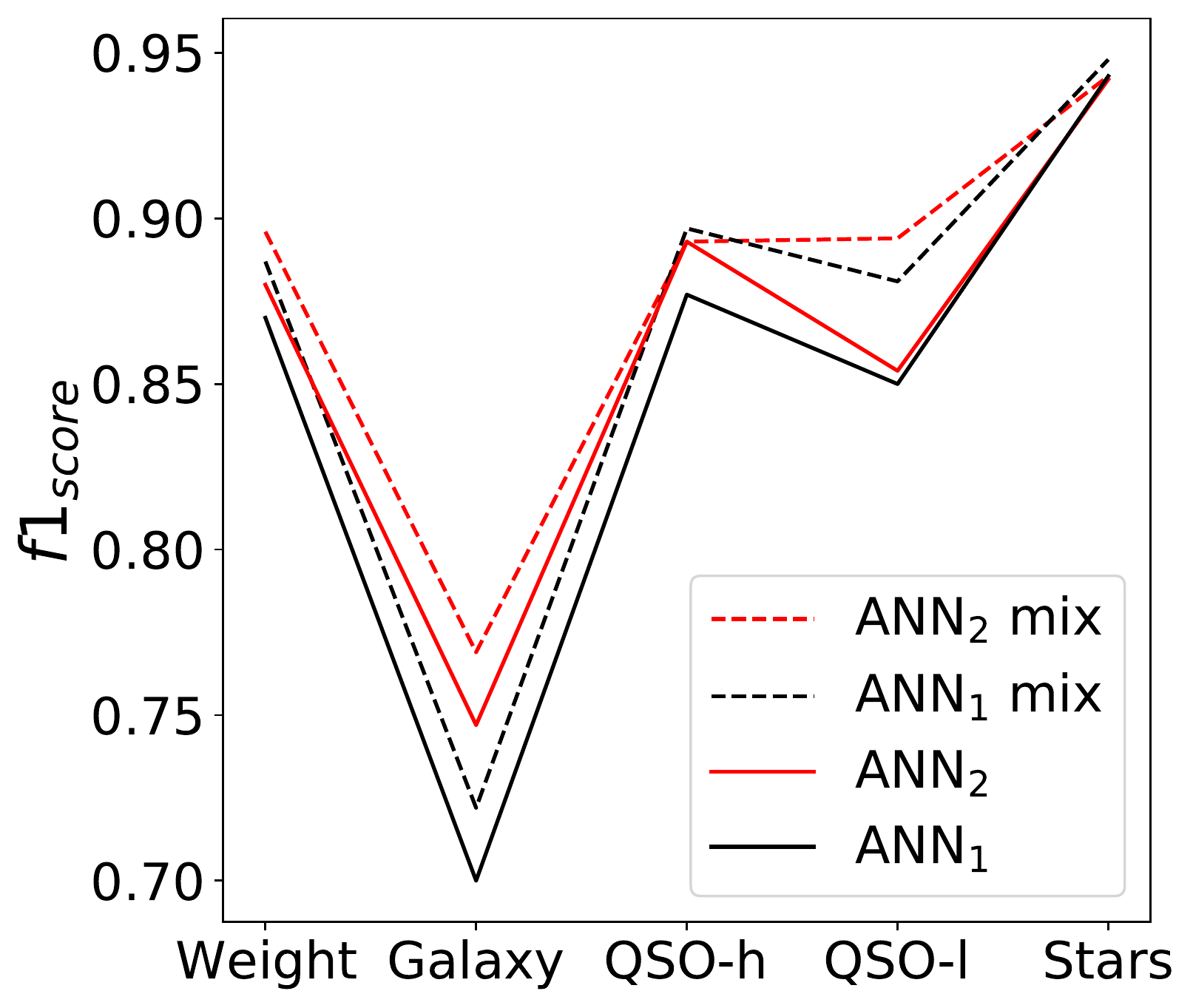}
        \caption{{$f^{W}_1$ score and $f_1$ score for each one of the classes obtained within the miniJPAS field observed and labelled according to the SDSS observations (see text in Sect. \ref{subsec:miniJPAS-SDSS}). Dashed (solid) lines represent the models trained with the hybrid (original) training set. ANN$_2$ and ANN$_2$ mix are trained with colours, while ANN$_1$ and ANN$_1$ mix are trained with fluxes (see Sect. \ref{subsec:ANN}).}}
    \label{fig:f1_score_SDSS}
\end{figure}

In this section, we test the ANN classifiers on the SDSS test sample. Fig. \ref{fig:f1_score_SDSS} shows the $f_1$ score for each class and the $f^{W}_1$ score. The performance of the algorithms that were trained with the hybrid set (ANN$_1$ mix and ANN$_2$ mix) are compared with the original training set (ANN$_1$ and ANN$_2$). Due to the limited number of objects, we did not separate these samples into magnitude bins. Most of the objects ($75\%$) belong to BIN 1, and only three are at the faint end (BIN 2). Therefore, the $f_1$ score is mostly representative of BIN 1. The results of the SSDS test sample are compatible with those obtained for the mock data ($ \sim 0.9$), suggesting that the simulations reproduce the miniJPAS observations fairly well at least for magnitudes brighter than 22.5. Unfortunately, we do not have enough labelled objects fainter than 22.5 within the miniJPAS field. We therefore needed to rely on the mock results for an expectation of the performance. As soon as WEAVE starts to observe the QSO target list provided by the J-PAS, we will be able to fully asses the performance of the algorithms for the full range of magnitudes.  
\par
\begin{figure}
    \centering
        \includegraphics[width=\hsize]{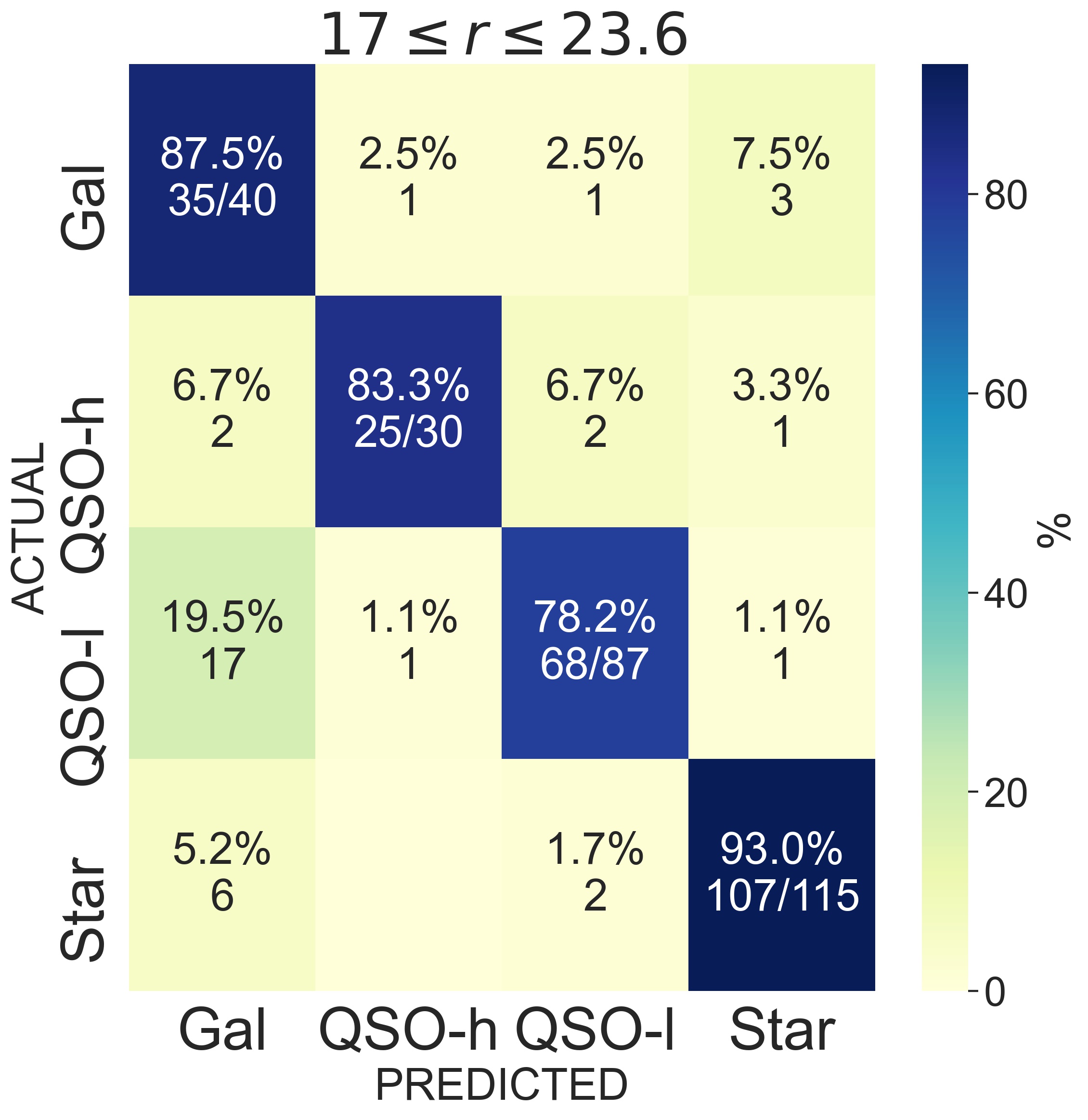}
        \caption{{Confusion matrix obtained with ANN1 for the SDSS test sample.}}
    \label{fig:CM_SDSS_ANN1}
\end{figure}

\begin{figure*}
    \includegraphics[width=5.9cm,height=5.6cm]{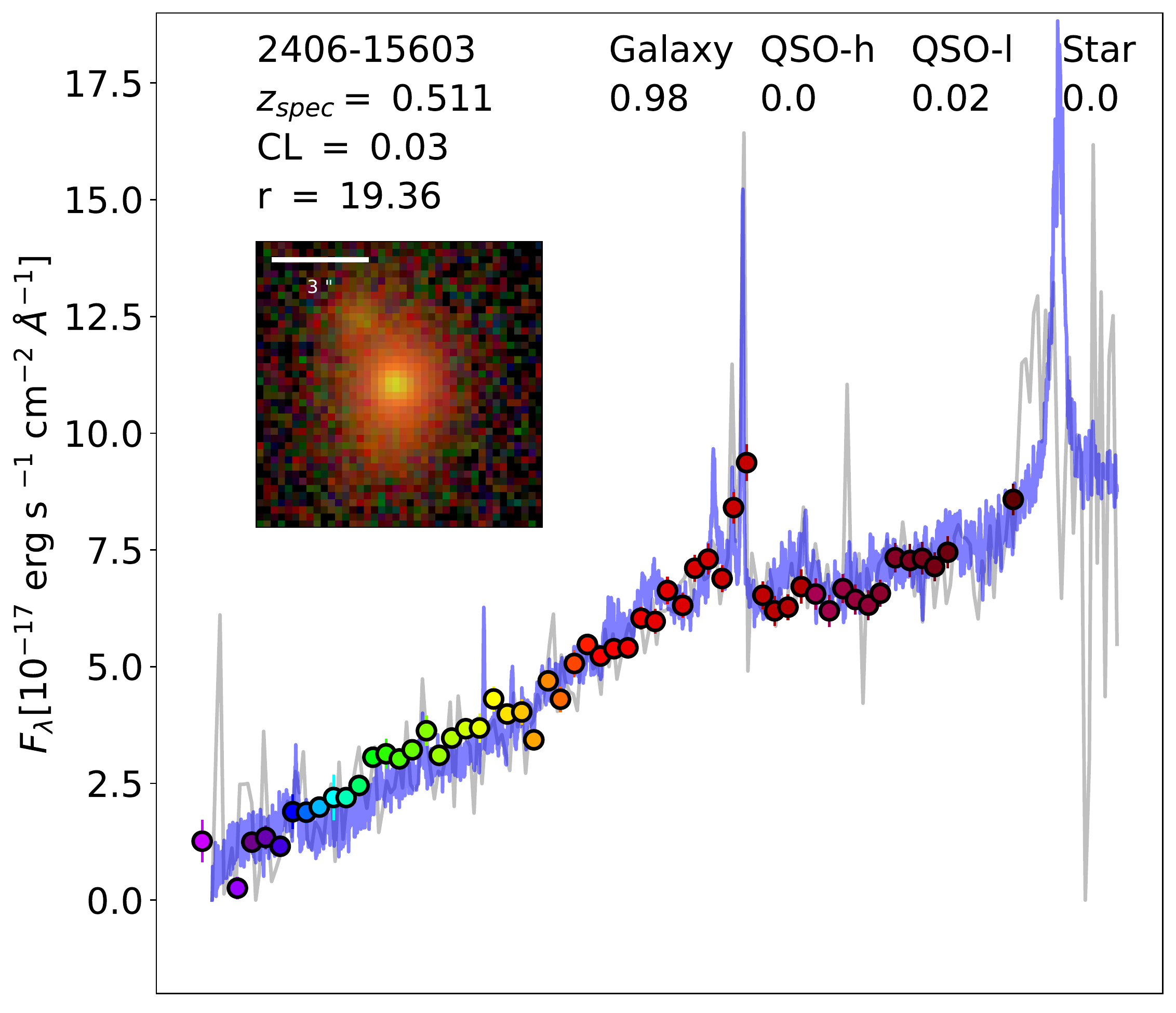}
    \includegraphics[width=5.9cm,height=5.6cm]{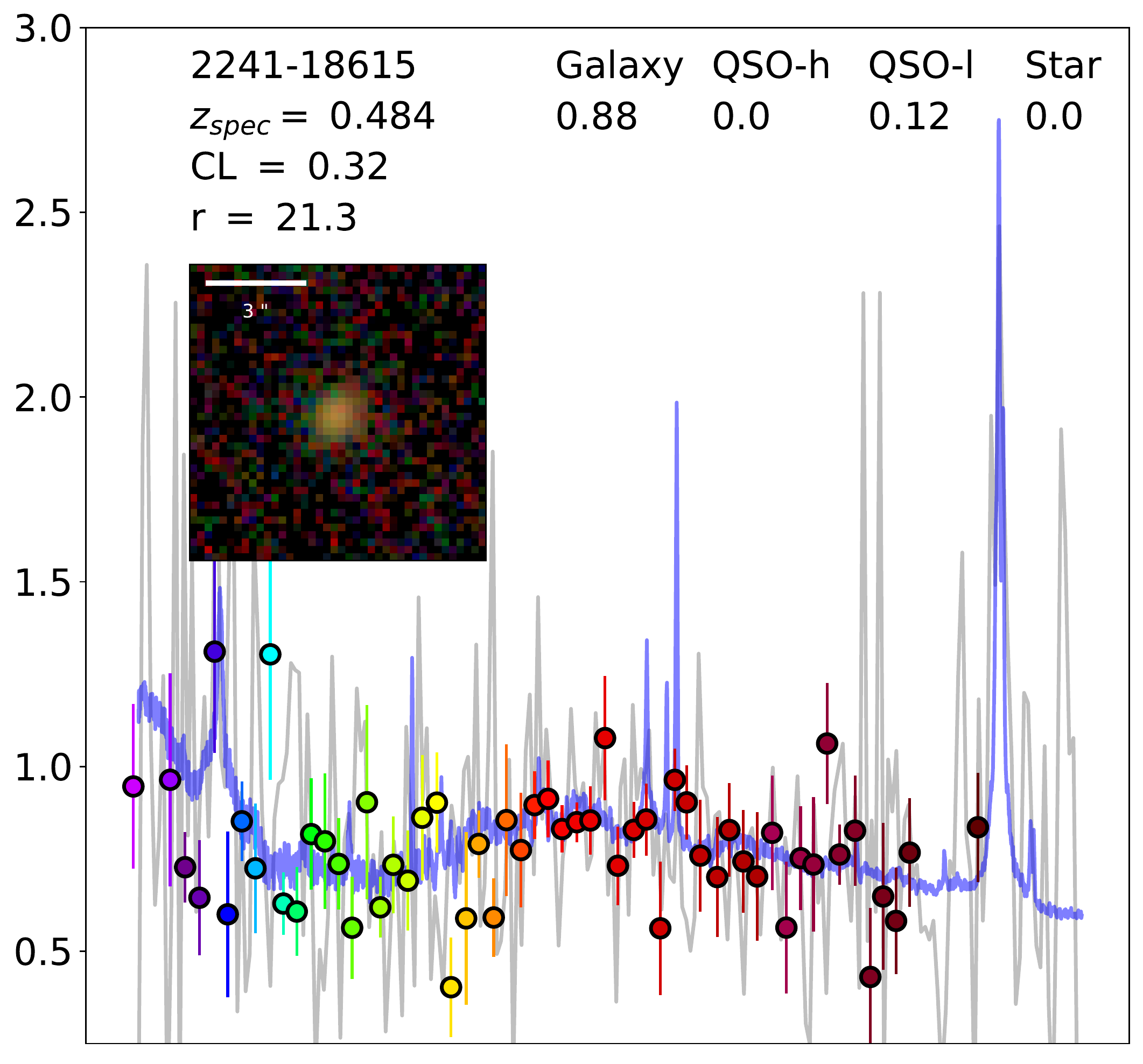}
    \includegraphics[width=5.9cm,height=5.6cm]{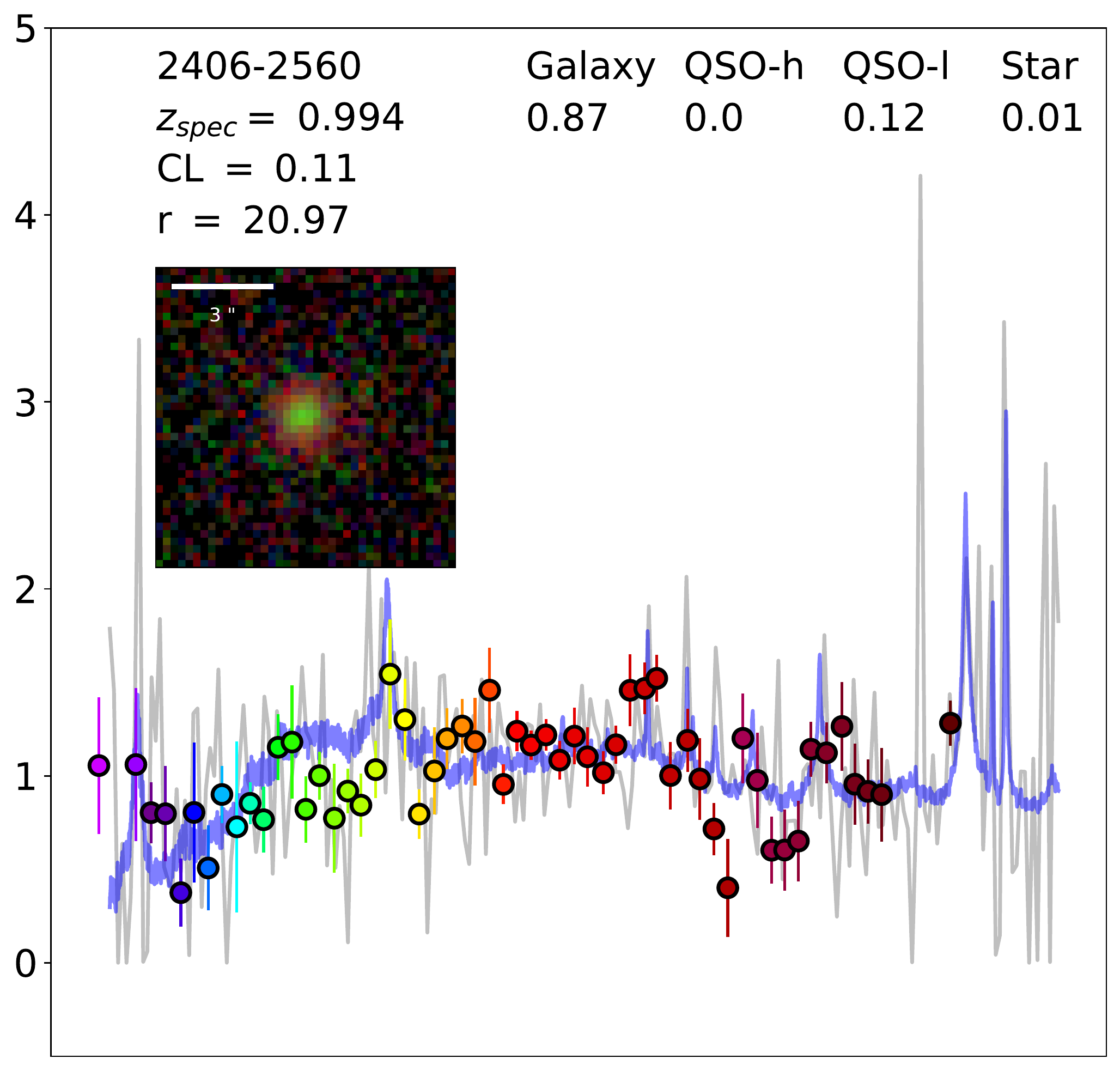}
    
    \includegraphics[width=5.9cm,height=5.6cm]{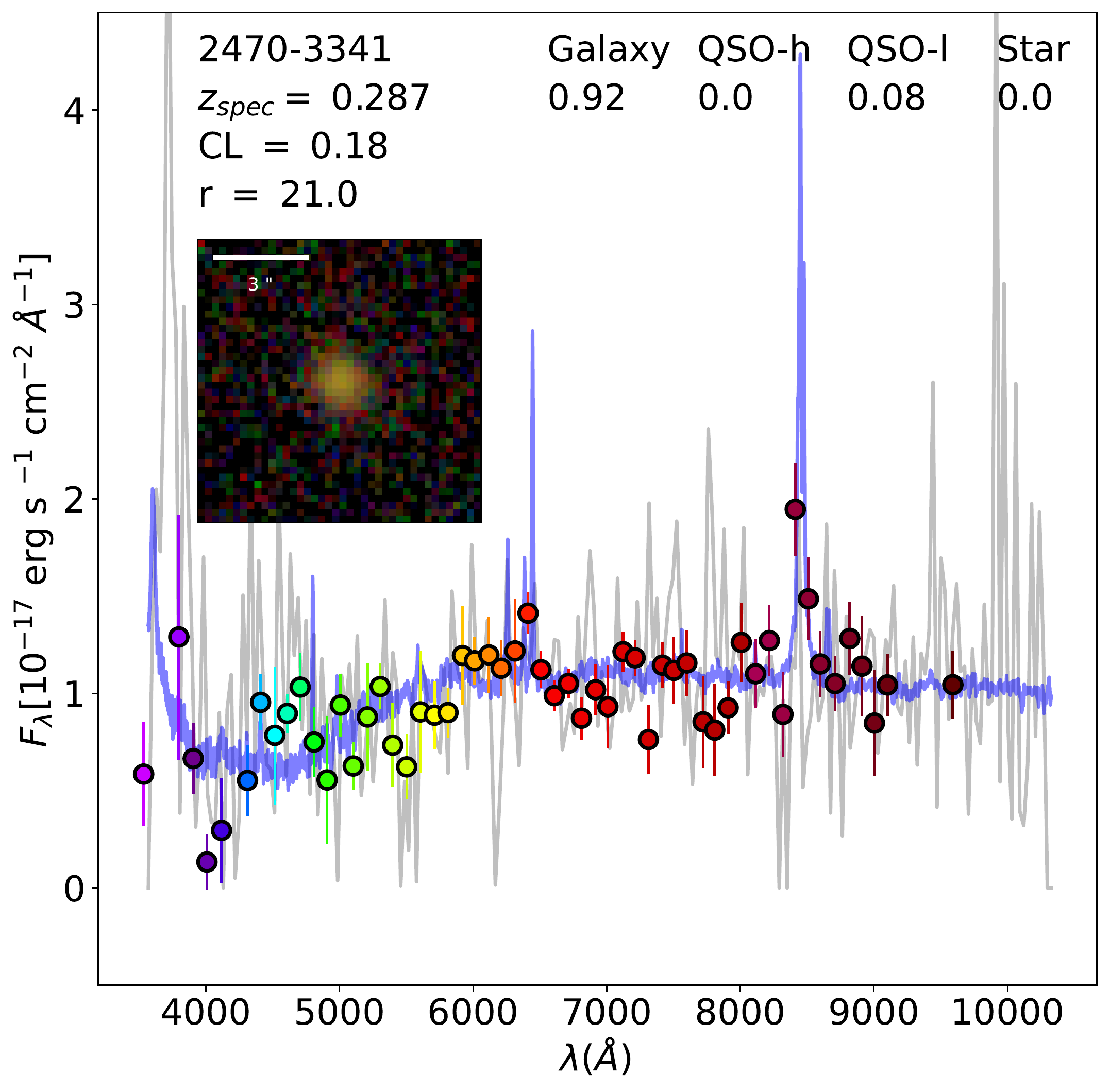}
    \includegraphics[width=5.9cm,height=5.6cm]{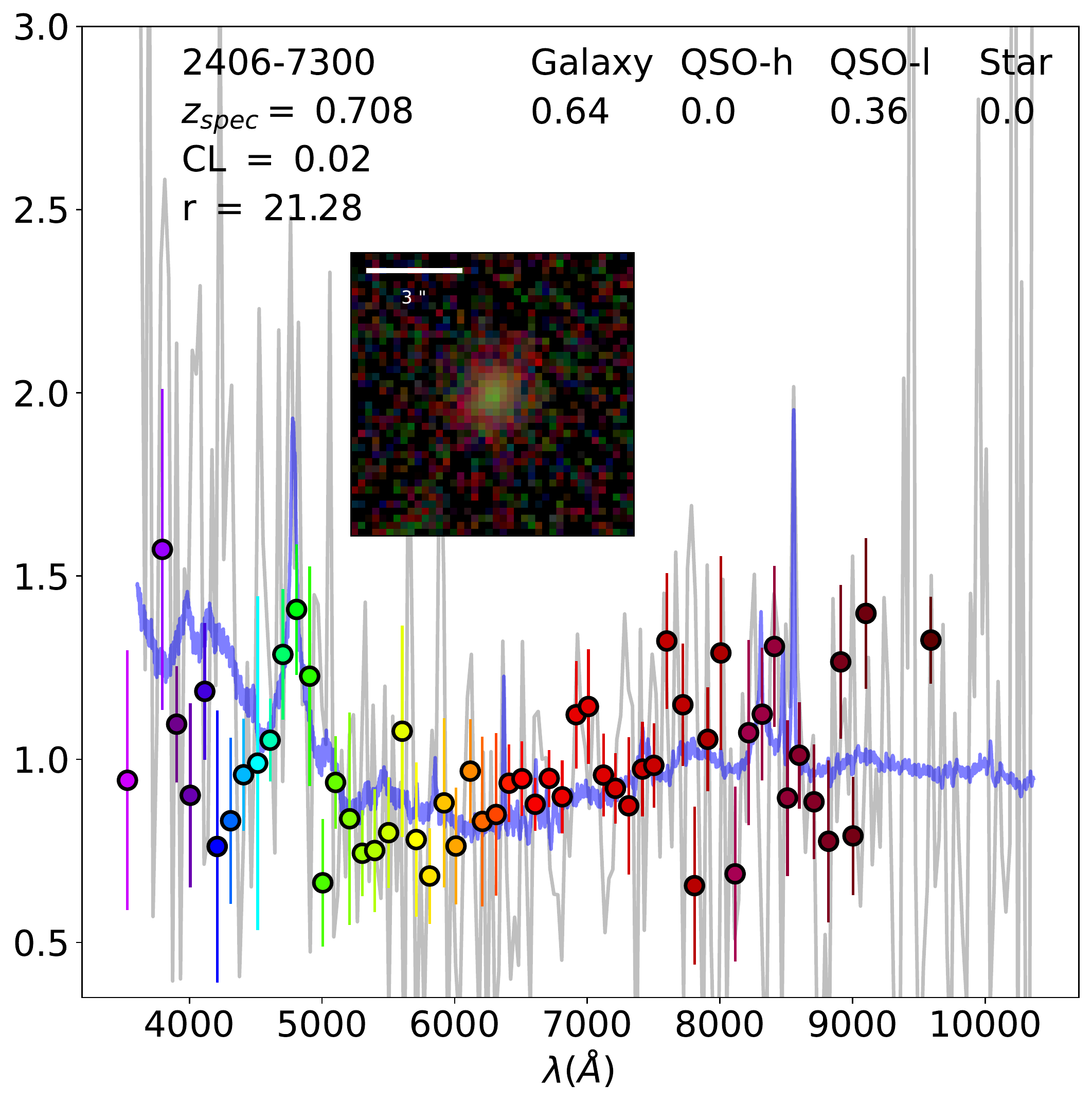}
    \includegraphics[width=5.9cm,height=5.6cm]{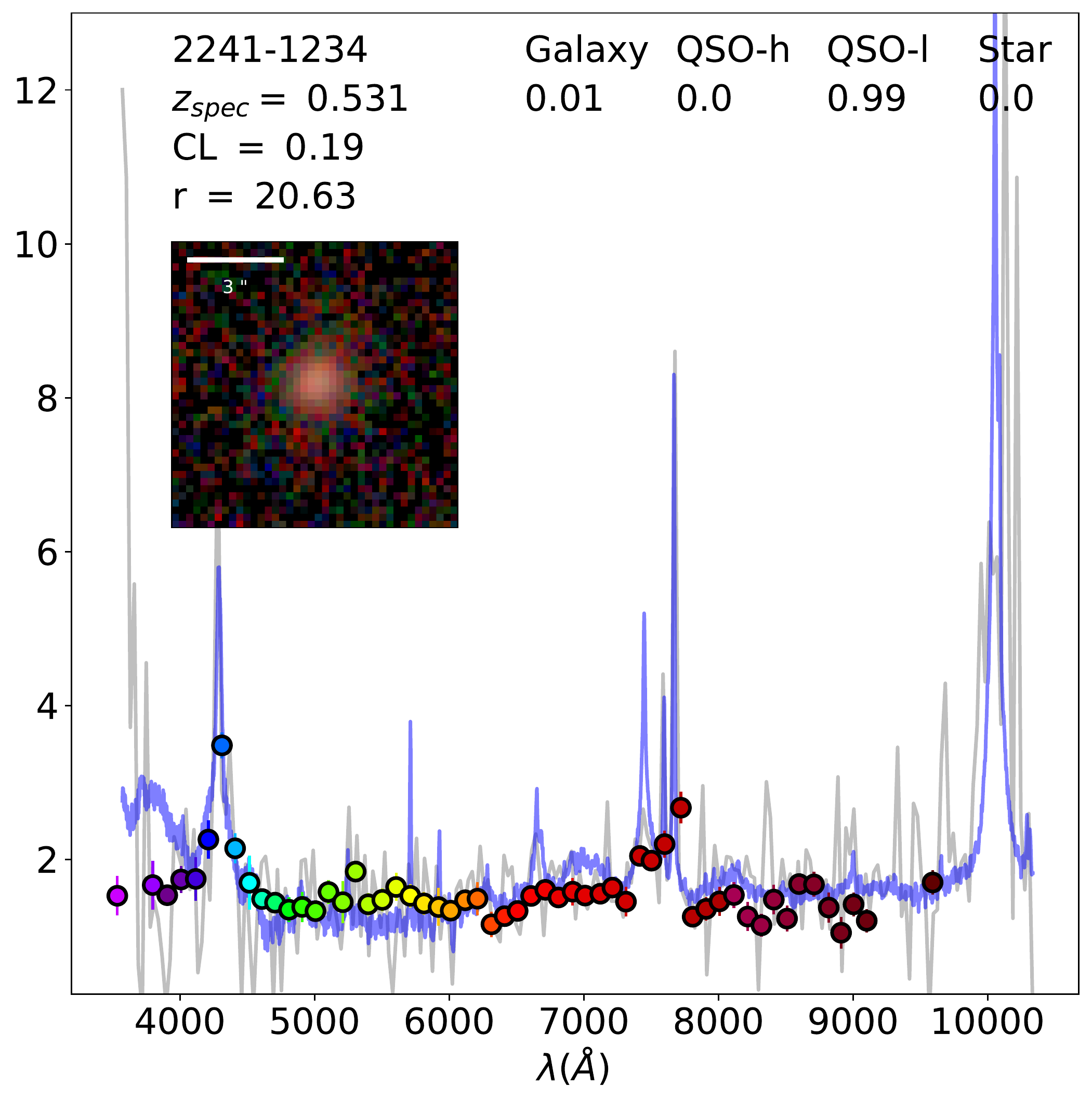}
    \caption{{Seyfert galaxies observed with miniJPAS and SDSS (see text in Sect. \ref{subsec:miniJPAS-SDSS}). The SDSS spectra are scaled to match the miniJPAS $r$ band. The solid grey line represents the actual SDSS observation, and the blue line is a model developed by the SDSS team. We indicate in the legend the miniJPAS ID, the spectroscopic redshift of the object, the class-star yielded by \texttt{SExtractor} (CL), and the AB magnitude in the $r$ band. We also show the probabilities obtained by the ANN$_1$ for each one of the classes, and we attach a multi-colour RGB image centred on the object covering 6.5 arcsec across. All objects except 2470-3341 are classified by the SDSS pipeline as QSOs.}}
    \label{fig:mixobj}
\end{figure*}

 In Fig. \ref{fig:CM_SDSS_ANN1} we show the confusion matrix obtained with ANN$_1$ for the SDSS test sample. The matrix for BIN 0 is purely diagonal and includes only one confusion between a low-redshift quasar and a galaxy (object ID: 2406-15603), which is shown in Fig. \ref{fig:mixobj} and is discussed below. Thus, the objects confused by the algorithms are all from BIN 1. The confusion matrices for the remaining models are listed in Appendix \ref{app:CM_SDSS}. The sample of QSO predicted by the ANN, and especially the sub-sample of QSO-h, contains very few false positives (QSO columns in the confusion matrix), meaning that the algorithms favour a pure rather than a complete sample. However, the sample of galaxies is more complete because $\sim~19~\%$ of them are classified as QSO-l, but only a few SDSS galaxies are missing. Finally, stars are identified with a very high accuracy, with only five false positives and eight true negatives. We recall that these results are partially biased due to the small number of objects in it and the selection criteria that were used by the SDSS team to select the sample of QSO targets. Hence, it can only give us a a glimpse of the actual performance of the ANN in real data. 
\par In Fig. \ref{fig:mixobj} we show some examples of objects that were observed simultaneously by SDSS and miniJPAS below redshift 1 that might present a spectrum composed of mixed features. In other words, the light coming from these objects has received contributions from both the AGNs and the stellar populations within the galaxies. All the objects except 2470-3341 are classified in SDSS as QSO-l. However, \texttt{SExtractor} identified them as extended sources with a class-star value below 0.35 (CL in Fig. \ref{fig:mixobj}). Following the SDSS classification criterion, only 2470-3341 and 2241-1234 are correctly classified by the ANN$_1$. Nevertheless, 2406-15603 is rather a Seyfert 1 galaxy with broad emission lines such as  H$\alpha$ and  H$\beta$. It also has a reddish spectrum, and the extended structure of the galaxy is very clearly visible in the image. Furthermore, while the SDSS spectrum detects the broad emission line of  H$\alpha$, the miniJPAS observation does not capture this emission, which is probably the most relevant feature for classifying this object as a QSO. Sources 2241-18615, 2406-2560, and 2406-7300 are classified as galaxies, but the second preferred class is QSO-l. These objects are indeed not very different from 2470-3341, which is a Seyfert 2 galaxy according to the SDSS pipeline. Once again, the J spectra miss two relevant features in 2241-18615 and 2406-2560, the H$\alpha$ and H$\beta$ emission lines, respectively. Finally, 2241-1234 is correctly classified. Although it is an extended source according to \texttt{SExtractor}, the emission of the AGN dominates the spectrum. The high S/N obtained in this object makes the classification more certain.

\section{Star, galaxy, and QSO classification in miniJPAS}\label{sec:minijpas_catalogue}
\begin{figure*}
    \includegraphics[width=6.2cm,height=5.6cm]{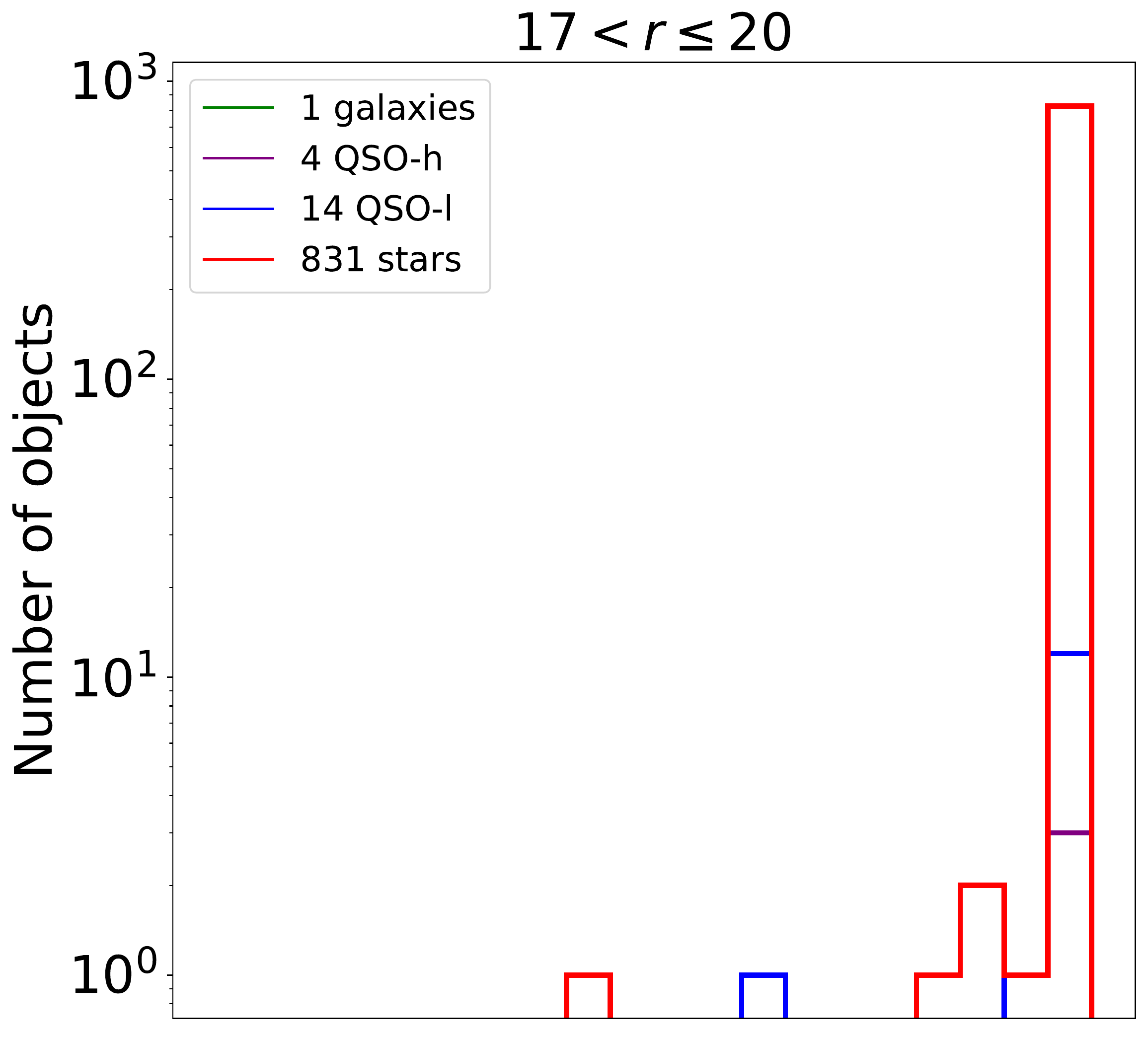}
    \includegraphics[width=5.6cm,height=5.6cm]{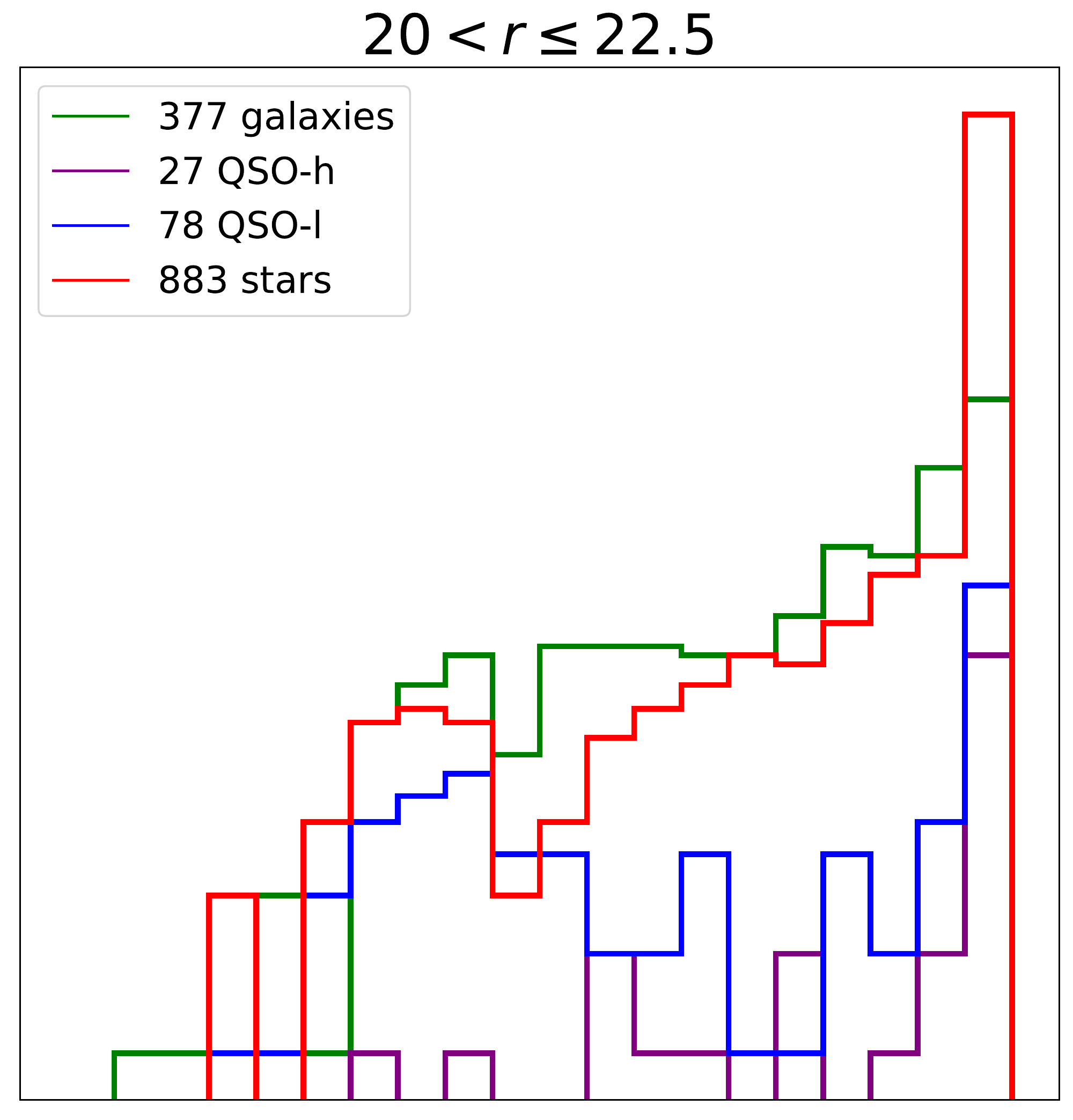}
    \includegraphics[width=5.6cm,height=5.6cm]{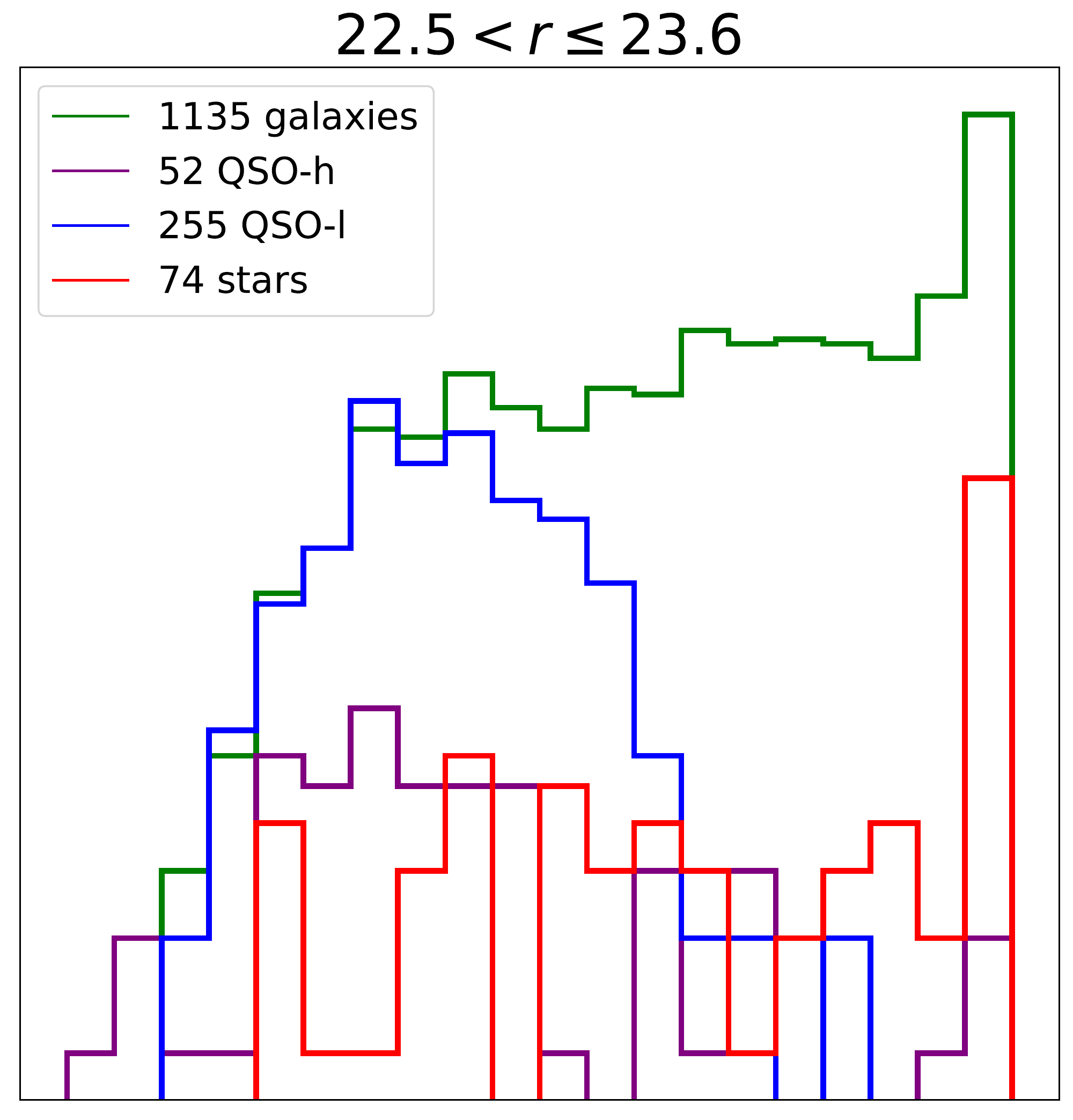}
    \\
    \includegraphics[width=6.2cm,height=5.6cm]{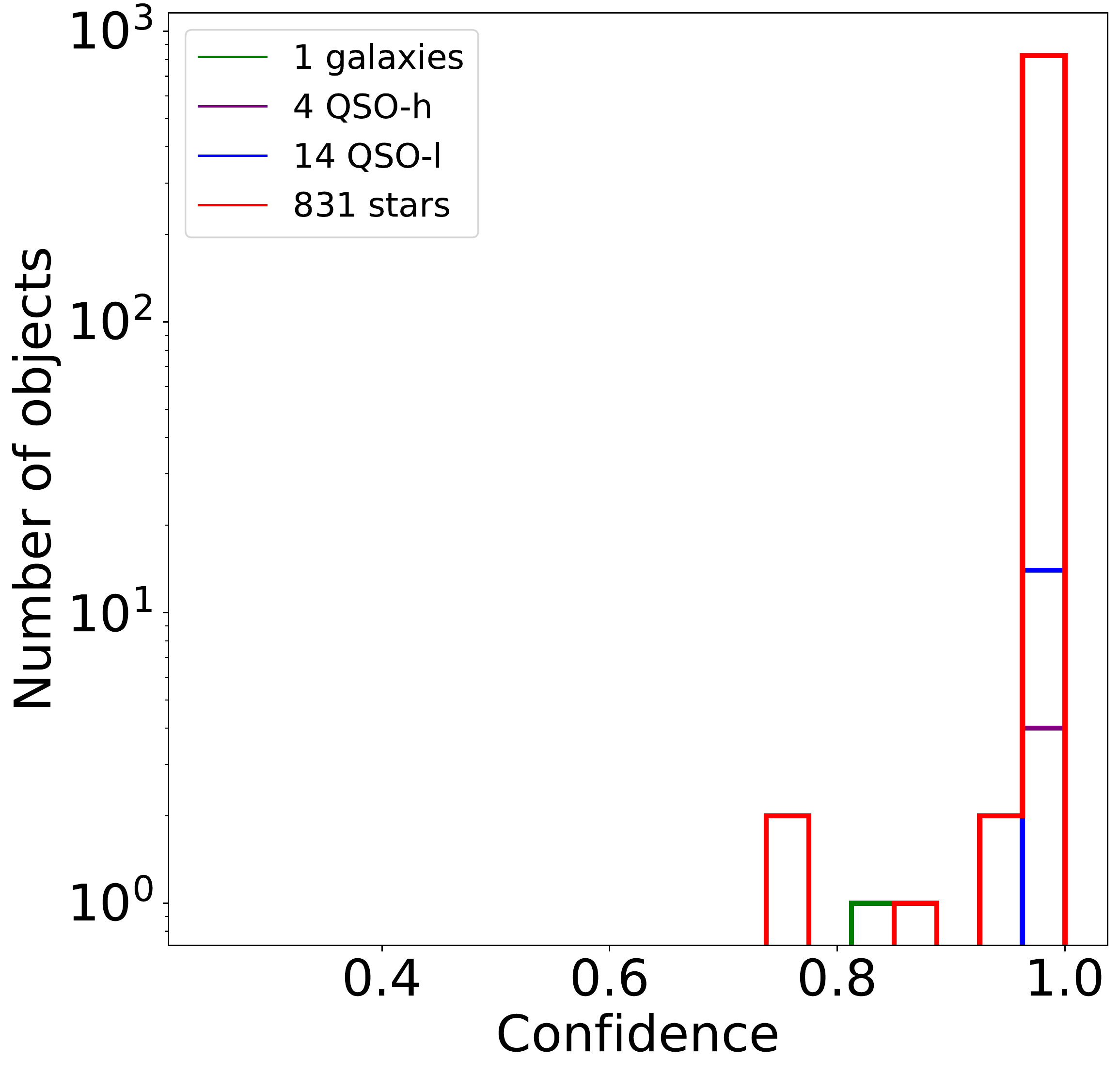}
    \includegraphics[width=5.6cm,height=5.6cm]{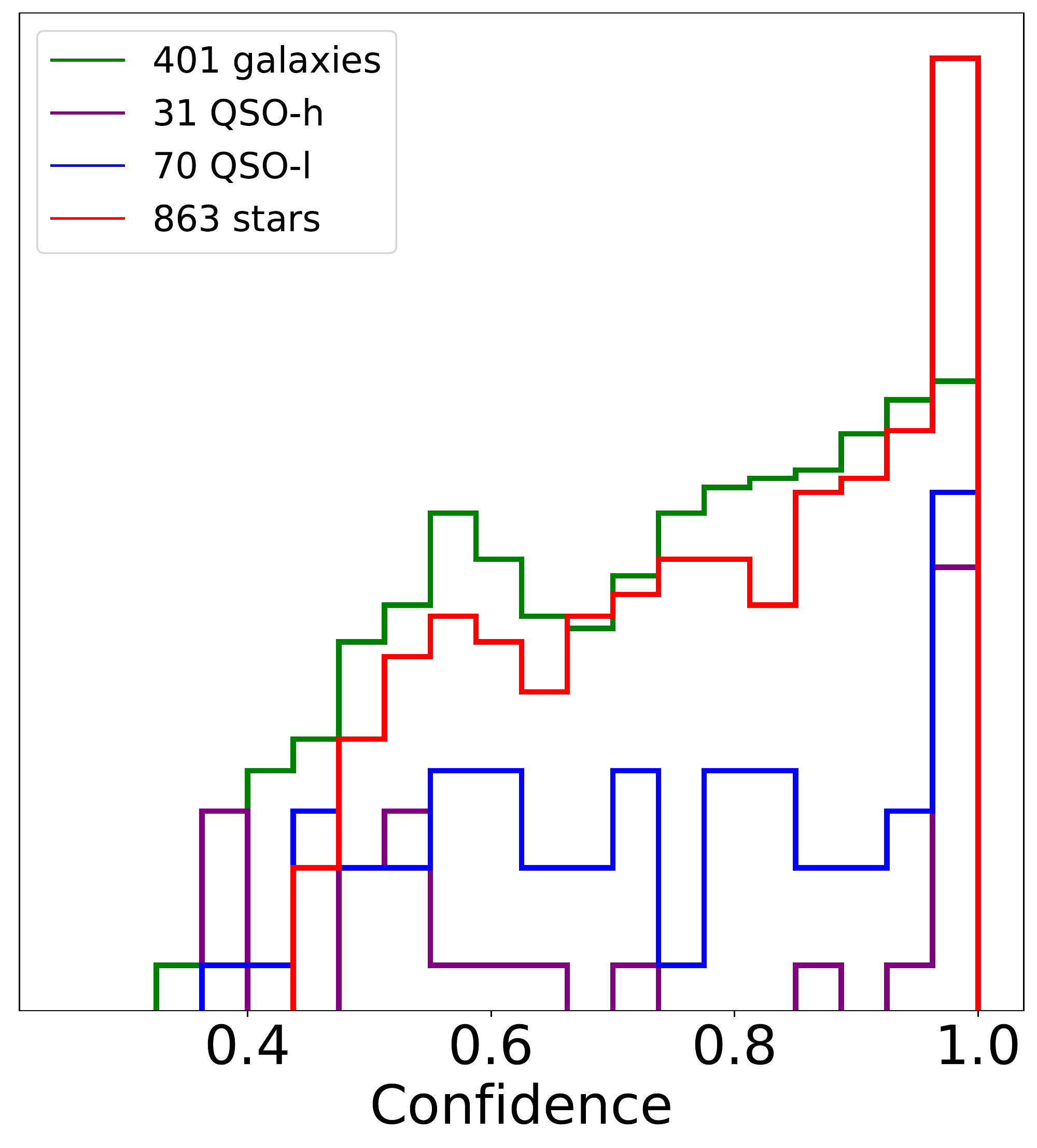}
    \includegraphics[width=5.6cm,height=5.6cm]{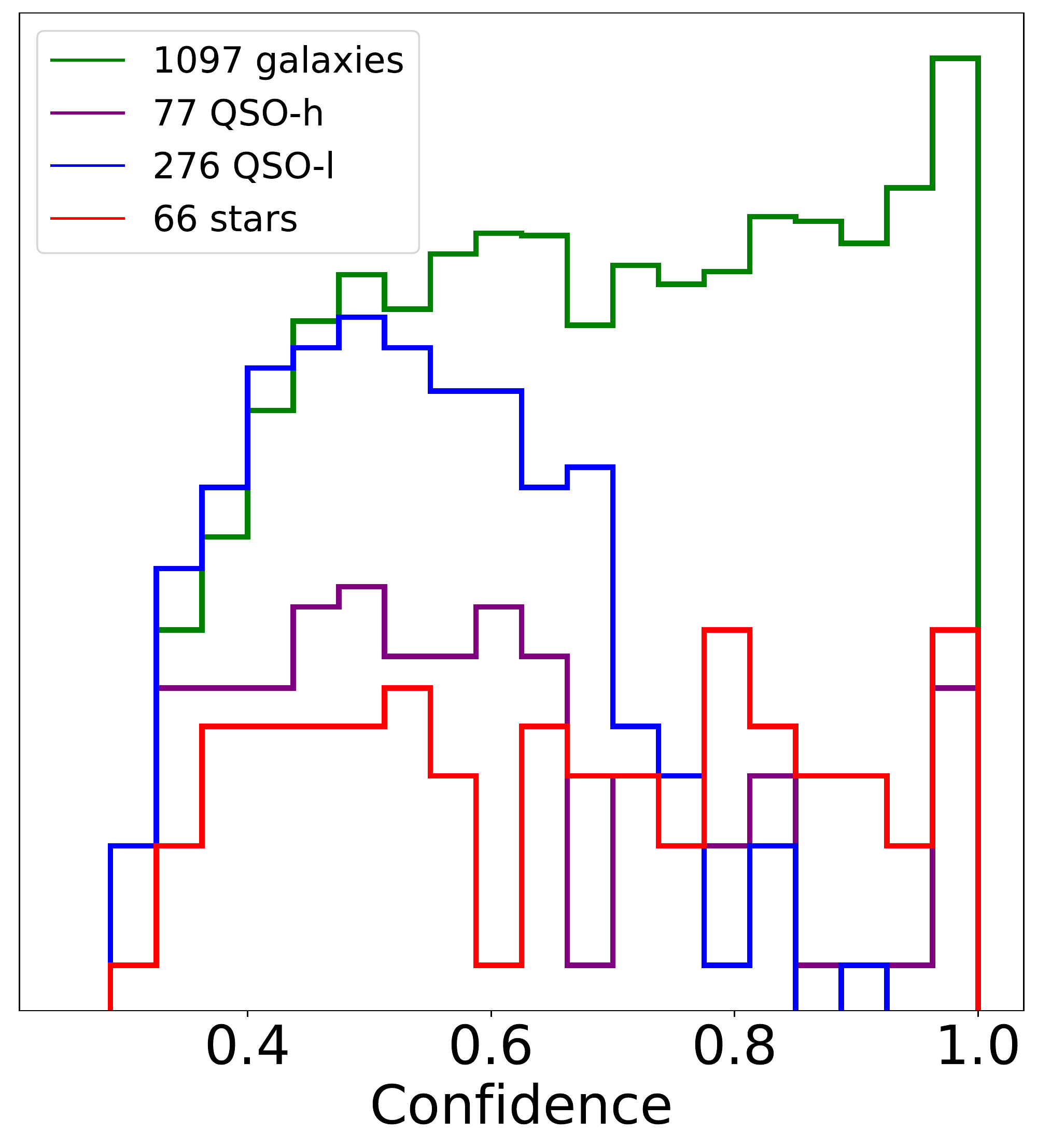}
    \caption{{Confidence (probability) yielded by the ANN$_1$ (top) and ANN$_2 $ (bottom)  classifiers for each class and magnitude BIN in miniJPAS observations for point-like sources (CL $> 0.5$). The numbers of classified objects are shown in the legend.}}
    \label{fig:Nobj_ANN1_miniJPAS}
\end{figure*}

We now focus our attention on the classifier predictions on miniJPAS data. In Fig. \ref{fig:Nobj_ANN1_miniJPAS} we show the distribution of the confidence (probability) levels yielded by the ANN$_1$ and ANN$_2$ classifiers for each class and each magnitude bin. We only predict the class for the objects that are considered point-like sources according to \texttt{SExtractor}, which used the detection band (rSDSS). Both ANN classifiers predict roughly the same number of objects, but they exhibit differences in the faintest magnitude bins, which are useful to later build a combined algorithm that uses information from several classifiers (P\'erez-R\`afols et al. in prep. b). 
\par The total number of QSOs predicted by ANNs for miniJPAS is compatible with previous estimates. \cite{2012MNRAS.423.3251A} expected $\sim 240$ QSOs per square degree with the J-PAS photometric system for a limiting magnitude of $g$=23, assuming that QSOs follow the luminosity function found by \cite{2009MNRAS.399.1755C} and the QSO selection is perfect. With the ANN$_1$ (ANN$_2$), we detect 163 (177) QSOs with a probability greater than $0.5$ and $g < 23$ in the point-like source catalogue (CL > 0.5). Even though $0.25$ is the threshold for an object to be classified as one particular class, we imposed that the probabilities of being a QSO have to be greater than the alternative ($P(\text{QSO-h}) + P(\text{QSO-l}) > P(\text{Galaxy}) + P(\text{Star}))$, which implies  $P(\text{QSO-h}) + P(\text{QSO-l}) > 0.5$. However, this is a conservative approach because we only considered high-probability objects. If instead we sum all over the probabilities of being QSO, we obtain 182.1 and 195.1 QSOs with the ANN$_1$, and ANN$_2$, respectively.
\par
\begin{figure*}
    \includegraphics[width=5.8cm,height=5.8cm]{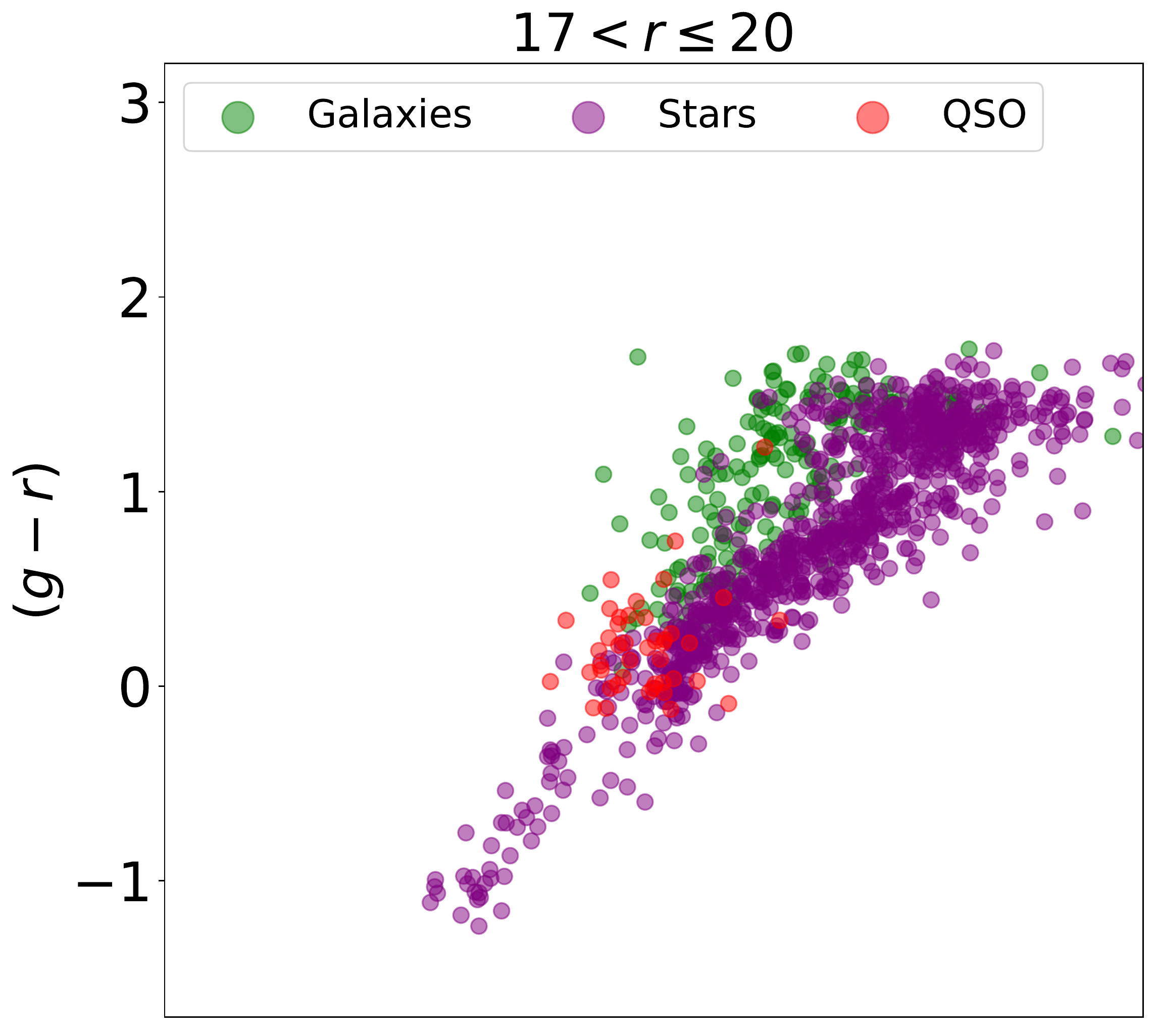}
    \includegraphics[width=5.2cm,height=5.8cm]{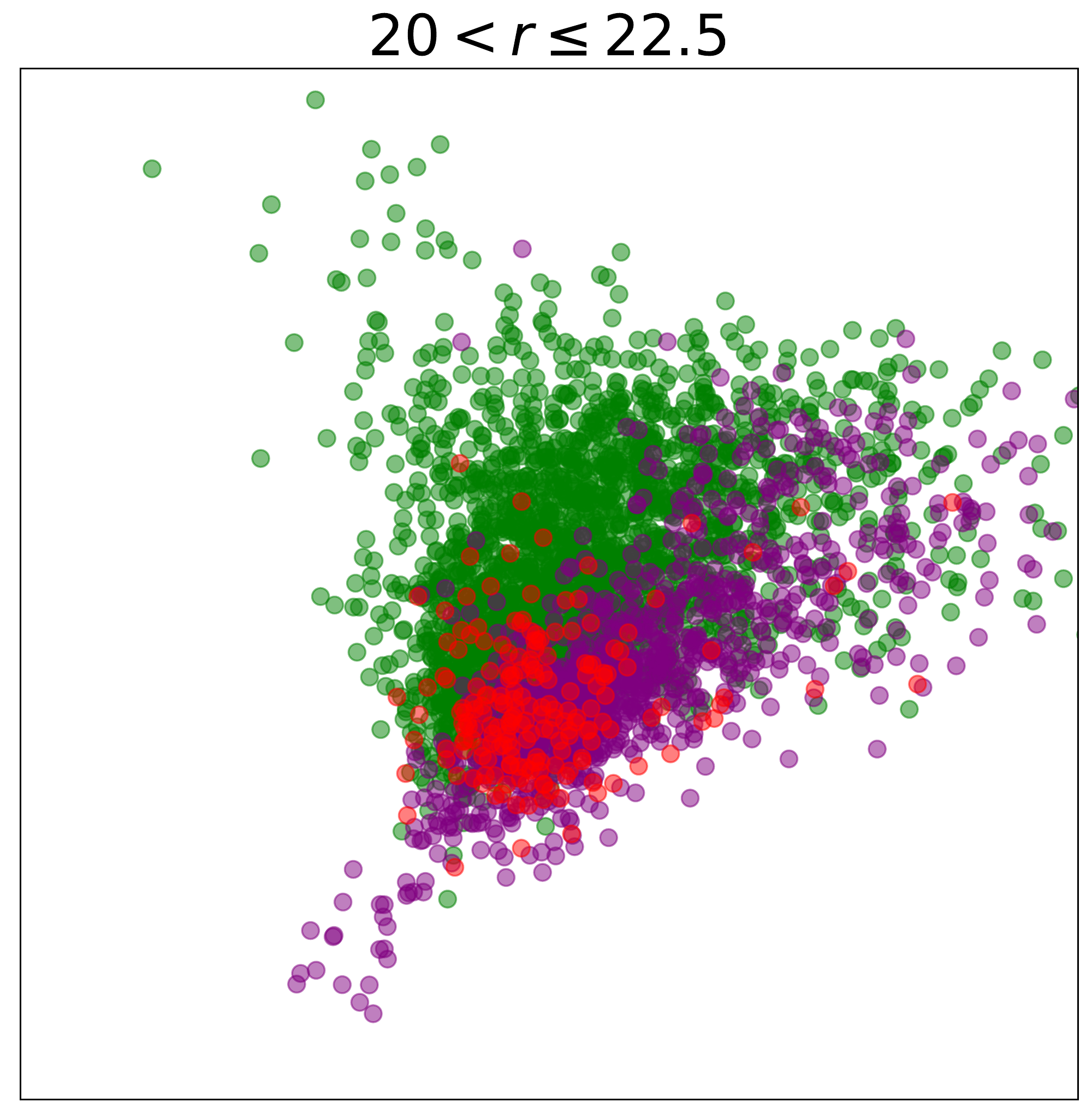}
    \includegraphics[width=5.2cm,height=5.8cm]{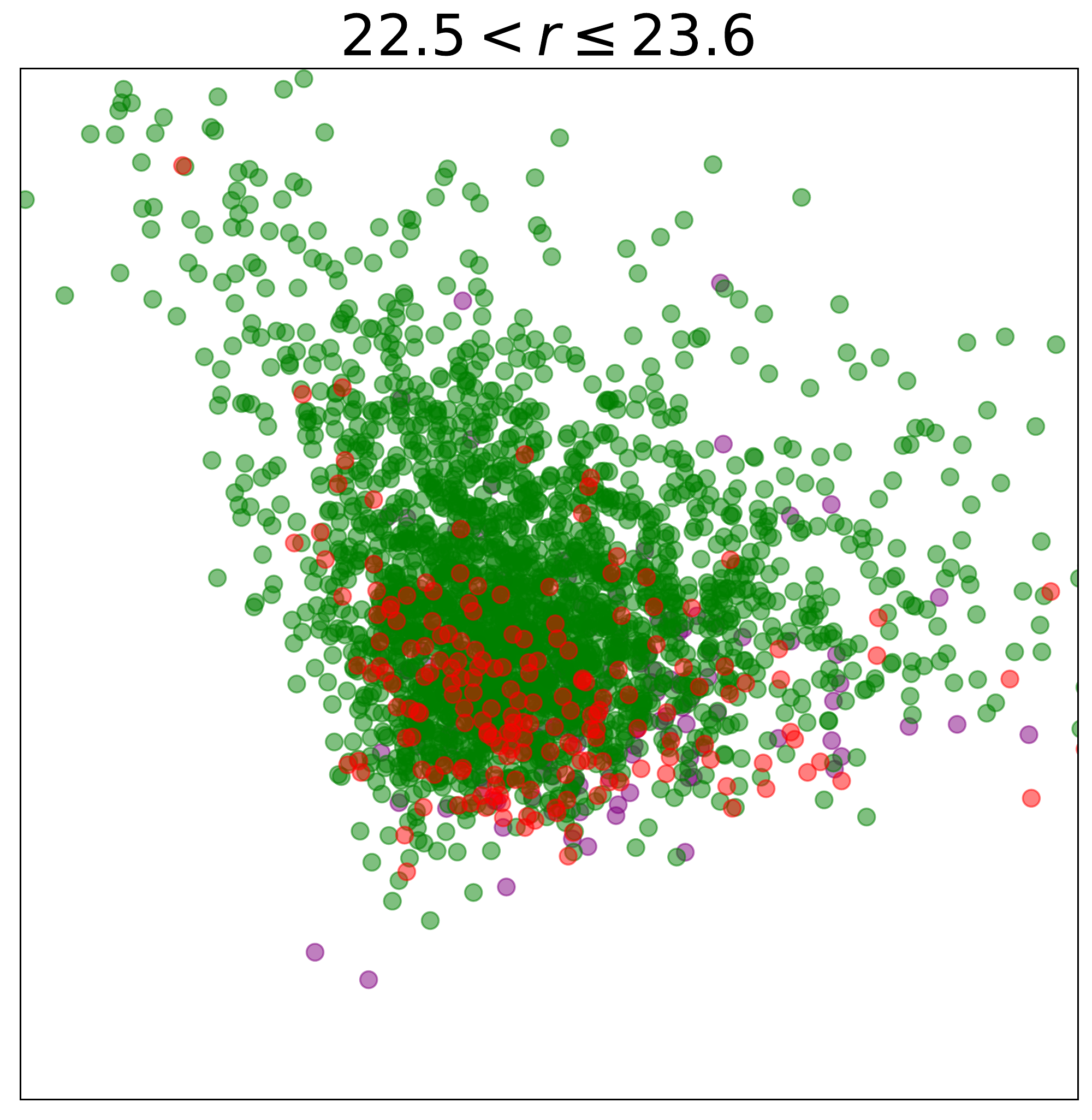}
    \includegraphics[width=1.3cm,height=5.6cm]{Images/barw.png}
    \\
    \includegraphics[width=5.8cm,height=5.6cm]{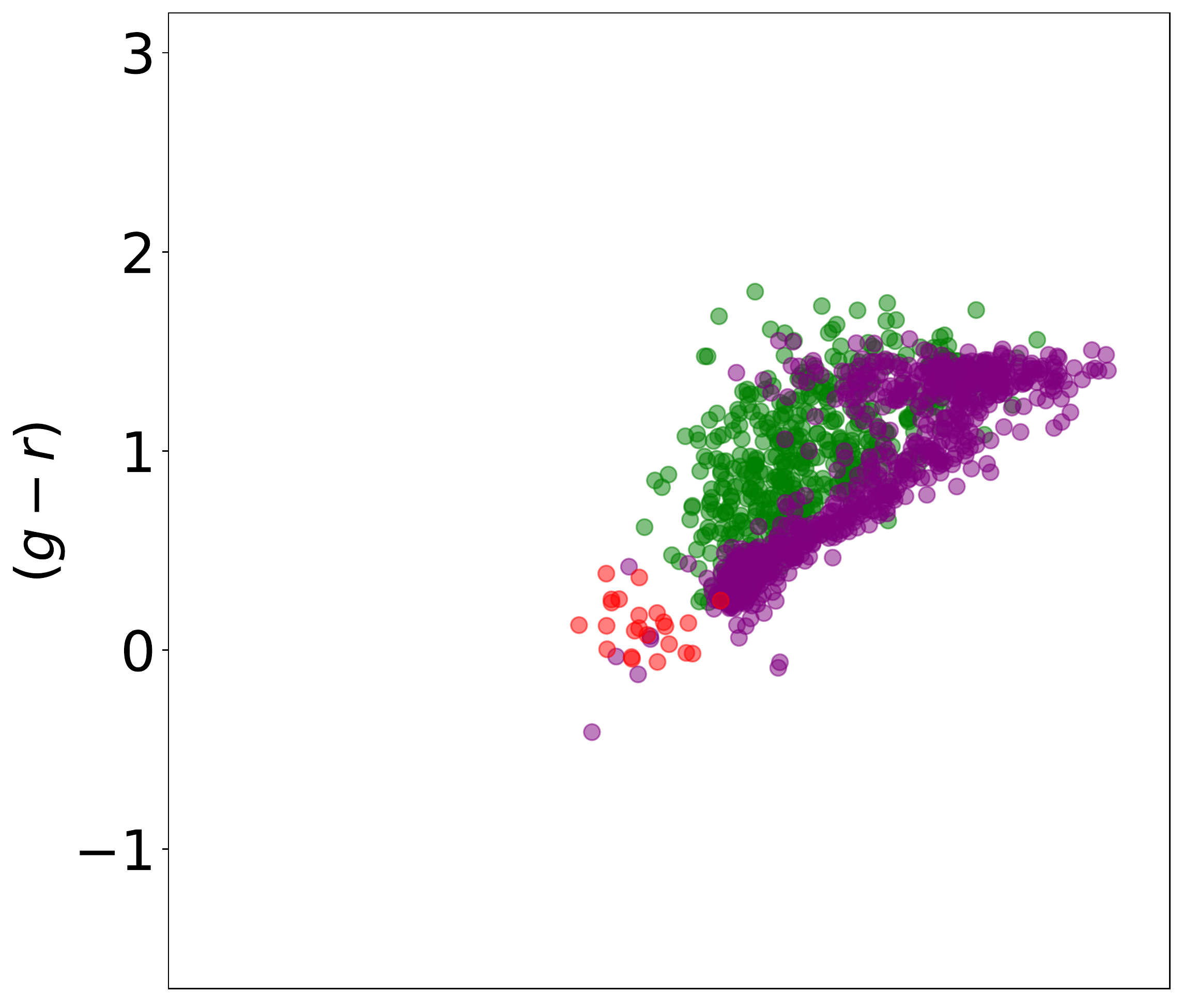}
    \includegraphics[width=5.2cm,height=5.5cm]{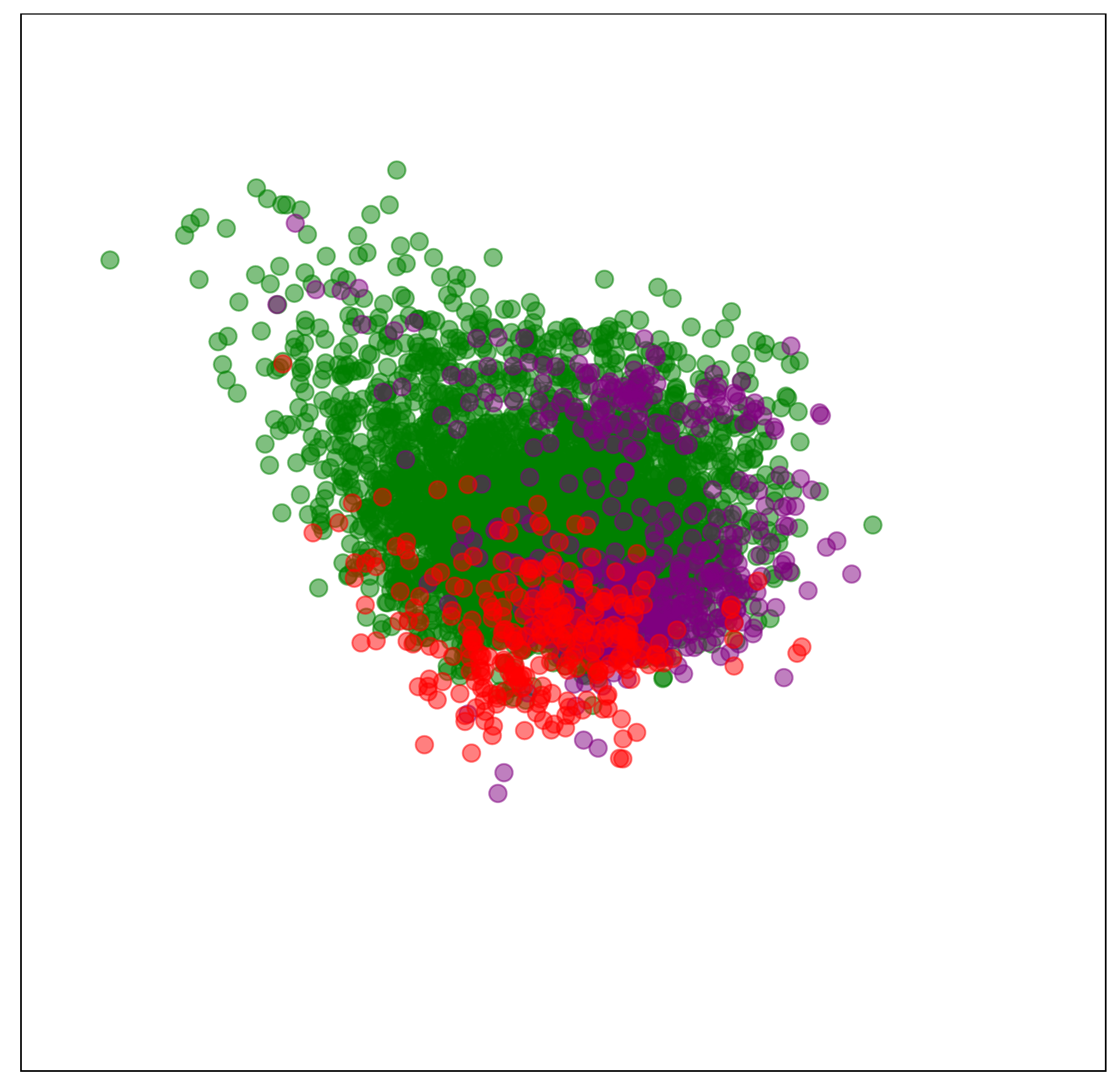}
    \includegraphics[width=5.2cm,height=5.5cm]{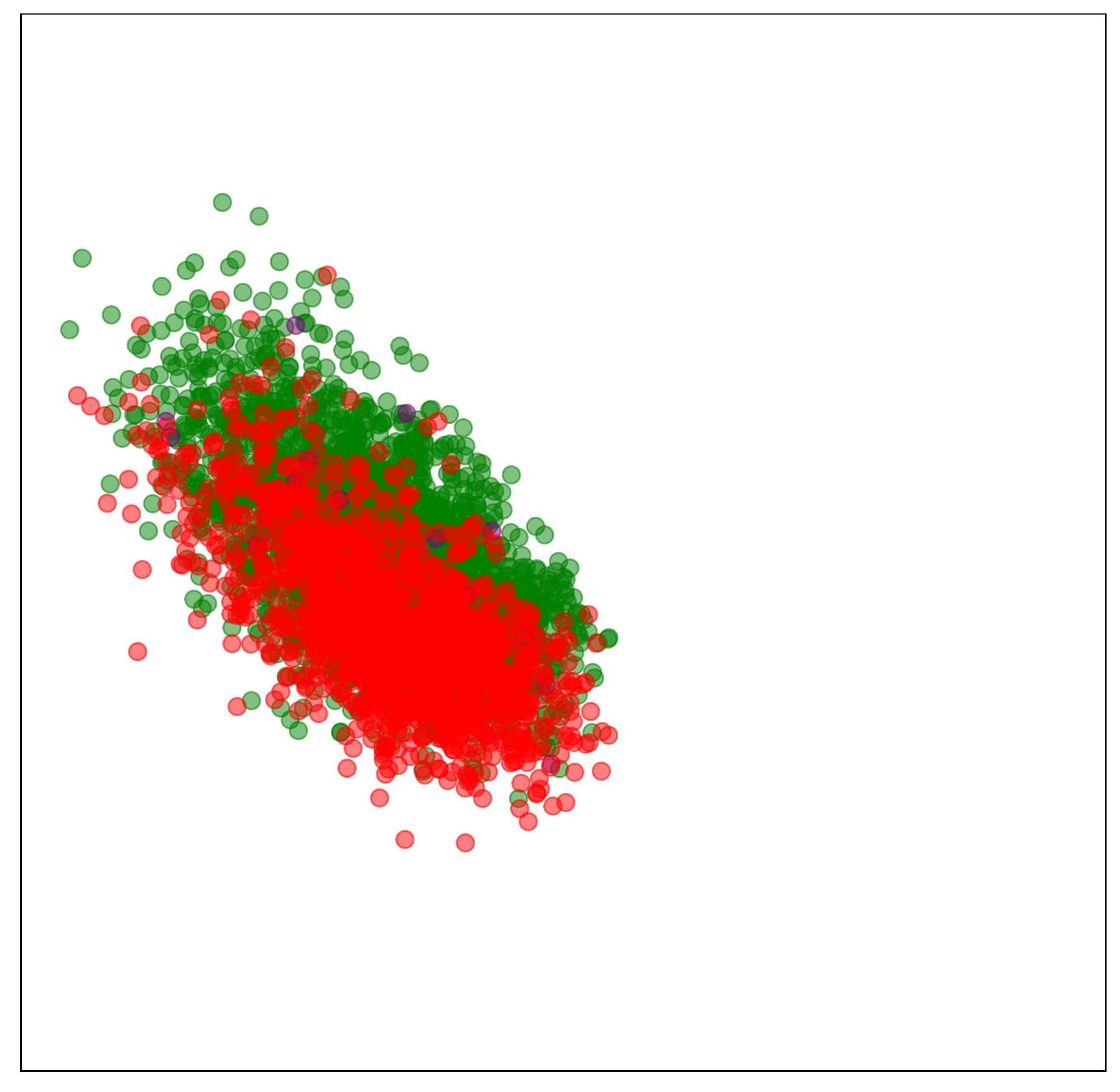}
    \includegraphics[width=1.3cm,height=5.6cm]{Images/barw.png}
    \\
    \includegraphics[width=5.8cm,height=5.6cm]{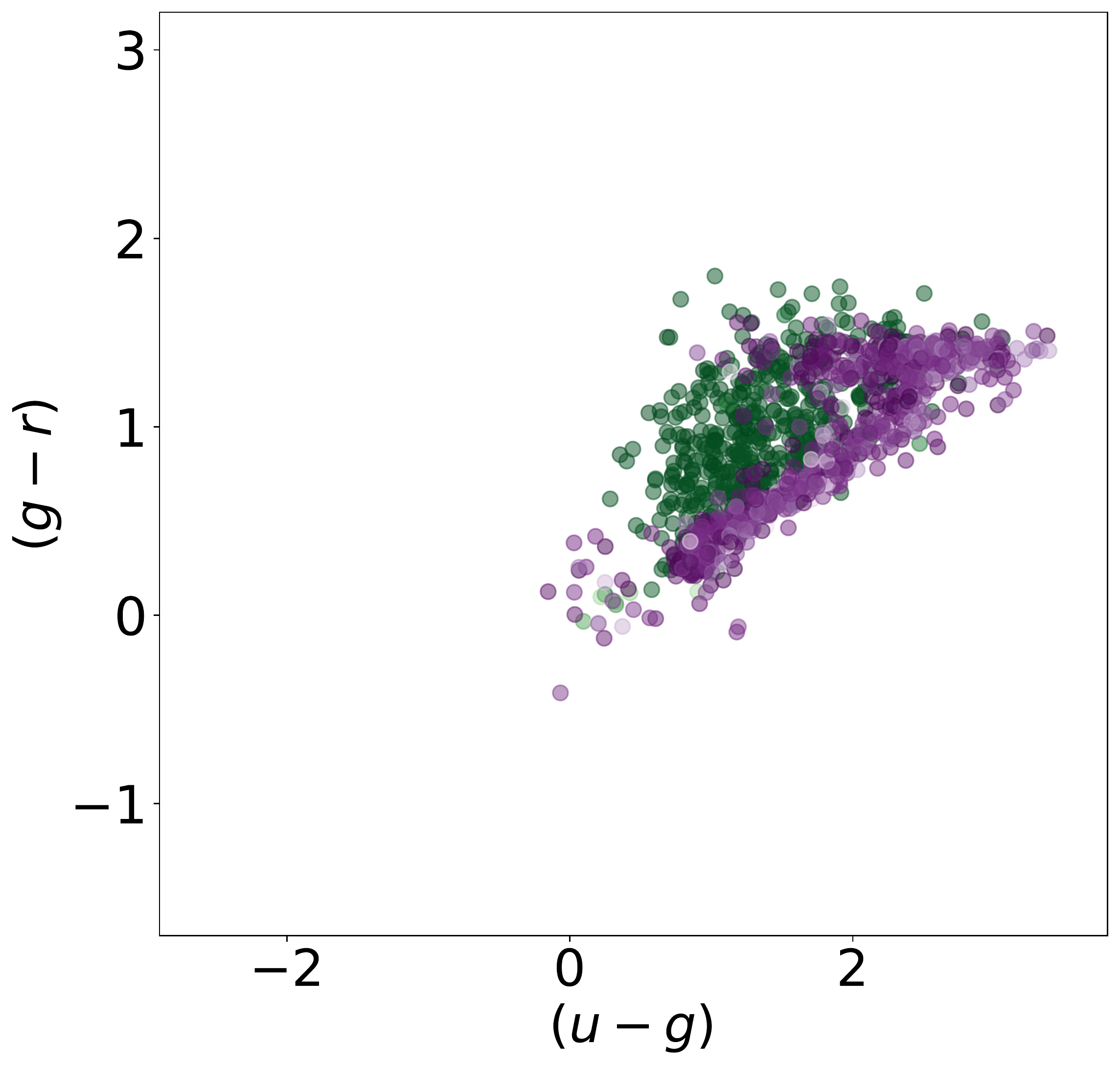}
    \includegraphics[width=5.2cm,height=5.5cm]{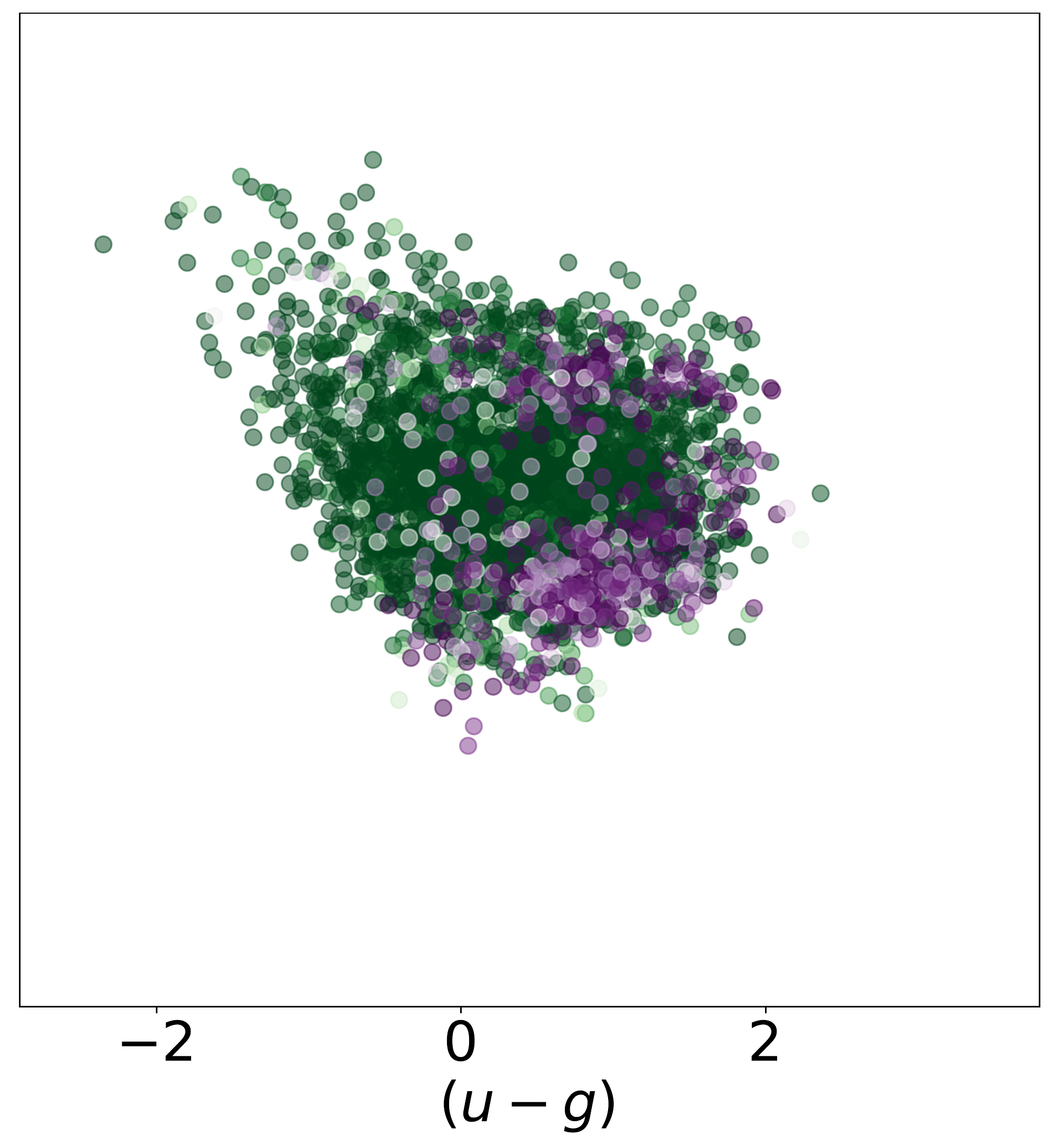}
    \includegraphics[width=5.2cm,height=5.5cm]{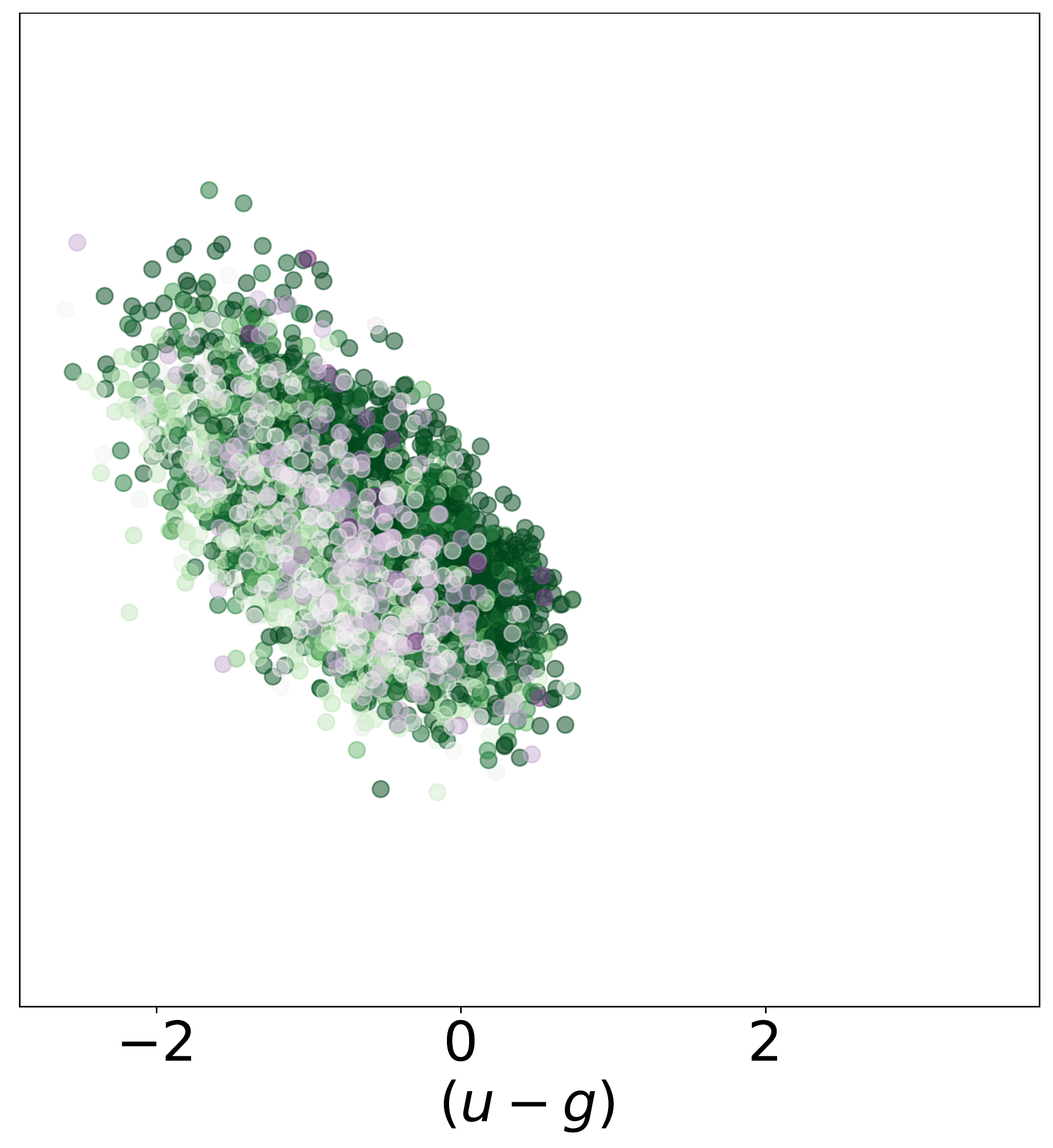}
    \includegraphics[width=1.3cm,height=5.6cm]{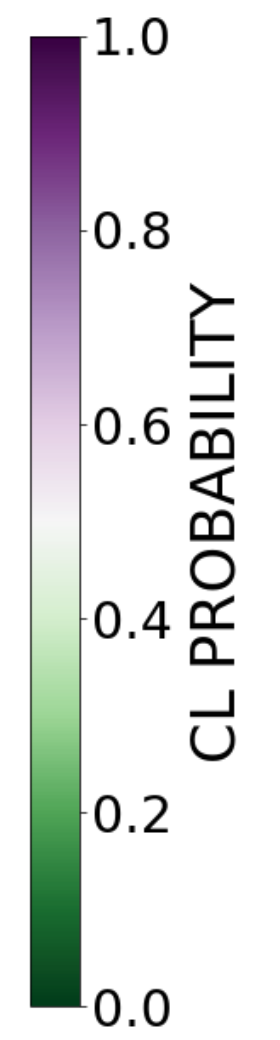}
    \caption{Observed $(g-r)$ vs. $(u-g)$ colour-colour diagram for the 1 deg$^2$ mock sample (first row) and miniJPAS observations (second and third rows). Stars, galaxies, and QSOs are predicted classes with the ANN$_1$ in miniJPAS observations, while they are true classes in the 1deg$^2$ mock sample. The dots in the third row are colour-coded according to the \texttt{SExtractor} probability developed to separate between point-like sources (CL $> 0.5$) and extended sources (CL $< 0.5$). Each column includes objects at different magnitude bins.}
    \label{fig:colour-diagram}
\end{figure*}

 In Fig. \ref{fig:colour-diagram} we show the observed $(g-r)$ versus $(u-g)$ colour-colour diagram for all the objects presented in the $\text{1 deg}^2$ mock sample (first row) and in the miniJPAS observations (second row). The positions of QSOs, stars, and galaxies are consistent in each magnitude bin. We included all objects in miniJPAS with \texttt{FLAGS$=0$} and \texttt{MASK\_FLAGS$=0$}. The 1 deg$^2$ mock sample includes a stellar pouplation that is absend from miniJPAS observations (bottom left side in BIN 0 and 1). These stars correspond to the most massive and bluest stars (O type) that are usually found in regions with a high activity of star formation. The luminosity functions for O and B stars were estimated by extrapolating the prediction of the Besançon model together with the stars within the miniJPAS SDSS superset sample. Therefore, these populations might be overestimated in the 1 deg$^2$ mock sample. However, even if that were the case, the fraction of these stars is still low compared with those on the main sequence. Therefore, the impact that this effect has on a classifier whose main goal is to identify QSO candidates is very limited. 
\par In the last row of Fig. \ref{fig:colour-diagram}, we colour-code the miniJPAS observations with the CL probability. QSOs and stars predicted by the ANN$_1$ are classified by \texttt{SExtractor} as point-like sources (CL$ > 0.5$), while galaxies are predicted as extended sources (CL$ < 0.5$). In the faintest magnitude bin extended and point-like sources are more difficult to distinguish in the colour space because they both overlap. The ratio of the number of point-like sources according to the CL and the ANN$_1$ ($R_{\text{point}} = N_{\text{point}}$ (CL)  / $N_{\text{point}}$ (ANN$_1$) for BIN 0, 1, and 2 are $R_{\text{point}} = 0.93,0.72,\text{and }0.18 $, respectively, and the ratio of the number of extended sources is $R_{\text{ext}} =1.12,1.06,\text{and }1.49$, respectively. We assumed that point-like sources are QSOs and stars, while galaxies were considered extended sources. In summary, our predictions agree with \texttt{SExtractor} considering that the performance of both classifiers decreases as a function of the observed magnitude. Concretely, \texttt{SExtractor} starts to degenerate around 21 magnitude because it is more difficult to distinguish between point-like and extended sources in the low-brightness regime. Furthermore, we predicted based on a sample (miniJPAs observations) that includes extended sources, while our training sample excluded this type of objects. 

\section{Summary and conclusions}\label{sec:conclusions}
We presented a method based on ANN to classify J spectra into four categories: stars, galaxies, QSO at high redshift ($z \geq 2.1$), and QSOs at low redshift ($z < 2.1$). The algorithms were trained and tested in mock data developed by \cite{10.1093/mnras/stac2962}. We employed two different representations of miniJPAS photometry in order to train the algorithms. ANN$_1$ used as input photometric fluxes normalised to the detection band (r), while ANN$_2$ employed colours plus the magnitude in the $r$ band. Therefore, ANN$_1$ only has information about the shape of the spectrum, while ANN$_2$ also has access to the observed luminosity.
\par We enlarged the training set by mixing features from four different classes by adapting the \texttt{mix up} technique. We do not observe significant differences in the performance of the algorithms when a hybrid set was used for the training. A fundamental difference between other works where \texttt{mix up} has been employed with success and this work is probably the complexity of astronomical data. Observations have associated errors that depend at first order on the luminosity of the observed objects. Therefore, features do not encode the same information when objects are brighter or fainter. In other words, mixing between classes appears as a natural outcome in the feature space as the errors increase, which makes faint objects indistinguishable from hybrid bright objects. Thus, hybridisation has an impact on the probabilities yielded by the ANN because they become less realistic when the level of mixing is increased in the training set. Well-calibrated algorithms are as important as obtaining a high performance because the outputs cannot be interpreted as a probability estimation otherwise. In this regard, we showed that the ANNs are better calibrated than the RF classifiers. Furthermore, ANN are better at classifying sources with J-PAS photometry, which was also proven in \cite{10.1093/mnras/stac2836}
\par We measured the performance of the algorithms in terms of the f1 score. In the case of quasars at high redshift, we obtain an f1 score with the ANN$_1$ (ANN$_2$) of $0.99$ ($0.99$), $0.93$ ($0.92$), and $0.63$ ($0.57$) for $17 < r \leq 20$, $20 < r \leq 22.5$, and $22.5 < r \leq 23.6$, respectively, where $r$ is the J-PAS rSDSS band. For low-redshift quasars, galaxies, and stars, the f1 score reached  $0.97$ ($0.97$), $0.82$ ($0.79$), and $0.61$ ($0.58$);  $0.94$ ($0.94$), $0.90$ ($0.89$), and $0.81$ ($0.80$); and $1.0$ ($1.0$), $0.96$ ($0.94$), and $0.70$ ($0.52$) in the same r bins. We also tested the algorithm on the SDSS test sample, and we obtained a performance compatible with the prediction in the mock test sample.  For quasars at high redshift, the f1 score with ANN$_1$ (ANN$_2$) reached 1.0 (1.0), and 0.84 (0.86) for $17 < r \leq 20$, and, $20 < r \leq 22.5$. For low-redshift quasars, galaxies, and stars, the f1 score reached 0.97 (0.97), and 0.81 (0.82);  0.91 (0.91), and 0.67 (0.73); and  1.0 (1.0), and 0.91 (0.91) in the same r bins. The main source of confusion appears to be between galaxies and low-redshift QSOs. We argue that these two classes are inherently physically mixed, and we provide some examples with SDSS spectra in which the host galaxy of the QSO contributes non-negligibly to the total observed light, showing its dual nature. In these cases, the classifiers mostly yield QSO and galaxy as the two preferred classes. Nevertheless, the SDSS test samples is a relatively small set, and it is only representative of objects brighter than 22.5. Therefore, the results should be treated with caution. The actual performance for fainter objects is still unknown, and the estimates we provided are based on our current physical knowledge of QSO, galaxies, and stars, together with our capability to generate simulated J spectra that mimic miniJPAS observations in the best possible way. 
\par A direct comparison of the results presented in this paper and other works in the literature is challenging because the data are divers and no standardised procedure is available to evaluate the algorithm performance. The capability of distinguishing between QSO, galaxies, and stars clearly depends on many factors, such as the S/N of the observations, the photometric filters, the wavelength coverage, and whether images or integrated photometry are used. For instance, \cite{2020A&A...633A.154L} used a clustering algorithm to find star, galaxy, and QSO clusters in a multi-dimensional colour space that included SDSS broad band photometry, Y, H, J, K bands, and WISE mid-infrared observations, together with spatial information such as the half-light radius (HLR). As expected, the performance is a function of the brightness of the source. For bright sources ($r < 21$), we obtain a similar performance (f1 score $\sim 0.9$). Nevertheless, there are not enough objects at magnitudes fainter that $21$ for which we could find significant differences. Although the spectral resolution is lower than ours, the spatial information encoded in the HLR and/or the use of infrared colours might improve their classification. \cite{2021MNRAS.508.2039H} instead used SDSS images in the u, g, i, and z bands and obtained an f1 score of $0.96$, $0.97$, and $0.99$ in quasars, stars, and galaxies, respectively, for a magnitude brighter than $20$. Once again, these results are very similar to ours, but they do not classify faint sources. Nevertheless, the use of images or channels to train convolutions neural networks is very promising as the algorithm can find more complex spatial patterns in the data that might be useful to improve the classification of sources. This approach might be implemented in the future as soon as J-PAS data are available. Finally, it is important to recall that our models were trained with a sample of galaxies and stars that is more likely to be confused with quasars, while the works mentioned above included all type of sources.
\par In the last section of this paper, we estimated the number of QSOs, galaxies, and stars in the miniJPAS observations, and we showed that our predictions are compatible with previous estimates as well as with \texttt{SExtractor}, which separates between point-like and extended sources. The algorithms presented in this work are part of a combined algorithm that unifies the outcomes of several classifiers (P\'erez-R\`afols et al. in prep. b).  
In the future, we will provide a QSO target list for a spectroscopic follow-up with the WEAVE survey \citep{2022arXiv221203981J}. This will give us valuable information about the strengths and weakness of our classifiers. WEAVE and the J-PAS collaboration will enter a feedback phase, where the knowledge acquired by one survey is to be transferred to the other in an interactive process. A natural extension of this work when enough spectroscopically confirmed sources are available is to use transfer learning and retraining the algorithms with observations in order to capture the structure of J-PAS data better. 
\section*{Acknowledgements}
This paper has gone through internal review by the J-PAS collaboration. G.M.S., R.G.D., R.G.B., L.A.D.G., and J.R.M. acknowledge financial support from the State Agency for Research of the Spanish MCIU through the "Center of Excellence Severo Ochoa" award to the Instituto de Astrof\'\i sica de Andaluc\'\i a (SEV-2017-0709), and to the AYA2016-77846-P and PID2019-109067-GB100. C.Q. acknowledges financial support from the Brazilian funding agencies FAPESP (grants 2015/11442-0 and 2019/06766- 1) and Coordenação de Aperfeiçoamento de Pessoal de Nível Su- perior (Capes) – Finance Code 001. I.P.R. was suported by funding from the European Union's Horizon 2020 research and innovation programme under the Marie Sklodowskja-Curie grant agreement No. 754510. M.P.P. and S.S.M. were supported by the Programme National de Cosmologie et Galax- ies (PNCG) of CNRS/INSU with INP and IN2P3, co-funded by CEA and CNES, the A*MIDEX project (ANR-11-IDEX-0001- 02) funded by the “Investissements d’Avenir” French Government program, managed by the French National Research Agency (ANR), and by ANR under contract ANR-14-ACHN-0021. N.R. acknowledges financial support from CAPES. R.A. was supported by CNPq and FAPESP. J.C.M. and S.B. acknowledge financial support from Spanish Ministry of Science, Innovation, and Universities through the project PGC2018-097585-B-C22. JAFO acknowledges the financial support from the Spanish Ministry of Science and Innovation and the European Union -- NextGenerationEU
through the Recovery and Resilience Facility project ICTS-MRR-2021-03-CEFCA. Based on observations made with the JST250 telescope and PathFinder camera for the miniJPAS project at the Observatorio Astrof\'isico de Javalambre (OAJ), in Teruel, owned, managed, and operated by the Centro de Estudios de Física del Cosmos de Arag\'on (CEFCA). We acknowledge the OAJ Data Processing and Archiving Unit (UPAD) for reducing and calibrating the OAJ data used in this work. Funding for OAJ, UPAD, and CEFCA has been provided by the Governments of Spain and Arag\'on through the Fondo de Inversiones de Teruel; the Aragonese Government through the Research Groups E96, E103, E16$\_$17R, and E16$\_$20R; the Spanish Ministry of Science, Innovation and Universities (MCIU/AEI/FEDER, UE) with grant PGC2018-097585-B-C21; the Spanish Ministry of Economy and Competitiveness (MINECO/FEDER, UE) under AYA2015-66211- C2-1-P, AYA2015-66211-C2-2, AYA2012-30789, and ICTS-2009- 14; and European FEDER funding (FCDD10-4E-867, FCDD13- 4E-2685). Funding for the J-PAS Project has also been provided by the Brazilian agencies FINEP, FAPESP, FAPERJ and by the National Observatory of Brazil with additional funding provided by the Tartu Observatory and by the J-PAS Chinese Astronomical Consortium. We also thank the anonymous referee for many useful comments and suggestions.

\bibliographystyle{aa}
\bibliography{aa}

\appendix

\section{Confusion matrices and mock test sample}\label{app:CM}
\begin{figure*}
    \centering
        \includegraphics[width=6cm,height=6.3cm]{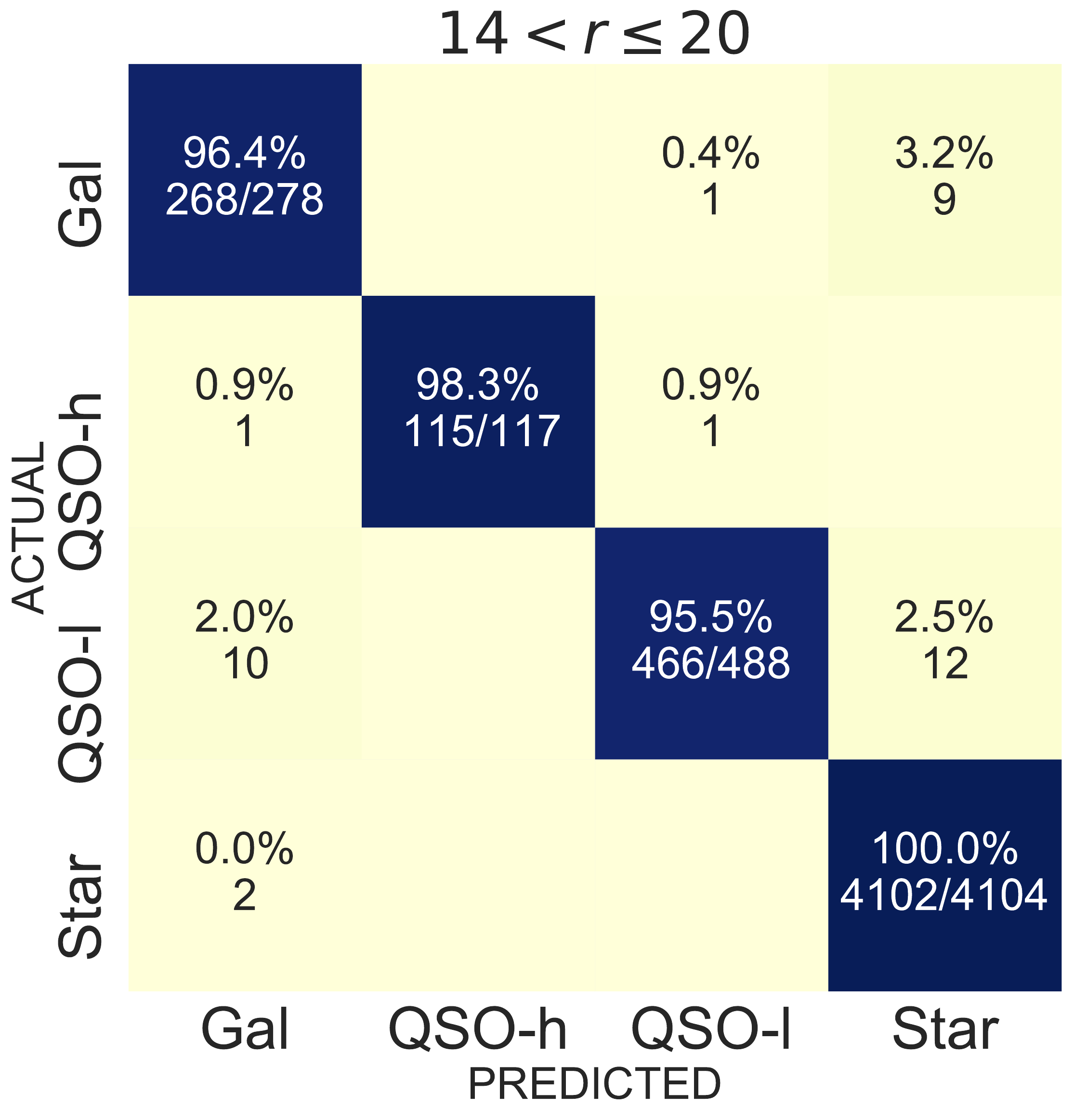}
        \includegraphics[width=5.6cm,height=6.3cm]{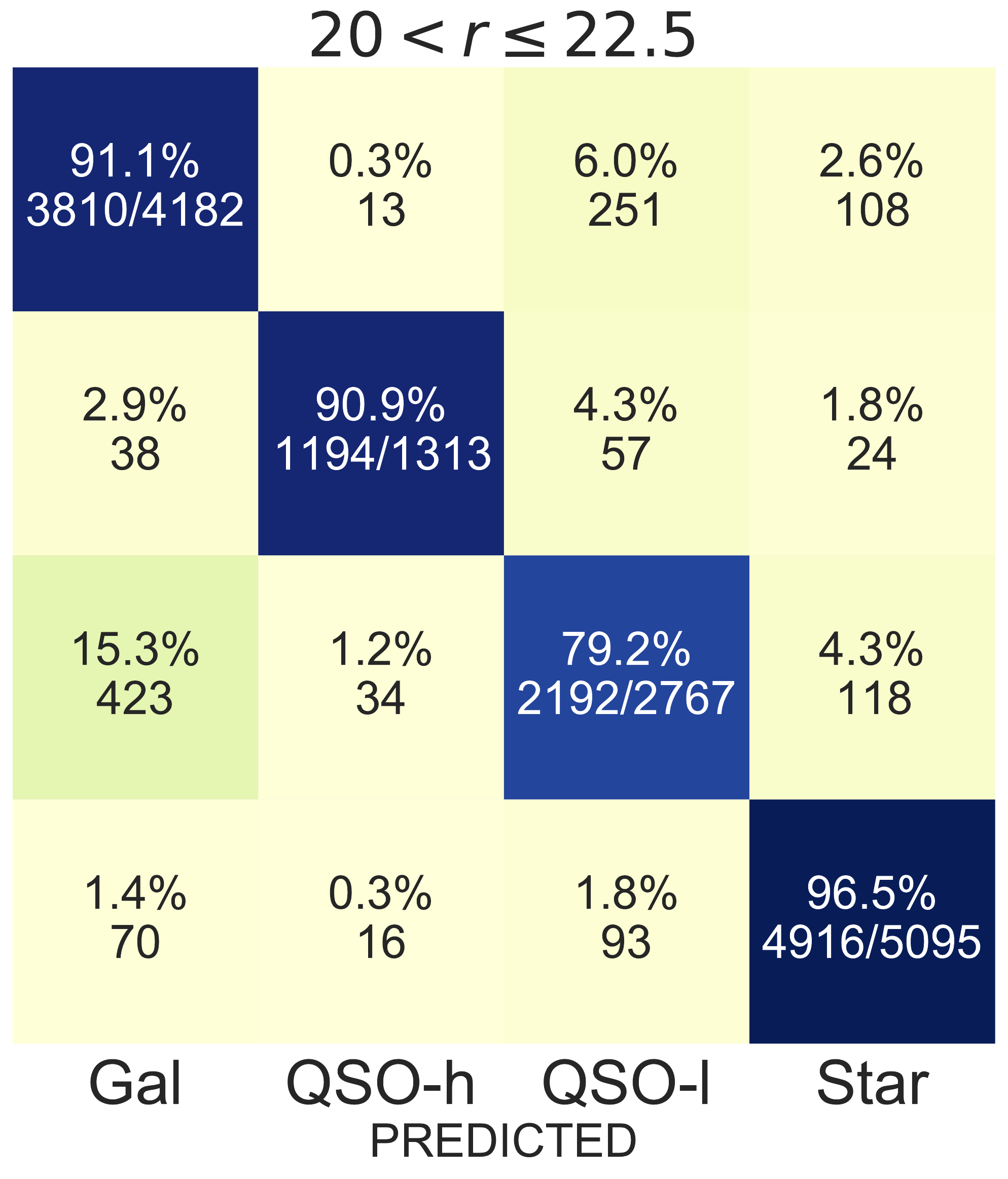}
        \includegraphics[width=5.6cm,height=6.3cm]{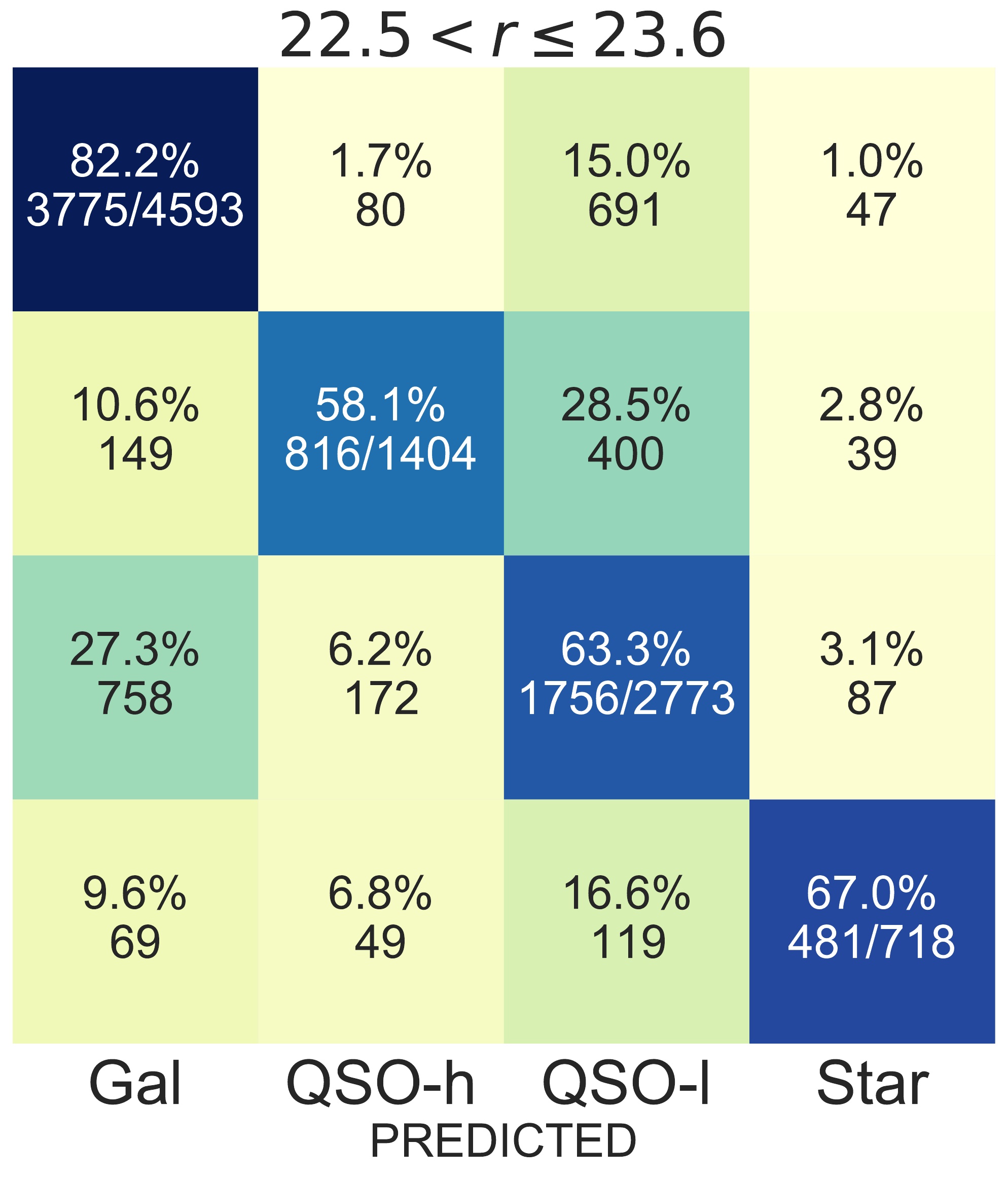}
        \caption{{Confusion matrices obtained with the ANN$_{1}$ mix in the test sample}.}
\label{fig:CM_ANN1mix_mock}
\end{figure*}

\begin{figure*}
    \centering
        \includegraphics[width=6cm,height=6.3cm]{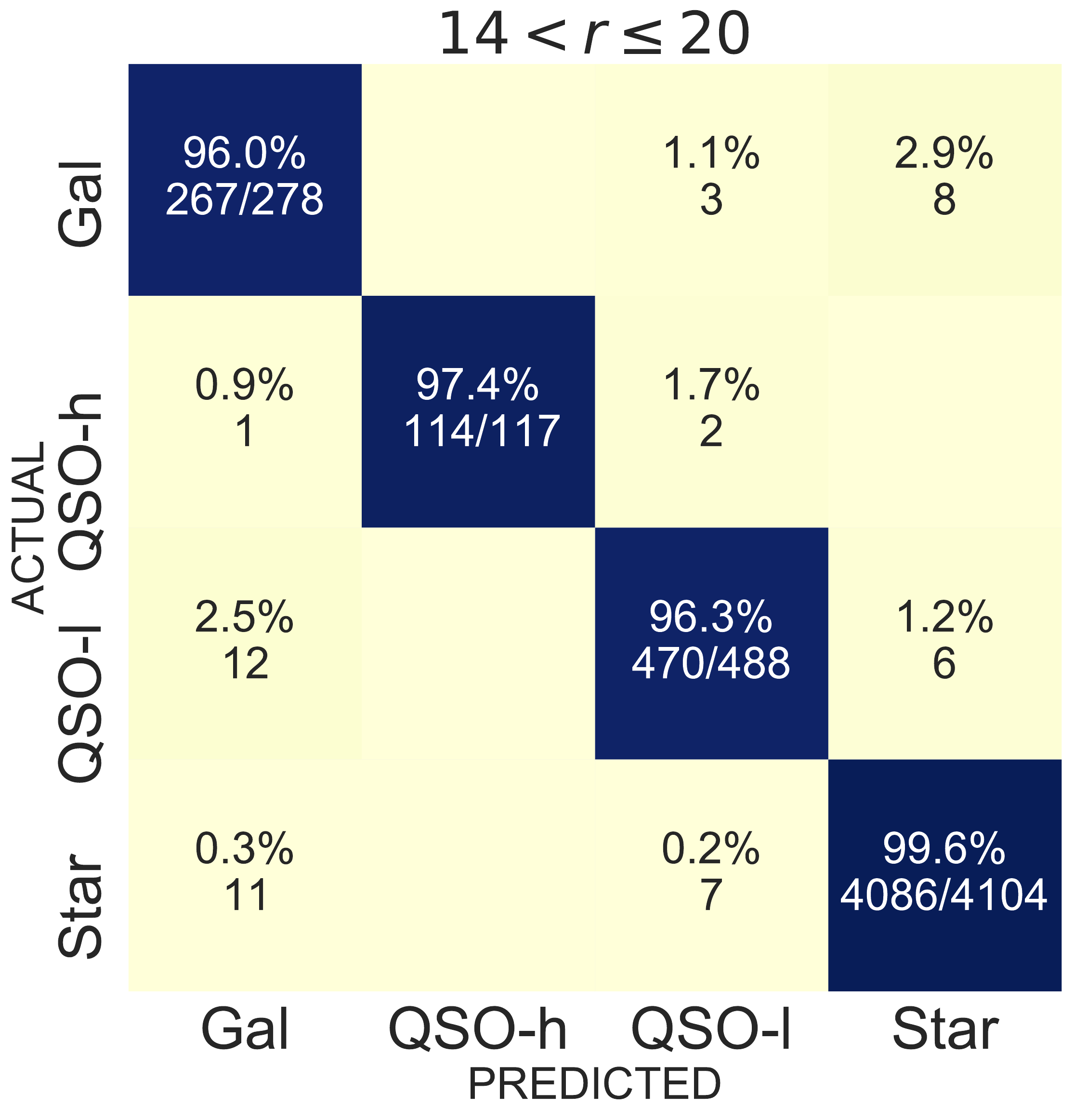}
        \includegraphics[width=5.6cm,height=6.3cm]{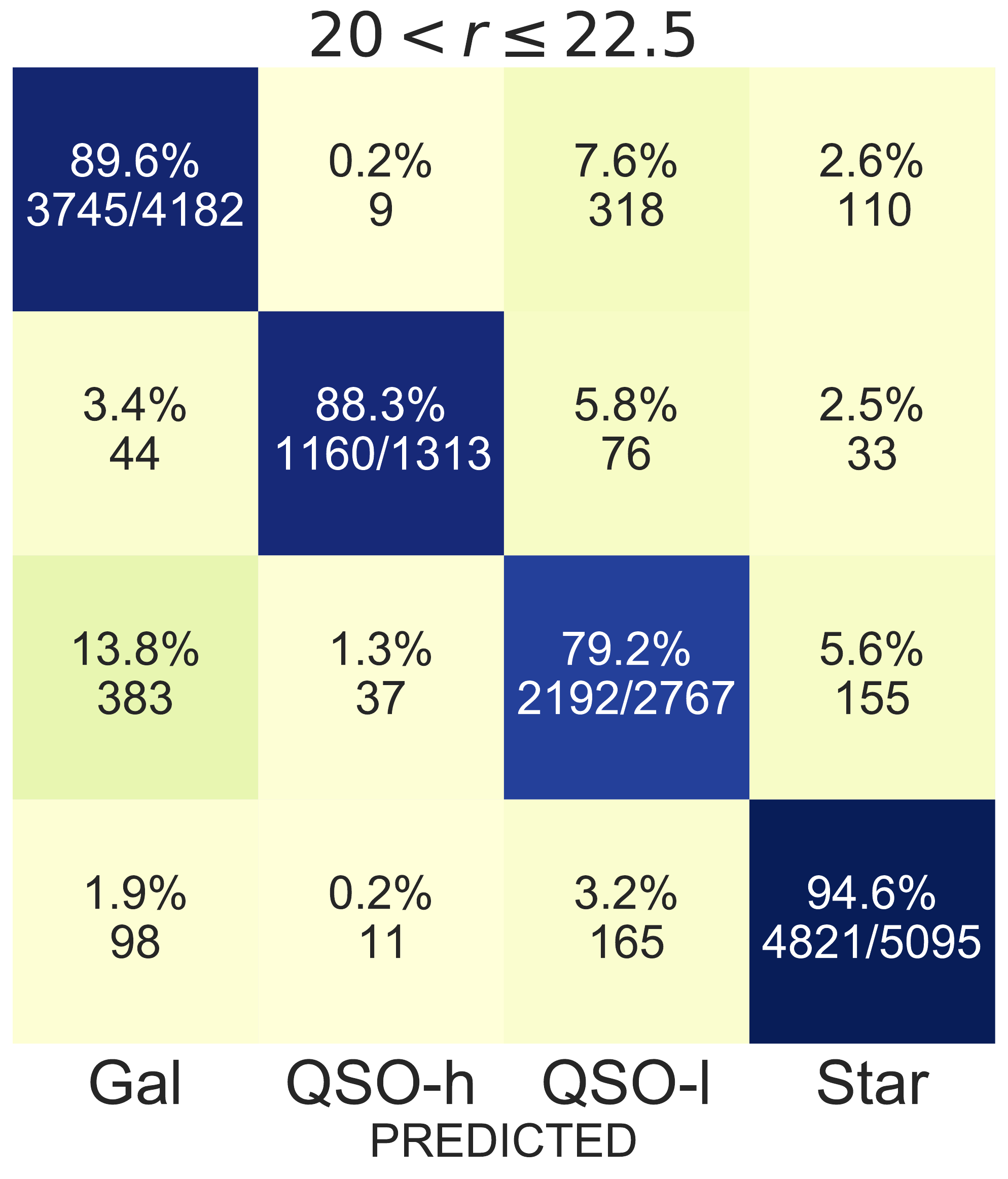}
        \includegraphics[width=5.6cm,height=6.3cm]{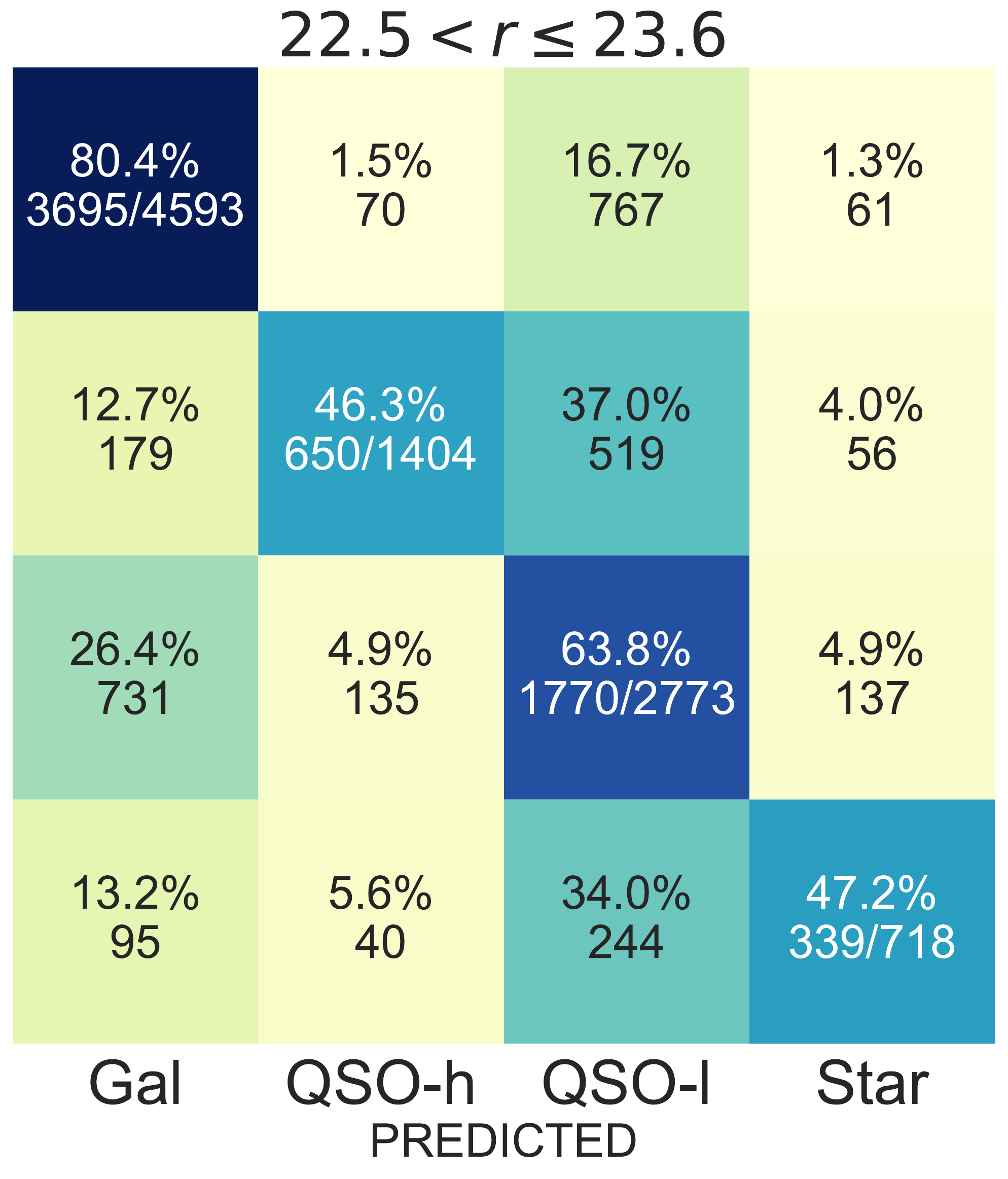}
        \caption{{Confusion matrices obtained with the ANN$_{2}$ in the test sample}.}
\label{fig:CM_ANN2_mock}
\end{figure*}

\begin{figure*}
    \centering
        \includegraphics[width=6cm,height=6.3cm]{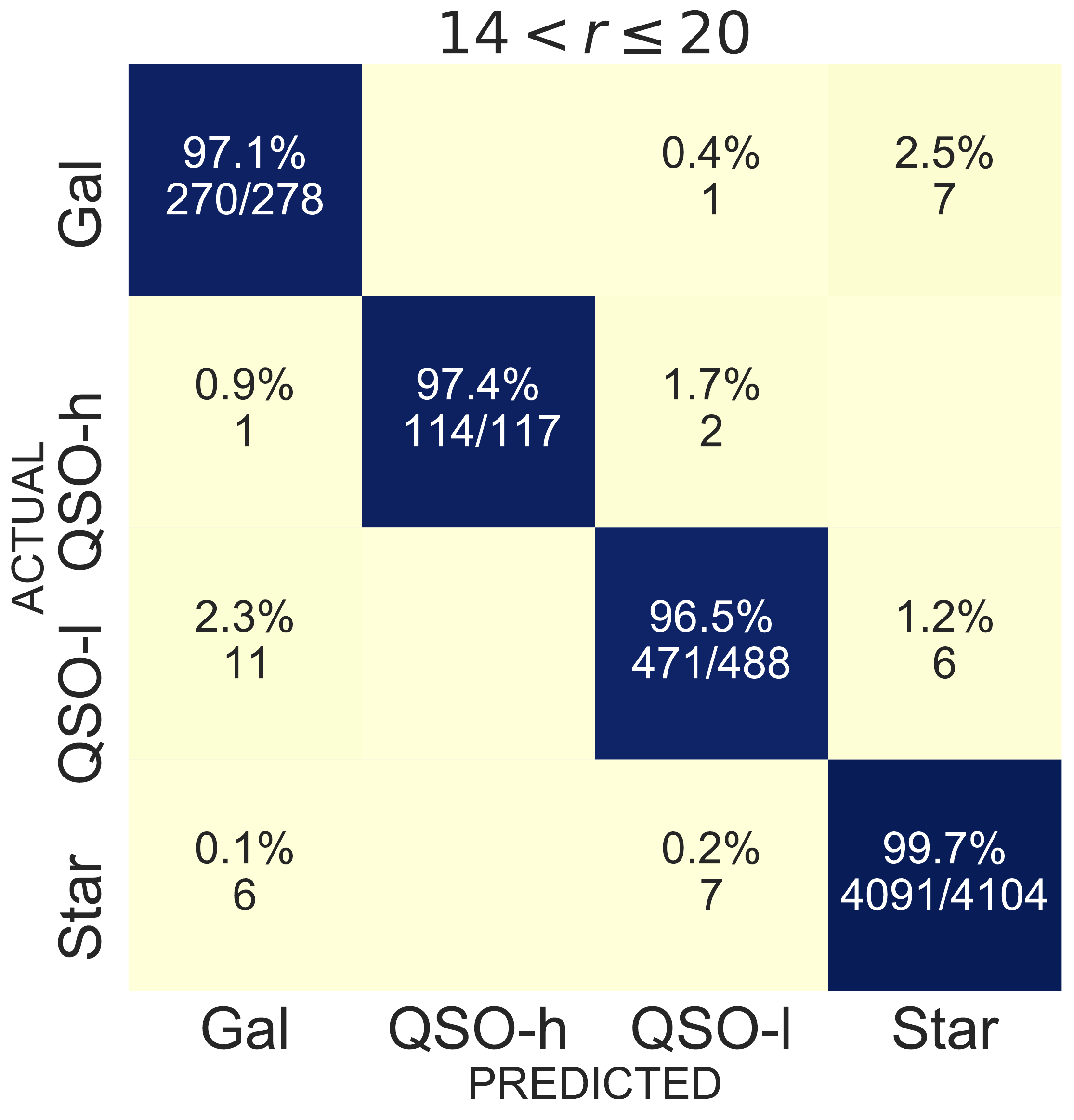}
        \includegraphics[width=5.6cm,height=6.3cm]{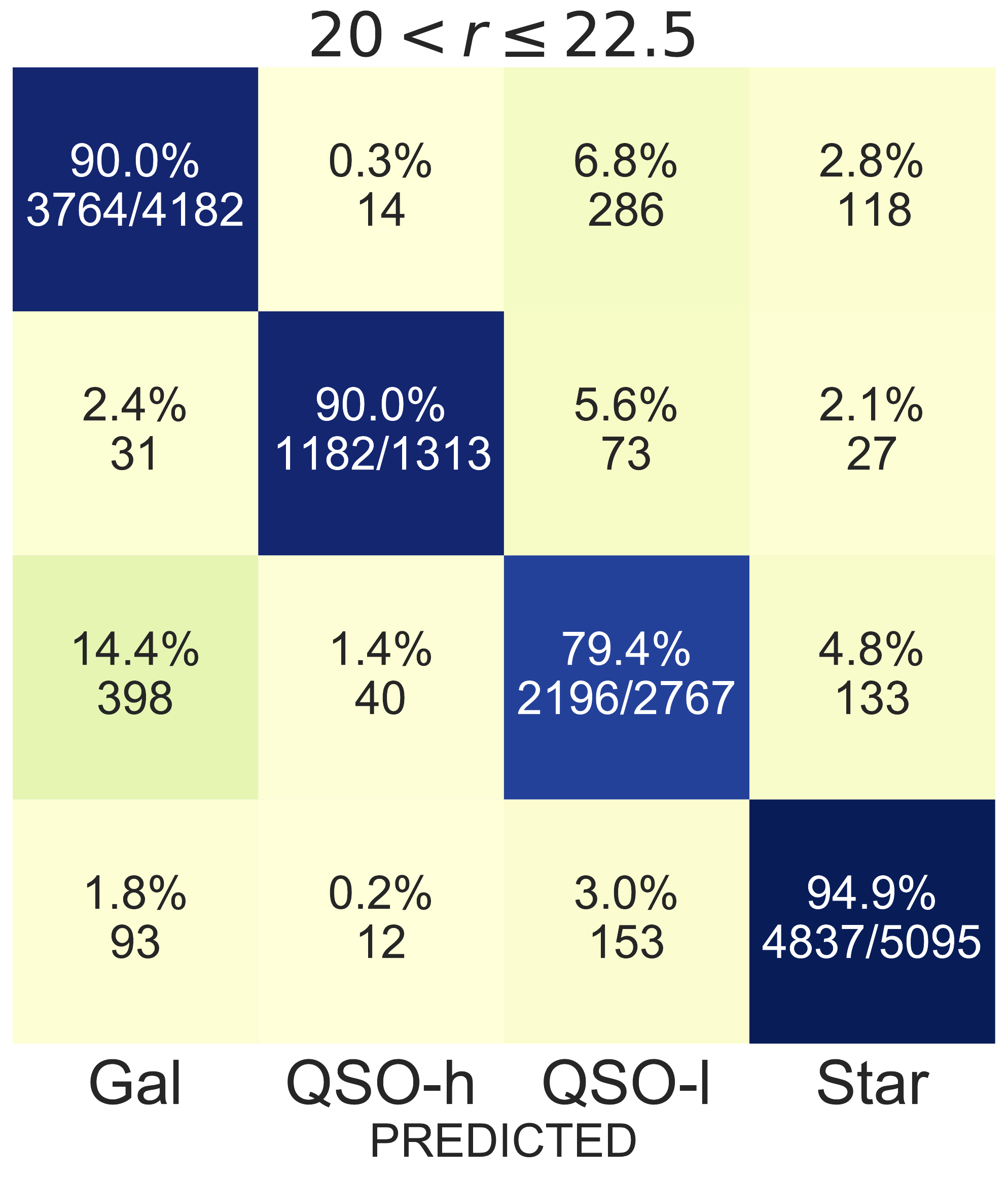}
        \includegraphics[width=5.6cm,height=6.3cm]{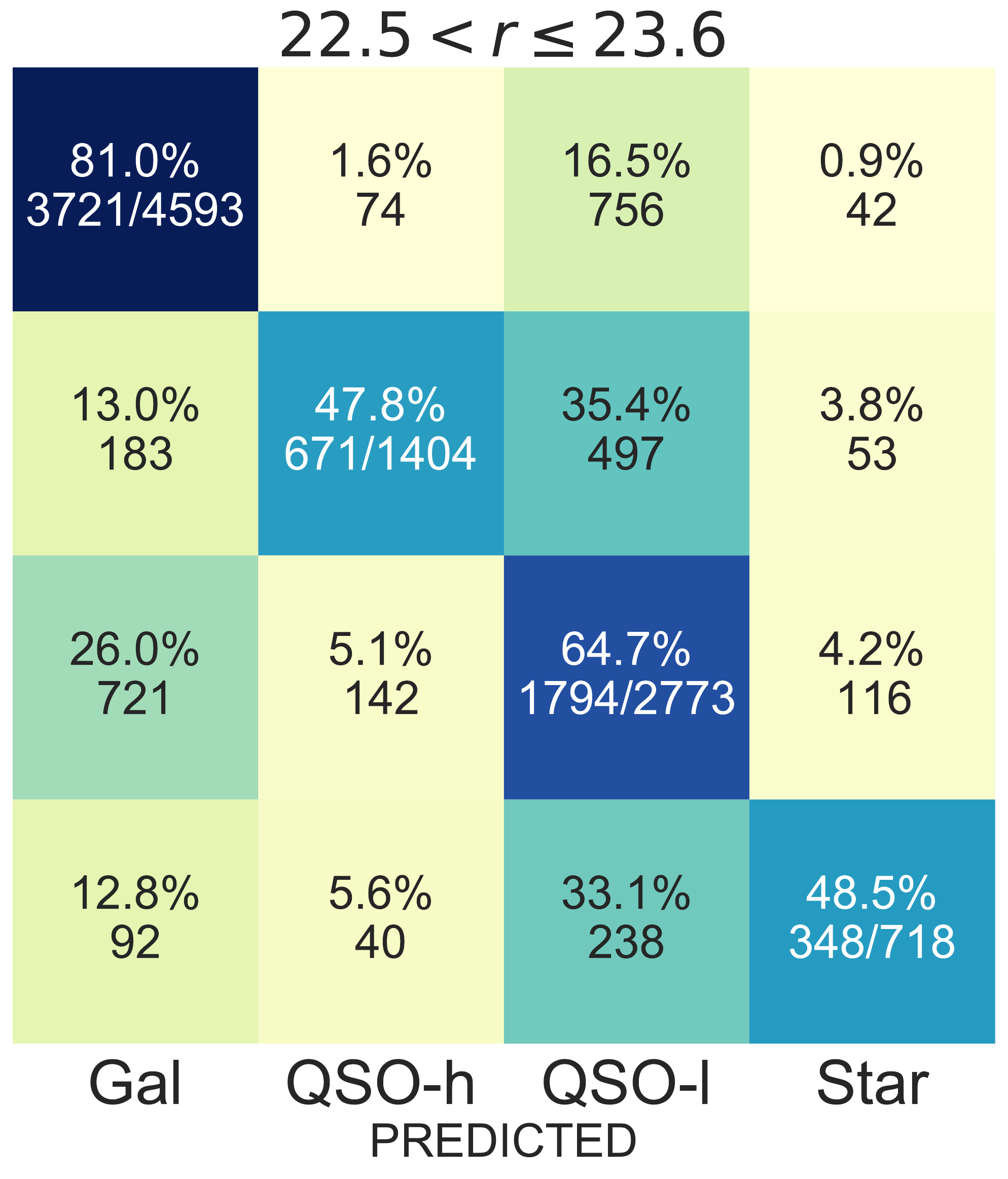}
        \caption{{Confusion matrices obtained with the ANN$_{2}$ mix in the test sample}.}
\label{fig:CM_ANN2mix_mock}
\end{figure*}

In this section, we show the confusion matrices obtained in the test sample with the ANN$_{1}$ mix (\ref{fig:CM_ANN1mix_mock}), ANN$_{2}$ (\ref{fig:CM_ANN2_mock}), and ANN$_{2}$ mix (\ref{fig:CM_ANN2mix_mock}). 

\section{Confusion matrices and SDSS test sample}\label{app:CM_SDSS}
\begin{figure}
    \centering
        \includegraphics[width=\hsize]{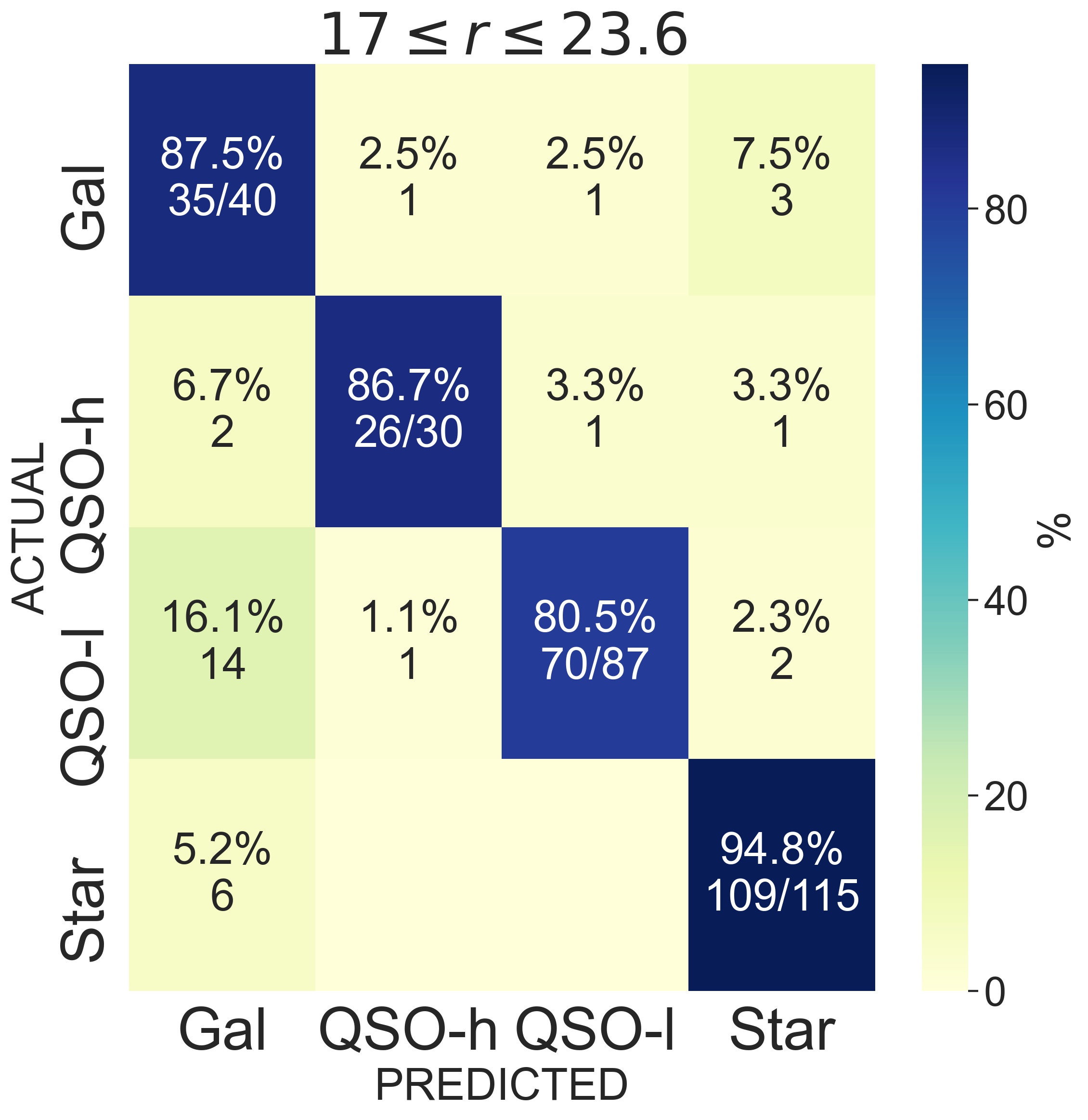}
        \caption{{Confusion matrix obtained with ANN1 mix in the SDSS test sample.}}
    \label{fig:CM_SDSS_ANN1mix}
\end{figure}

\begin{figure}
    \centering
        \includegraphics[width=\hsize]{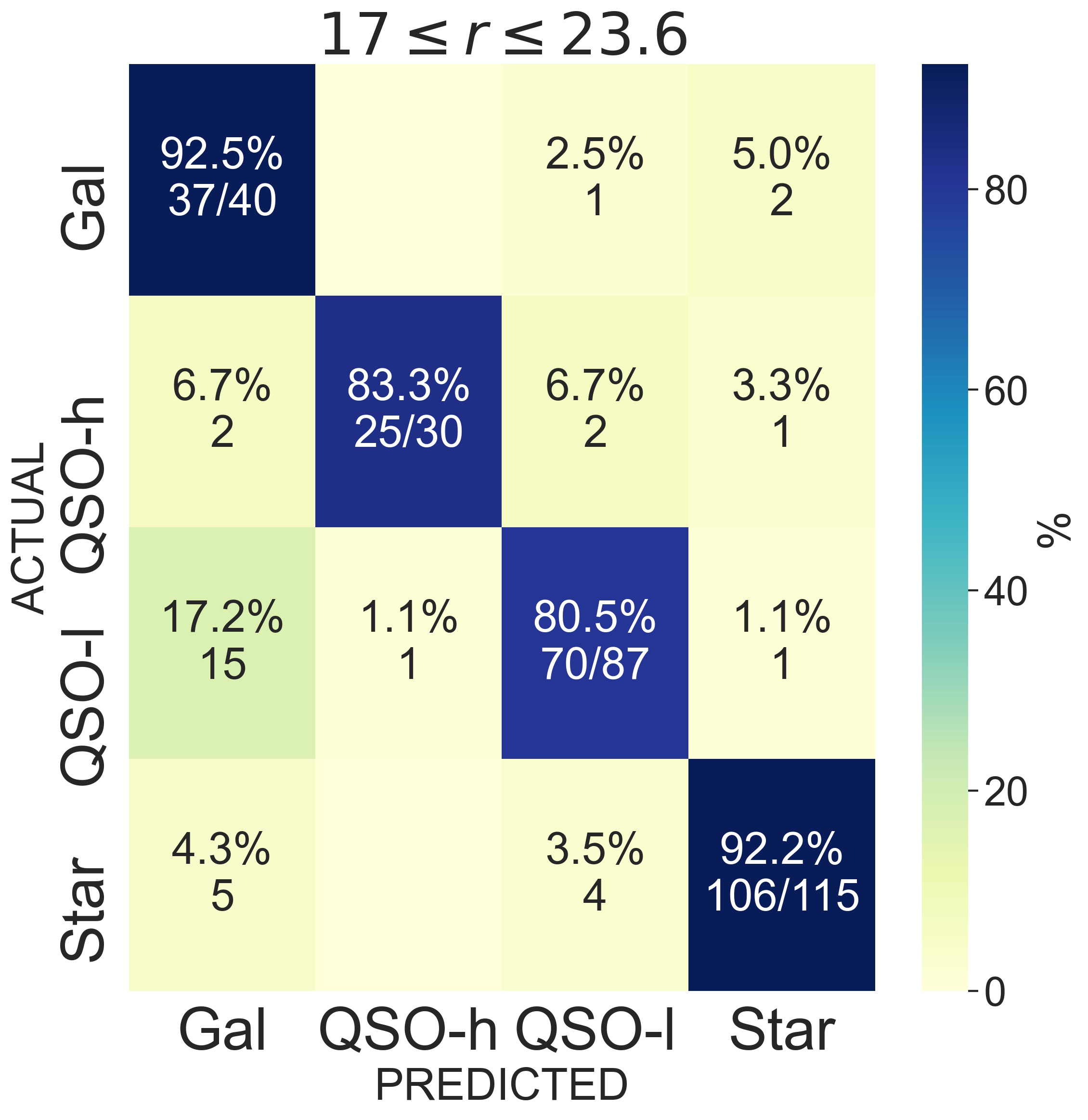}
        \caption{{Confusion matrix obtained with ANN2 in the SDSS test sample.}}
    \label{fig:CM_SDSS_ANN2}
\end{figure}

\begin{figure}
    \centering
        \includegraphics[width=\hsize]{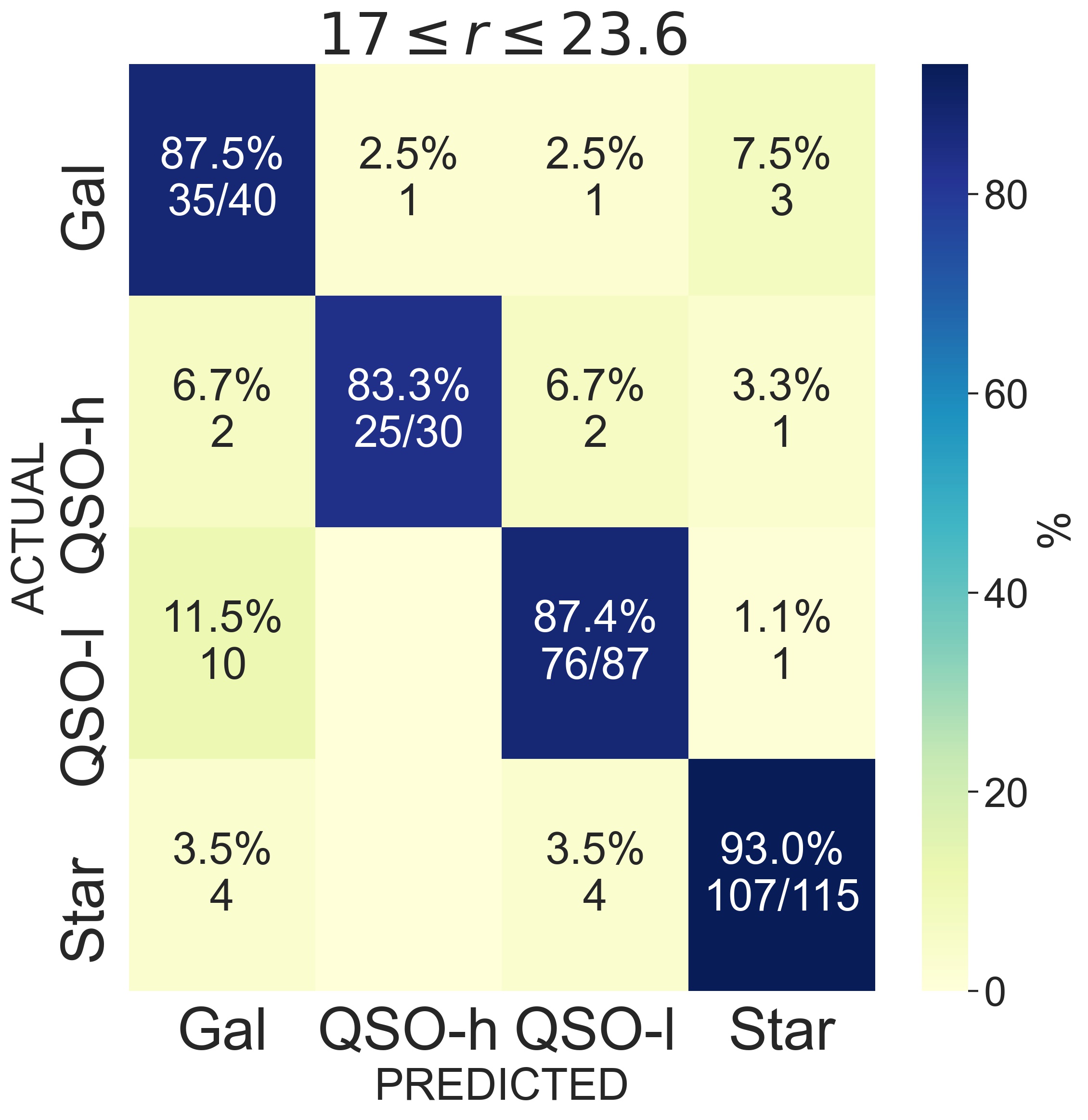}
        \caption{{Confusion matrix obtained with ANN2 mix in the SDSS test sample.}}
    \label{fig:CM_SDSS_ANN2mix}
\end{figure}

In this section, we show the confusion matrices obtained in the SDSS test sample with the ANN$_{1}$ mix (\ref{fig:CM_SDSS_ANN1mix}), ANN$_{2}$ (\ref{fig:CM_SDSS_ANN2}), and ANN$_{2}$ mix (\ref{fig:CM_SDSS_ANN2mix}). 


\end{document}